\documentclass[12pt]{article}
 \pdfoutput=1
\usepackage[table]{xcolor}
\usepackage{amsmath,amssymb,exscale}
\usepackage{graphicx}
\usepackage{epsfig}
\usepackage{multicol}
\usepackage{color}
\usepackage{mathrsfs}
\usepackage{blindtext}
 \usepackage{fancyhdr}
\usepackage{hyperref}
\usepackage{cite}
\usepackage{mathtools}
\usepackage{amsmath}
\usepackage{rotating,slashed,amsmath,charter,xcolor,catchfilebetweentags,ifluatex}

\usepackage{graphicx}
\usepackage{sidecap}

\usepackage[latin1]{inputenc} 
\usepackage{multicol}  
 
 \usepackage{titlesec}
 
 \usepackage{rotating,slashed,xcolor,amsfonts,expdlist,charter}

\numberwithin{equation}{section}

\usepackage{xcolor}
\usepackage{sectsty}

\usepackage{mdframed}
\usepackage{titletoc}

\definecolor{secnum}{RGB}{13,151,225}
\definecolor{ptcbackground}{RGB}{212,237,252}
\definecolor{ptctitle}{RGB}{0,177,235}

\titlecontents{lsection}
  [5.8em]{\sffamily}
  {\color{secnum}\contentslabel{2.3em}\normalcolor}{}
  {\titlerule*[1000pc]{.}\contentspage\\\hspace*{-5.8em}\vspace*{5pt}%
    \color{white}\rule{\dimexpr\textwidth-15.5pt\relax}{1pt}}

\usepackage{hyperref}
\hypersetup{colorlinks,bookmarksopen,bookmarksnumbered,citecolor=blus,
linkcolor=redy,pdfstartview=FitH,urlcolor=blus}
\usepackage{slashed}

\definecolor{blus}{cmyk}{1,0.9,0,0.1}
\definecolor{verdes}{cmyk}{0.99,0,0.59,0.65}
\definecolor{rossos}{cmyk}{0,1,1,0.55}
\definecolor{redy}{cmyk}{0,1,1,0.7}
\definecolor{greeny}{cmyk}{0.99,0,0.59,0.98}
\definecolor{green-go}{cmyk}{0.79,0,0.59,0.5}

\usepackage{titlesec}

\def\Lag{\mathscr{L}}

\newcommand{\beq}{\begin{equation}}
\newcommand{\eeq}{\end{equation}}

\def\hhref#1{\href{http://arxiv.org/abs/#1}{arXiv:#1}} 

 \def\Lag{\mathscr{L}}
 
\newcommand{\tmtextbf}[1]{{\bfseries{#1}}}
\newcommand{\tmtextrm}[1]{{\rmfamily{#1}}}
\def\bp{M_{P}}

\def\be{\begin{equation}}
\def\ee{\end{equation}}
\def\ba{\begin{array} }

\def\bac{\begin{array} {c}}
\def\bacc{\begin{array} {cc}}
\def\baccc{\begin{array} {ccc}}
\def\bacccc{\begin{array} {cccc}}
\def\ea{\end{array}}
\def\bea{\begin{eqnarray}}
\def\eea{\end{eqnarray}}

\definecolor{red}{rgb}{1,0,0}

\def\psl{\hbox{\hbox{${p}$}}\kern-1.9mm{\hbox{${/}$}}}
\def\dsl{\hbox{\hbox{${\partial}$}}\kern-2.2mm{\hbox{${/}$}}}
\def\Dsl{\hbox{\hbox{${D}$}}\kern-2.6mm{\hbox{${/}$}}}

\def\Lag{\mathscr{L}}

\newcommand{\gappeq}{{\rlap{{\raise}.5ex\text{\ensuremath{>}}}{{\lower}.5ex\text{\ensuremath{\sim}}}}}
\newcommand{\lappeq}{{\rlap{{\raise}.5ex\text{\ensuremath{<}}}{{\lower}.5ex\text{\ensuremath{\sim}}}}} 
\newcommand{\I}{\tmtextrm{1{\kern}-.24em l}}

\begin{document}
\topmargin -1.0cm
\oddsidemargin 0.9cm
\evensidemargin -0.5cm

{\vspace{-1cm}}
\begin{center}

\vspace{-1cm}


 {\huge \tmtextbf{ 
\color{rossos} (Multi-field) Natural Inflation and Gravitational Waves}} {\vspace{.5cm}}\\
 

\vspace{1.9cm}

{\large  {\bf Alberto Salvio} and {\bf Simone Sciusco}
{\em  

\vspace{.4cm}

 Physics Department, University of Rome Tor Vergata, \\ 
via della Ricerca Scientifica, I-00133 Rome, Italy\\

\vspace{0.6cm}

I. N. F. N. -  Rome Tor Vergata,\\
via della Ricerca Scientifica, I-00133 Rome, Italy\\ 

\vspace{0.4cm}

\vspace{0.2cm}

 \vspace{0.5cm}
}

\vspace{0.2cm}

}
\vspace{0.cm}

%
%
 %
%
%

\end{center}

%
 
\noindent --------------------------------------------------------------------------------------------------------------------------------

\begin{center}
{\bf \large Abstract}
\end{center}
We provide a detailed study of natural inflation with a periodic non-minimal coupling, which is a well-motivated inflationary model that admits an explicit UV completion. We demonstrate that this construction can satisfy the most recent observational constraints from Planck and the BICEP/Keck collaborations. We also compute the corresponding relic gravitational wave background  due to tensor perturbations and show that future space-borne interferometers, such as DECIGO, BBO and ALIA, may be able to detect it. Next, we  extend this analysis and establish the validity of these results in a multi-field model featuring an additional $R^2$  term in the action, which allows us to interpolate between natural and scalaron (a.k.a.~Starobinsky) inflation. We investigate the conditions under which the aforementioned future interferometers will have the capability to differentiate between pure natural inflation and natural-scalaron inflation. The latter analysis could open the door to distinguishing between single-field and multi-field inflation through gravitational wave observations in more general contexts.

\noindent

  \vspace{0.4cm}

\noindent --------------------------------------------------------------------------------------------------------------------------------

\vspace{1.1cm}

\vspace{2cm}

\tableofcontents

\vspace{0.5cm}

\section{Introduction}\label{Introduction}

 Gravitational Waves (GWs) are ripples in the fabric of space-time predicted by Einstein's theory of general relativity; they travel at
the speed of light with very weak (Planck-suppressed) interactions with matter. The observation of a GW event in 2015, known as GW150914, resulting from binary black hole mergers~\cite{Abbott:2016blz,TheLIGOScientific:2016wyq1},  marked the beginning of the era of GW astronomy. Since then, interest in this branch of physics has steadily increased, and other sources of GWs, such as binary neutron stars, have been identified~\cite{LIGOScientific:2017ync}. More recently, interest in GW astronomy has been further amplified by the evidence for a background of GWs provided by pulsar timing arrays, including the North American Nanohertz Observatory for Gravitational Waves (NANOGrav), the Chinese Pulsar Timing Array (CPTA), the European Pulsar Timing Array (EPTA) and the Parkes Pulsar Timing Array (PPTA)~\cite{NANOGrav:2023gor,Antoniadis:2023ott,Reardon:2023gzh,Xu:2023wog}.

There are several interesting phenomena that admit a particle physics (microscopic) description and, at the same time, can generate GWs within the reach of present or future detectors. This can be used to test the Standard Model of particle physics (SM) and its extensions. One example is given by phase transitions that admit field-theoretic descriptions both perturbatively (see~\cite{Salvio:2023qgb,Salvio:2023ynn} for a model-independent analysis) and non-perturbatively (see e.g.~\cite{Kajantie:1993ag,Kajantie:1995kf,Kajantie:1996mn,Gould:2019qek,Gould:2022ran}). If strong enough, these phenomena can generate a GW spectrum within the reach of current or future detectors. Another example is given by cosmic strings (see~\cite{Vilenkin:2000jqa} for a detailed textbook introduction).

In this paper we will focus on yet another important example: the quantum tensor fluctuations generated during inflation. Detecting the corresponding GW spectrum would offer further evidence for inflation and, notably, would provide direct observational confirmation of the quantum nature of gravity. Additionally, since inflation is often realized through quantum fields and typically involves extremely high energies, the detection of these GWs would provide valuable insights into the more fundamental theory beyond the Standard Model governing particle interactions. Some future space-borne interferometers can probe such GW spectra. An example is  the Japanese mission  DECi-hertz Interferometer Gravitational wave Observatory (DECIGO)~\cite{DECIGO} that is planned to have the best sensitivity  around the frequency $\nu\sim 0.1$~Hz. Indeed, the primary objective of DECIGO is to observe  the early universe. Another proposed space-borne detector that can probe the inflationary GW spectrum is the  Big Bang Observer (BBO)~\cite{Crowder:2005nr,BBO2},  intended as a follow-on mission to the Laser Interferometer Space Antenna (LISA).  The Advanced Laser Interferometer Antenna (ALIA)~\cite{Crowder:2005nr,Gong:2014mca} is yet another proposed space-borne interferometer with similar characteristics, albeit with a slightly lower frequency scale where the best sensitivity is achieved.

Numerous inflationary models predict an amplitude of quantum tensor fluctuations that is large enough to be within the reach of DECIGO, BBO and ALIA. For this reason we think it is important to focus on those that are best motivated from the particle-physics perspective. The SM Higgs field is a well-motivated inflaton candidate~\cite{Bezrukov:2007ep,Hamada:2014iga,Bezrukov:2014bra,Hamada:2014wna,Salvio:2018rv}  (being the sole fundamental scalar within the SM) and the comparison between its theoretical predictions for the inflationary GW spectrum and the expected sensitivities of future detectors has been performed in~\cite{Salvio:2021kya}. 

This paper focuses  on another well-motivated type of inflatons, namely a pseudo-Nambu-Goldstone boson (PNGB)~\cite{Freese:1990rb}. Indeed, Goldstone's theorem protects the mass (actually the full potential) of a Nambu-Goldstone boson (NGB) in such a way that the required potential flatness 
becomes a natural feature. Eventually, explicit symmetry-breaking terms (which transform the NGB into a PNGB and introduce a potential slope) are necessary to terminate inflation, but these terms can be small, preserving the naturalness of PNGB-driven inflation. This scenario is consequently referred to as natural inflation\footnote{PNGBs also appear frequently in beyond-the-SM constructions. A popular example is an axion(-like) particle, namely a scalar $\phi_A$ that corresponds to a  spontaneously broken approximate axial U(1) symmetry and  can feature an interaction of the form $\sim \phi_A F_{\mu\nu}\tilde F^{\mu\nu}$, where $F_{\mu\nu}$ is some gauge field strength and $\tilde F_{\mu\nu}$ its dual. A particular type of natural inflation is the axion(-like) particle driven inflation (simply known as axion inflation); for a review on axion inflation see~\cite{Pajer:2013fsa}.}.  This possibility has attracted and still attracts a lot of interest in the particle-cosmology community.  The natural inflaton has a periodic potential that can be derived from UV complete (e.g. QCD-like) field theories.

UV completions generically lead not only to a periodic potential but also to a periodic non-minimal coupling between the PNGB field and the Ricci scalar. This has been explicitly shown in Ref.~\cite{Salvio:2021lka} (see in particular Appendix A there), where a concrete UV-complete QCD-like theory was considered. On the contrary, non-periodic non-minimal couplings are not known to have a UV completion. This is unsurprising because, as a compact field variable, a PNGB is expected to exhibit periodicity in all terms in the action. Nevertheless, Refs.~\cite{AlHallak:2022gbv,Bostan:2022swq,AlHallak:2022haa,Bostan:2023ped}
 studied ``natural" inflation with a non-periodic non-minimal coupling. 
 
 One of the main purposes of this work is to perform a complete study of natural inflation with a periodic non-minimal coupling  to understand whether  the most recent cosmic microwave background (CMB) observations performed by the Planck, BICEP and Keck collaborations~\cite{Ade:2015lrj,BICEP:2021xfz}, allow for this  intriguing possibility. This requires an exhaustive study of the theory's parameter space and the corresponding inflationary predictions. In this work we will consider standard Einstein gravity, i.e.~the pure gravitational term in the action will be the Einstein-Hilbert one\footnote{For a study of natural inflation without non-minimal coupling in a modified gravity scenario see~\cite{Salvio:2019wcp,Salvio:2020axm}. For previous partial studies with older observational data see~\cite{Ferreira:2018nav}. Also, for an analysis of a variant of natural inflation (which, regrettably, results in an unacceptably small number of e-folds) see~\cite{AlHallak:2022haa}.}.
 In the possible viable region of parameter space we can then assess  whether future GW detectors, such as DECIGO, BBO and ALIA, will be able to detect the corresponding primordial spectra. Notably, this is a study that has never been attempted for natural inflation (neither with nor without non-minimal coupling). Here the question we aim to address is ``can we test (primordial) Nambu-Goldstone bosons with gravitational-wave detectors?".
 
 While the inflaton potential is protected from quantum corrections, other terms in the action, such as the pure gravitational part, are not. In any phenomenologically viable model, which must include matter fields, quantum corrections generate quadratic-in-curvature terms, with the simplest being $R^2$~\cite{Utiyama:1962sn,Salvio:2014soa,Salvio:2015kka,Salvio:2017qkx,Salvio:2021lka}. This fact has been used by Starobinsky as an inspiration to construct the first inflationary model~\cite{Starobinsky:1980te}, in which the Einstein-Hilbert action  is extended to include an $R^2$ term. Such term is equivalent to a scalar $z$, known as the scalaron, with a quasi-flat potential at large enough $z$ that naturally allows for slow-roll inflation. For this reason a well-motivated multi-field extension of natural inflation is natural-scalaron inflation~\cite{Salvio:2021lka}. 
 
 Another important aim of this work is, therefore, to extend the above-mentioned analysis of natural inflation to this well-motivated multi-field version: is natural-scalaron inflation with a periodic PNGB potential and non-minimal coupling consistent with the most recent\footnote{For a preliminary analysis using previous (no longer updated) observations, refer to~\cite{Salvio:2021lka}.} inflationary observations? In the region of the parameter space where this consistency takes place, can we detect the corresponding primordial GW spectrum with future interferometers?
 
 One reason multi-field inflationary models are interesting in this regard is that they generically predict a different frequency dependence of the GW spectra. This is due to the fact that such dependence is mainly given by the tensor spectral index $n_T$, which is related to the tensor-to-scalar ratio $r$ by the relation $n_T=-r/8$ in single-field models, while $n_T< -r/8$ in multi-field scenarios. Therefore, GW detectors could provide insights into whether multiple inflatons were actively driving inflation. In this paper we hope to pave the way for distinguishing between single-field and multi-field inflation in this manner.

 The paper is structured as follows.
\begin{itemize} 
\item  In Sec.~\ref{Natural inflation with a periodic non-minimal coupling}, after introducing natural inflation (with a periodic non-minimal coupling) and studying in detail slow-roll in this context, we will perform a complete analysis of the current  observational constraint 
 on the corresponding parameter space. Moreover, in the same section we will investigate the relic GW background due to natural inflation and the potential for future interferometers to detect such signals. 
 \item In Sec.~\ref{A multi-field natural inflation: the natural-scalaron case} we will extend the analysis of Sec.~\ref{Natural inflation with a periodic non-minimal coupling} to the natural-scalaron model. Furthermore, in Sec.~\ref{A multi-field natural inflation: the natural-scalaron case}  we will also explore the possibility of distinguishing between the considered single-field and multi-field natural  models through future GW interferometers  capable of detecting relic GW backgrounds produced during the inflationary epoch.
 \item Finally, in Sec.~\ref{Conclusions} we will offer a detailed summary of our results and the  concluding remarks.
 \end{itemize}


\section{Natural inflation with a periodic non-minimal coupling}\label{Natural inflation with a periodic non-minimal coupling}

We start with the single-field natural inflation featuring a periodic non-minimal coupling.
\subsection{The model}\label{The model}
The part of the action responsible for natural inflation is
\begin{equation}
    S_\mathrm{infl} = \int d^4x\sqrt{-g}\left[ M_P^2\frac{F(\phi)}{2}R \,-\, \frac{1}{2}(\partial \phi)^2 \,-\, V(\phi) \right]
    \label{eq:action}
\end{equation}
where $M_P$ is the reduced Planck mass, $\phi$ is the inflaton PNGB field, $(\partial \phi)^2\equiv g^{\mu\nu}\partial_\mu\phi\,\partial_\nu\phi$,
\begin{equation}
    F(\phi) \equiv 1 + \alpha\, \left[ 1 + \cos\left( \frac{\phi}{f} \right)\right]
    \label{eq:F(phi)}
\end{equation}
is the non-minimal coupling, and
\begin{equation}
    V(\phi) \equiv \Lambda^4 \left[ 1 + \cos\left(\frac{\phi}{f}\right) \right]+\Lambda_\mathrm{cc}
    \label{eq:V(phi)}
\end{equation}
is the natural-inflaton potential~\cite{Freese:1990rb}. A microscopic origin of both these functions in terms of a fundamental QCD-like field theory has been provided in Ref.~\cite{Salvio:2021lka}.
Here $\Lambda$ and $f$ are two energy scales, $\alpha$ is a real parameter that must satisfy $\alpha>-1/2$ in order for the effective Planck mass to be real for all $\phi$, i.e.~$M_{P,\mathrm{eff}}^2\equiv M_P^2F(\phi)>0$. The constant $\Lambda_{\rm cc}$ accounts for 
the (tiny and positive) cosmological constant responsible for the observed dark energy and is negligible during inflation, which occurs at a much larger energy scale. The functions $F(\phi)$ and $V(\phi)$ are both even and periodic with period $2\pi f$, so we can restrict ourselves to the interval
\begin{equation}
    \phi \in [0,\pi f].    
\end{equation}
This model depends on three dimensionless parameters: $\alpha, f/M_P, \Lambda/M_P$.

\begin{figure}[!t]
    \centering
    \includegraphics[scale=0.34]{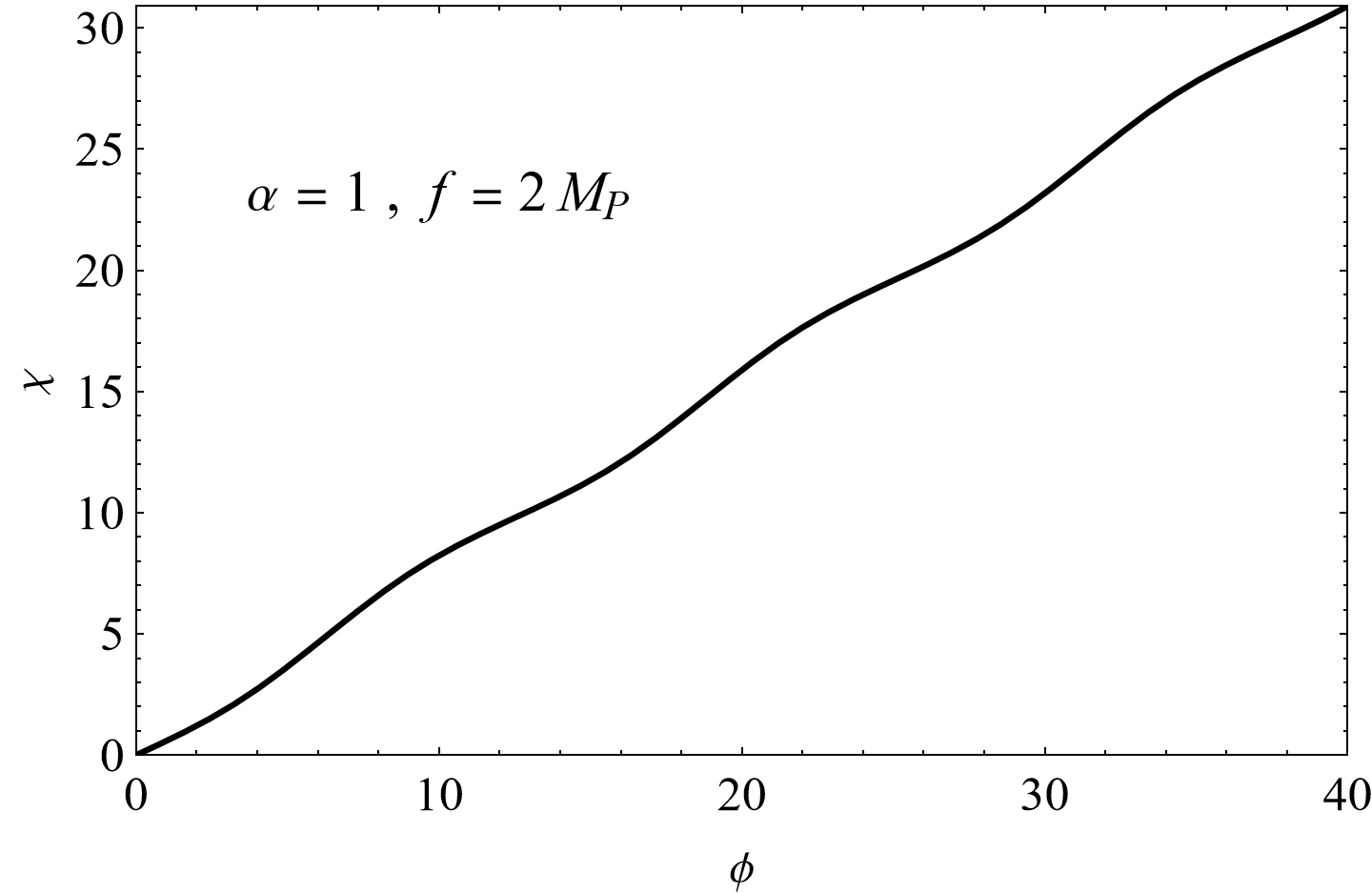}
\hspace{1cm}
\includegraphics[scale=0.31]{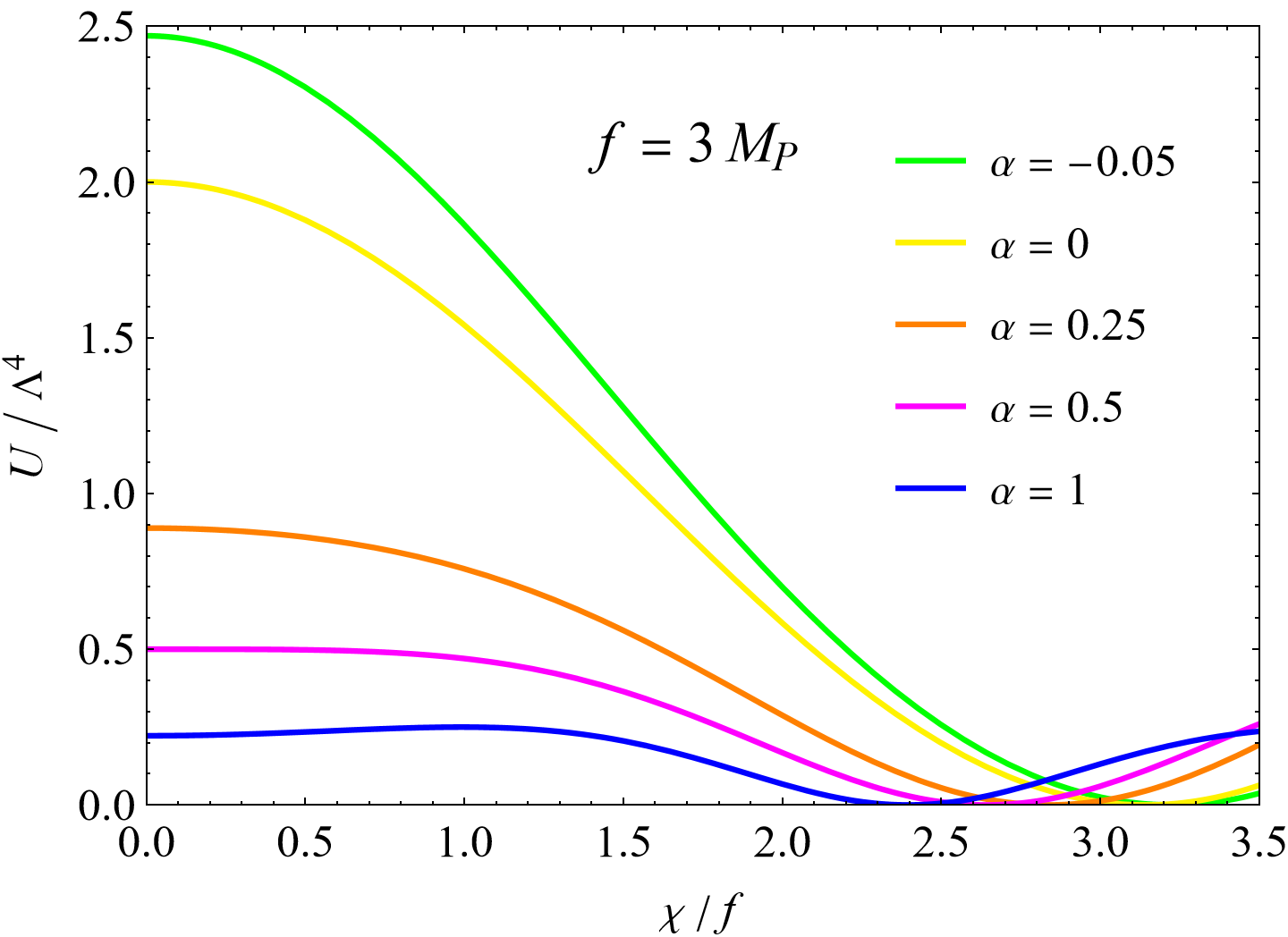}
    \caption{\textbf{Left}: {\em the new canonical field $\chi(\phi)$ for $\alpha=1$ and $f=2M_P$.} \textbf{Right}: {\em the normalized effective potential $U/\Lambda^4$ as a function of $\chi/f$ for some values of $\alpha$ and $f=3M_P$.}}
    \label{fig:Chi.norm.U}
\end{figure}

The first term of Eq.~(\ref{eq:F(phi)}) gives us the ordinary Einstein-Hilbert action, while the second, proportional to $\alpha$, provides the non-minimal coupling between the PNGB and gravity. In order to use the standard formul\ae~for single-field slow-roll inflation we need to move to the Einstein frame (where  we have a canonical Einstein gravitational term $M_P^2R/2$ instead of the non-canonical one $M_P^2F(\phi)R/2$). To do so, we perform a conformal transformation of the metric:
\begin{equation}
    g_{\mu\nu} \to \frac{g_{\mu\nu}}{F(\phi)}.
\end{equation}
With this transformation we obtain  a canonical Einstein gravitational term, but also a non-canonical contribution to the scalar kinetic term 
\begin{equation}
    S_\mathrm{infl} = \int d^4x\sqrt{-g}\, \left\{\,
    \frac{1}{2}M_P^2R \,-\, 
    \frac{1}{2}
    K(\phi)
    (\partial \phi)^2
    \,-\,
    \frac{V(\phi)}{F^2(\phi)} \right\},
    \label{eq:Action.Einstein.Frame}
\end{equation}
where
\begin{equation}
    K(\phi) \equiv \frac{2F\,+\,3M_P^2F'^2}{2F^2}
    \label{eq:K(phi)}
\end{equation}
and $F'(\phi)\equiv dF/d\phi$. The scalar kinetic term can be brought into a canonical form through a field redefinition $\phi\to \chi$ with the property 
\begin{equation}
    \frac{d\chi}{d\phi} \,\equiv\, \sqrt{K(\phi)}, \qquad \chi(\phi=0) = 0,
    \label{eq:define.Chi}
\end{equation}
which leads to
\begin{equation}
    S_\mathrm{infl} = \int d^4x\sqrt{-g}\, \left\{\,
    \frac{1}{2}M_P^2R \,-\, 
    \frac{1}{2}
   g^{\mu\nu}\partial_\mu\chi\partial_\nu\chi
    \,-\,
    U(\chi) \right\},
\end{equation}
where we have defined the effective potential of $\chi$ as 
\begin{equation}
    U(\chi) \,\equiv\, \frac{V(\phi(\chi))}{F^2(\phi(\chi))}.
    \label{eq:U(chi)}
\end{equation}
It is generically very difficult, given a non-minimal coupling $F(\phi)$ and inflaton potential $V(\phi)$, to derive an analytical expression for the new canonical field $\chi(\phi)$ and its inverse $\phi(\chi)$ (which is needed to write $U(\chi)$). We then proceed numerically to obtain $\chi(\phi)$, $\phi(\chi)$ and hence  $U(\chi)$ (see Fig.~\ref{fig:Chi.norm.U} for some examples).

\begin{figure}[!t]
    \centering
    \includegraphics[scale=0.37]{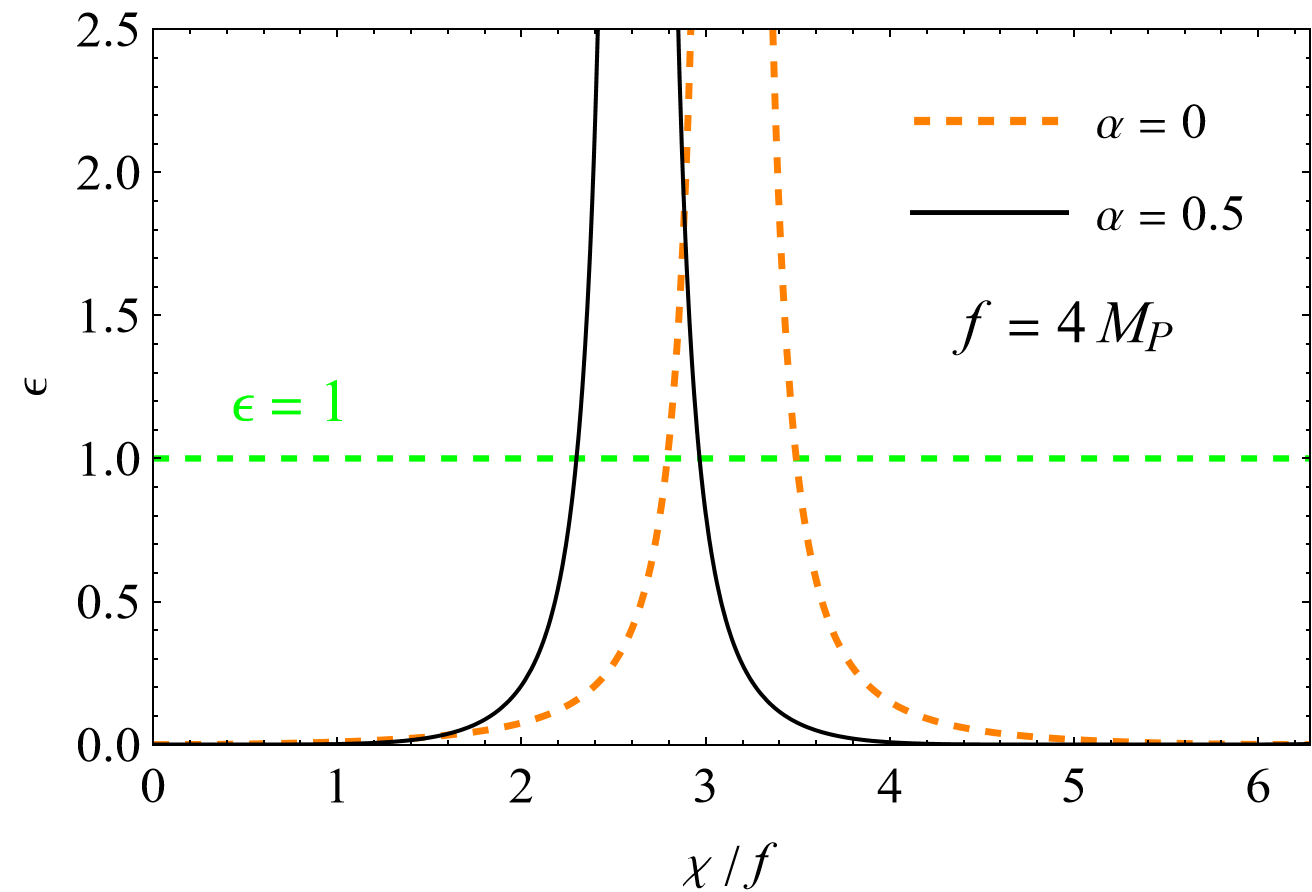}
    \includegraphics[scale=0.37]{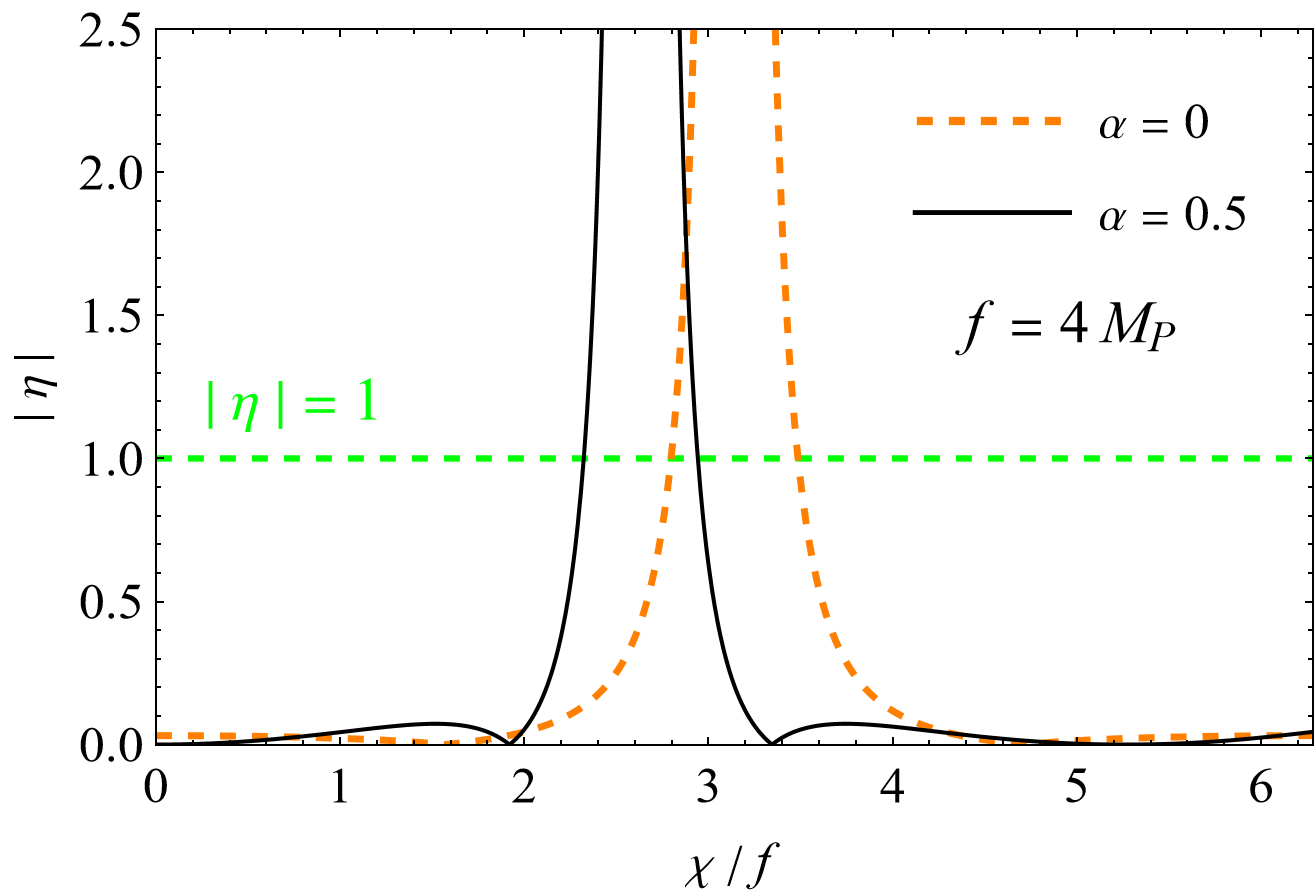}
    \medskip
    \includegraphics[scale=0.34]{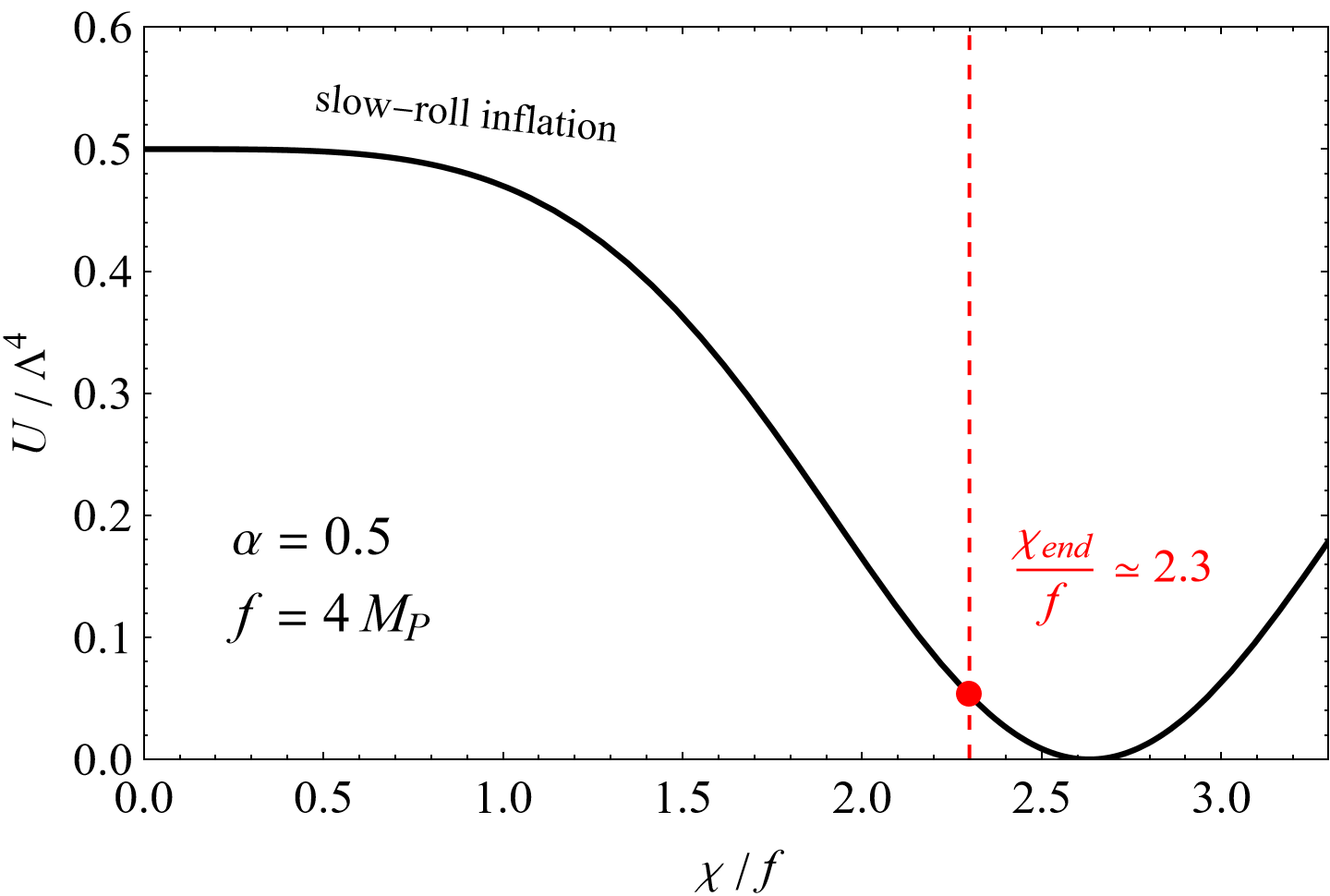}
    \includegraphics[scale=0.34]{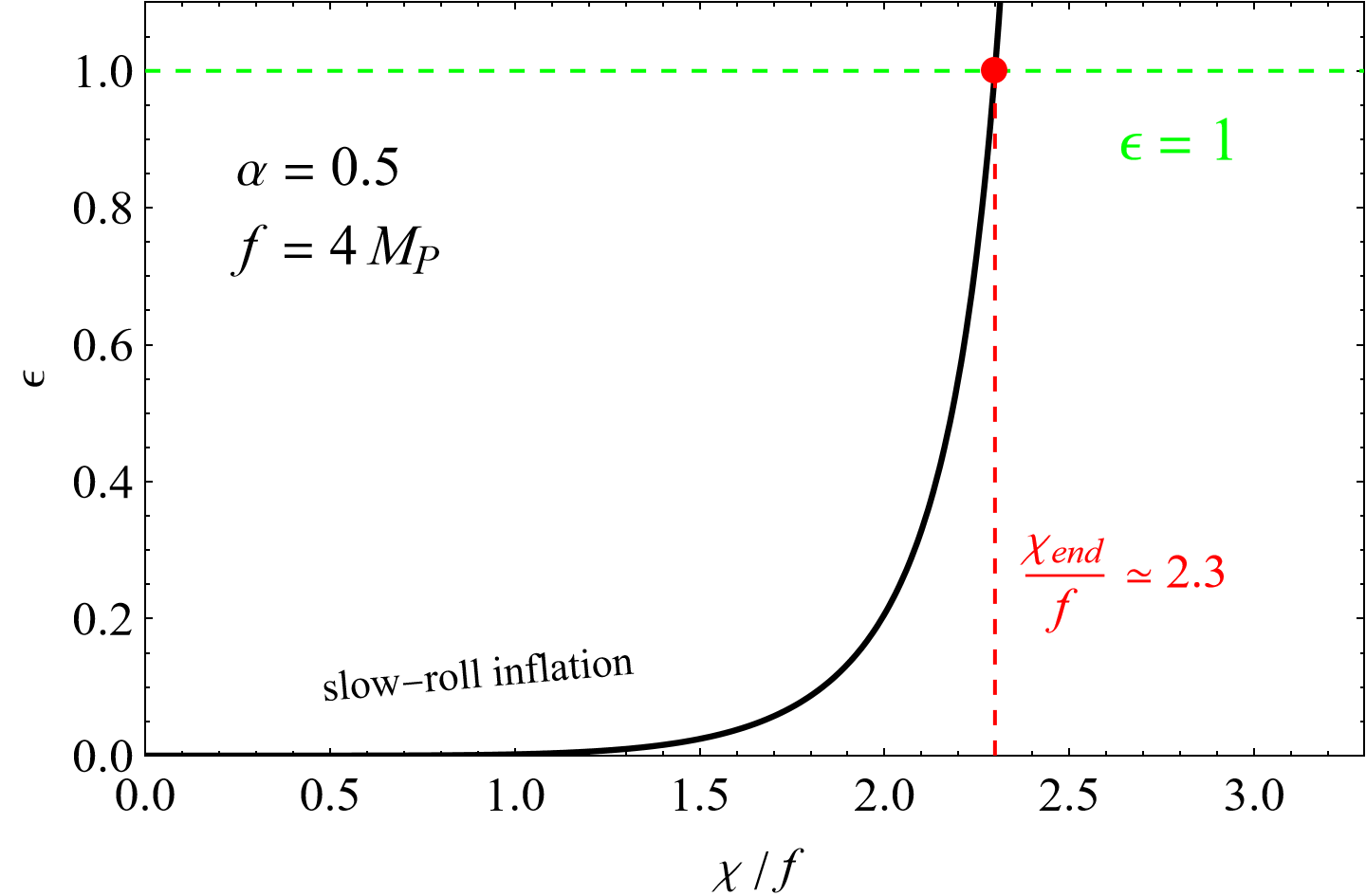}
    \caption{\textbf{Top left}: 
    {\em the first slow-roll parameter $\epsilon$ as a function of $\chi/f$ for $\alpha=\{0,\,0.5\}$ and $f=4M_P$.}  \textbf{Top right}: {\em the second slow-roll parameter $|\eta|$ as a function of $\chi/f$ for $\alpha=\{0,\,0.5\}$ and $f=4M_P$.}  \textbf{Bottom row}: {\em here we can see the dynamics of inflation for $\alpha=0.5$ and $f=4M_P$. The field $\chi$ slowly-rolls along $U$ from $\chi=0$ towards the minimum of the potential (bottom left image). Correspondingly, the slow-roll parameter $\epsilon$ increases until it reaches unity at a certain $\chi_{end}$ (bottom right image), and inflation ends there.}
    }
    \label{fig:EpsilonandEta}
\end{figure}

As always, however, the numerical analysis is simplified by some analytical understanding of the problem, which we now present. For $\alpha>-1/2$ the kinetic factor $K(\phi)$ is always strictly positive with no singularities, and therefore we will face no problem in solving Eq.~(\ref{eq:define.Chi}) numerically. As far as $U(\chi)$ is concerned, for $-1/2<\alpha\leq1/2$ the effective potential has a maximum in $\chi(\phi=0)$ and a minimum in $\chi(\phi=\pi f)$. This means that the inflaton rolls from the origin to $\chi(\pi f)$. For $\alpha>1/2$ instead, the effective potential $U$ exhibits a maximum in 
\begin{equation}
    \hat{\phi}=f\arccos\left(\frac{1-\alpha}{\alpha}\right)
    \label{eq:phi.hat}
\end{equation}
 that moves from $0$ to $\pi f$ as $\alpha$ increases, whilst the origin becomes a relative minimum. This means that if the inflaton evolution begins right in $\hat{\phi}$ the field remains in that unstable equilibrium point. Furthermore, if we give an initial condition before $\hat{\phi}$, the evolution would bring the inflaton to the new relative minimum in $\phi=0$, instead of bringing it to the absolute minimum in $\pi f$. In short, the new minimum in the origin develops a basin of attraction of the field. For $\alpha>1/2$ we then take care to give initial conditions after the maximum in $\hat{\phi}$, such as the field evolves towards the absolute minimum of the effective potential, which we identify with the field configuration we live with.

\subsection{Slow-roll natural inflation} \label{ch:Slow-roll.inflation}

Now that we have a canonical field $\chi$ in the Einstein frame, we can use the standard formul\ae~for single-field slow-roll inflation. The two slow-roll parameters are 
\begin{equation}
    \epsilon(\chi) \equiv \frac{M_P^2}{2} \left( \frac{U'(\chi)}{U(\chi)} \right)^2
    \quad,\quad
    \eta(\chi) \equiv M_P^2 \frac{U''(\chi)}{U(\chi)}
    \label{eq:epsilon.and.eta}
\end{equation}
being $U'\equiv dU/d\chi$. These parameters must be small in slow-roll inflation. In the top row of Fig.~\ref{fig:EpsilonandEta} we can see the numerical evaluation of $\epsilon(\chi)$ and $|\eta(\chi)|$ for $\alpha=\{0,0.5\}$ and $f=4 M_P$. 
Looking at the top-left figure there, we see that the period of $\epsilon(\chi/f)$ (i.e.~$\chi(2\pi f)/f$, that is exactly $2\pi$ for $\alpha=0$) decreases as $\alpha$ increases, so that inflation ends earlier. 
The top-right figure shows instead $|\eta|$. From numerical evaluation it is possible to see that $\epsilon$ reaches unity before $|\eta|$ for any value of $\alpha$ and $f$, so that the field value $\chi_{end}$ corresponding to the end of inflation is always determined by $\epsilon(\chi_{end})=1$. In the bottom row of Fig.~\ref{fig:EpsilonandEta} we show the value of the potential $U$ when inflation ends.

\begin{figure}[t]
    \centering
    \includegraphics[scale=0.32]{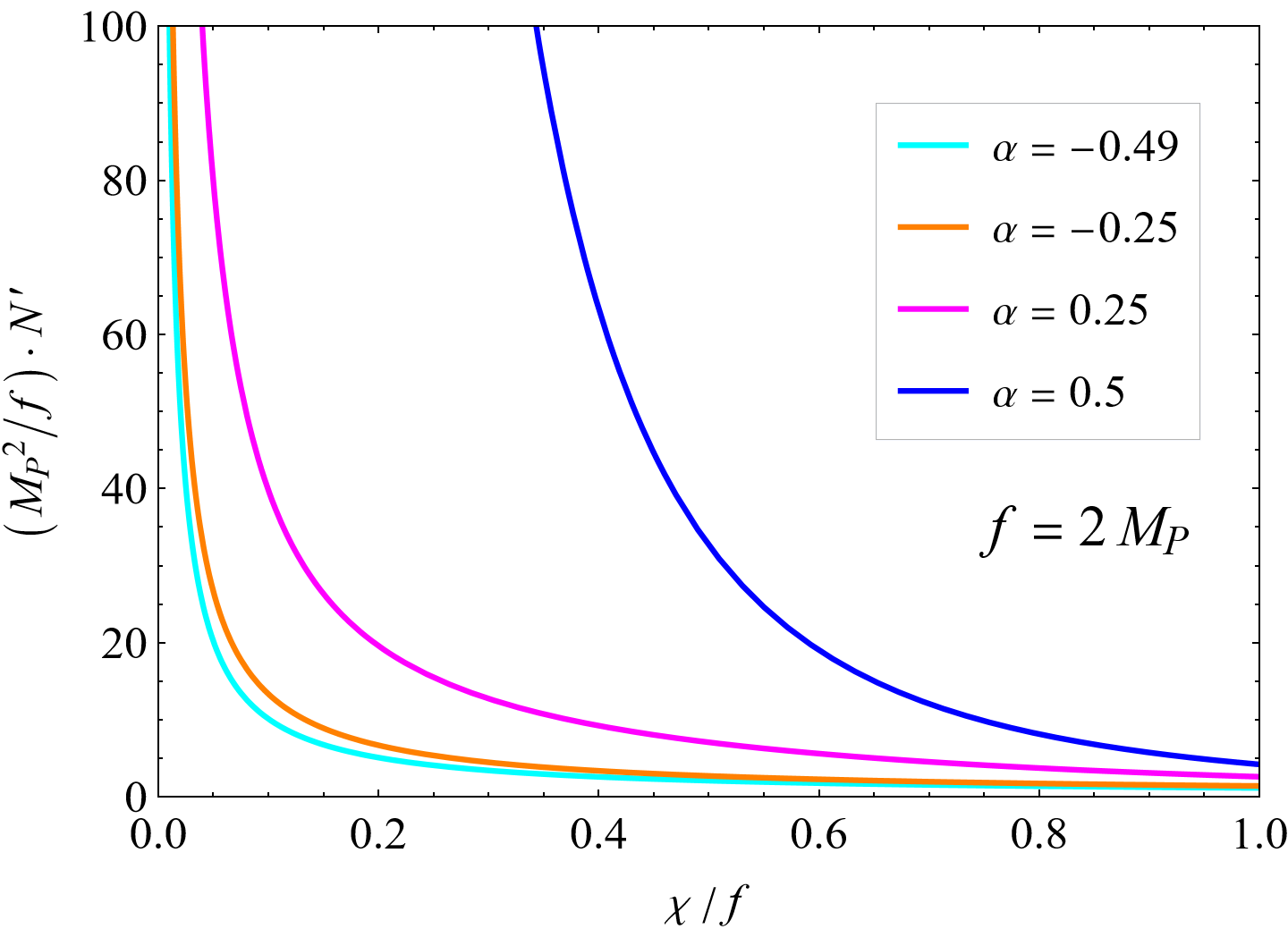}
    \hspace{0.5cm}
    \includegraphics[scale=0.35]{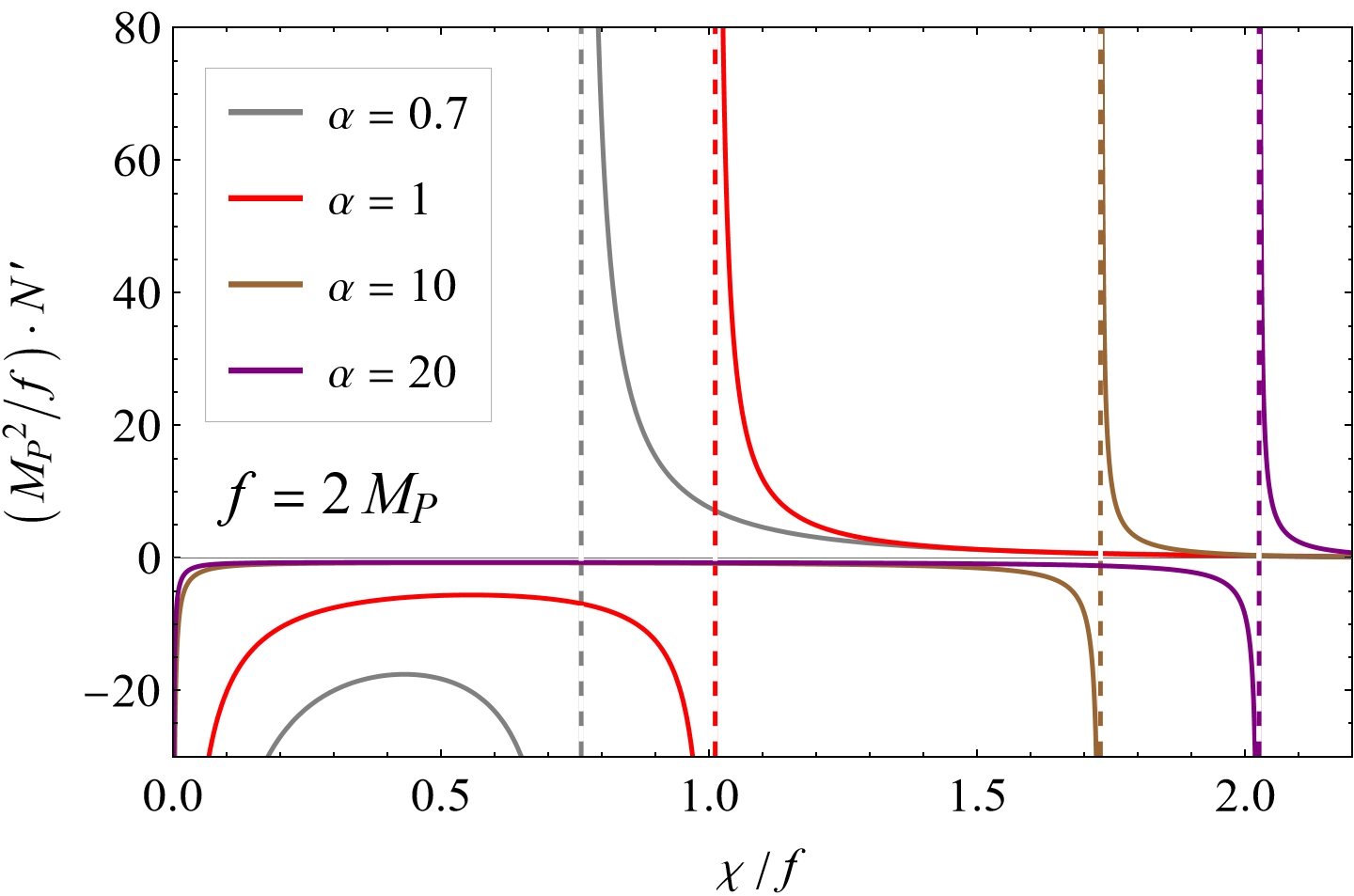}
    \caption{{\em the integrand function of the number of e-folds $N'$ (see Eq.~(\ref{eq:N.e-folds})) multiplied by the overall factor $M_P^2/f$ as a function of $\chi/f$ for some values of $\alpha$ and $f=2M_P$.} \textbf{Left}: {\em for $-1/2<\alpha\leq1/2$ we have only one singularity at the origin.} \textbf{Right}: {\em for $\alpha>1/2$ a new singularity, indicated by the dashed vertical lines, appears at $\chi(\hat{\phi})$ (see Eq.~(\ref{eq:phi.hat})). }}
    \label{fig:N'(chi/f)}
\end{figure}

When the slow-roll conditions are satisfied, the 00 component of the Einstein field equations and the conservation equation of the energy-momentum tensor respectively reads
\begin{equation}
    3M_P^2H^2 \,\simeq\, U, \qquad
    3H\dot{\chi} \,\simeq\, -\,U',
    \label{eq:slow-roll.motion.equations}
\end{equation}
where $H\equiv \dot a/a$,  a dot represents a derivative with respect to (cosmic) time  $t$ and   $a$ is the cosmological scale factor.
 Also, the number of e-folds between times $t_1$ and $t_2$ (corresponding to field values $\chi_1$ and $\chi_2$) is
\begin{equation}
    N(t_2,t_1) 
    \,\,\,\simeq\,\,\,
    -\,\frac{1}{M_P^2} \int_{\chi_1}^{\chi_2} d\chi\,\frac{U(\chi)}{U'(\chi)}.
    \label{eq:slow-roll.N.e-folds}
\end{equation}
Equivalently, we can express  the number of e-folds $N_e$ before the end of inflation in terms of the field $\chi$:
\begin{equation}
    N_e(\chi) \,\,\simeq\,\, 
    -\,\frac{1}{M_P^2} \int_{\chi}^{\chi_{end}} d\xi\,\frac{U(\xi)}{U'(\xi)}.
    \label{eq:N.e-folds}
\end{equation}
 We have performed an in-depth analytical study of the integrand function $N'(\xi)\equiv dN_e(\xi)/d\xi\,=\,-\,U/M_P^2U'$, in order to render the numerical evaluation of the integral in~(\ref{eq:N.e-folds}) easier.  
 For $-1/2<\alpha\leq1/2$ the integrand function $N'(\chi(\phi))$ has one positive vertical asymptote in the origin and we can safely integrate it from $\chi>0$ to $\chi_{end}$ to obtain any number of e-folds. Conversely, for $\alpha>1/2$ a new singularity appears in $\chi(\hat{\phi})$ because of the new maximum of the effective potential (if the initial condition is given right in $\chi(\hat{\phi})$ the field remains there and the system makes an infinite number of e-folds). Moreover, for $\phi\in(0,\hat{\phi})$ the integrand function is negative: the field is evolving back towards the relative minimum in $\chi(0)$. In order to let the field correctly evolve towards the absolute minimum $\chi(\pi f)$, in
  Eq.~(\ref{eq:N.e-folds}) we must choose an extreme of integration such that $\chi>\chi(\hat{\phi})$, i.e.~beyond the singularity. Only in this way we are able to obtain a positive number of e-folds towards the absolute minimum. 
  Furthermore, as $\alpha$ increases the singularity moves towards $\chi(\pi f)$, meaning that we have less and less values of $\chi$ in which we can integrate Eq.~(\ref{eq:N.e-folds}). This means that for too large values of $\alpha$ the system  is not  able to perform, say, $60$ e-folds, being the integral of $N'$ from $\chi(\hat{\phi}^+)$ to $\chi(\pi f)$ less than $60$. 
  That is, we expect that for too large values of $\alpha$ the system will not be able to describe a field that evolves towards the absolute minimum of the potential in an acceptable way, because the field is more and more driven towards the relative minimum as the potential peaks and squeezes more and more. In Fig.~\ref{fig:N'(chi/f)} we can see the numerical evaluation of $N'(\chi)$ for some values of  $\alpha$ and $f=2M_P$.

 In the slow-roll regime we also have the following formul\ae~for the scalar spectral index $n_s$, the tensor-to-scalar ratio $r$ and the curvature power spectrum $\mathcal{P}_\mathcal{R}$ (here we evaluate the power spectra  at horizon exit, $k=a H$)
\begin{equation}
    n_s(\chi) = 1 - 6\epsilon(\chi) + 2\eta(\chi)
    \quad,\quad
    r(\chi) = 16\epsilon(\chi),
    \label{eq:ns.and.r}
\end{equation}
\begin{equation}
    \mathcal{P}_\mathcal{R}(\chi) \,=\, \frac{U(\chi)}{24\,\pi^2\,M_P^4\,\epsilon(\chi)}.
    \label{eq:power.spectrum}
\end{equation}
Observe that  none of the functions $\epsilon(\chi)$, $\eta(\chi)$, $N_e(\chi)$, $n_s(\chi)$, and $r(\chi)$ depends on the energy scale $\Lambda$, which appears in the numerator of the effective potential $U$. This is because those quantities are generally invariant under rescaling of $U$. This means that a value for $\Lambda$ is not needed neither to find the end of inflation $\chi_{end}$ nor  to find the field value $\chi$ that corresponds to a certain $N_e(\chi)$. The only slow-roll function above that depends on the choice of $\Lambda$ is the scalar power spectrum $\mathcal{P}_\mathcal{R}(\chi)$, and we will use this fact to fix $\Lambda$ from observations~\cite{Ade:2015lrj}:
\be \mathcal{P}_\mathcal{R}(k_*) =(2.10 \pm 0.03) 10^{-9}, \label{PRobserved}\ee
where the pivot scale  $k_* =0.05~{\rm Mpc}^{-1}$  is used as in~\cite{Ade:2015lrj}.

\subsection{The natural inflation parameter space and observational constraints}\label{ch:the.natural.inflation.parameter.space}

For a given number of e-folds $N_e$ the observable predictions of inflation, for example for $n_s$ and $r$, as well as the slow-roll parameters $\epsilon$ and $\eta$ depend on $\alpha$ and $f$, having fixed $\Lambda$  with (\ref{PRobserved}). We will present our results as functions of $f$ for several values of $\alpha$ and focusing on the e-fold interval $55\leq N_e\leq 65$. 

\begin{figure}[t!]
\begin{center}
    \includegraphics[width=0.495\textwidth]{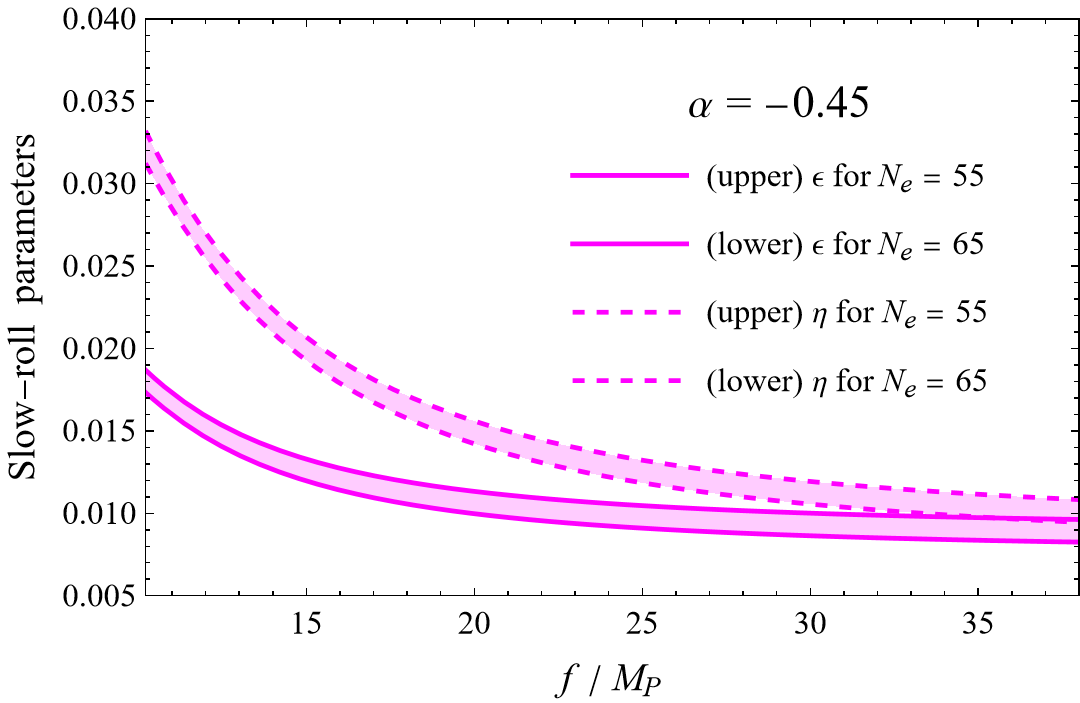}
    \includegraphics[width=0.495\textwidth]{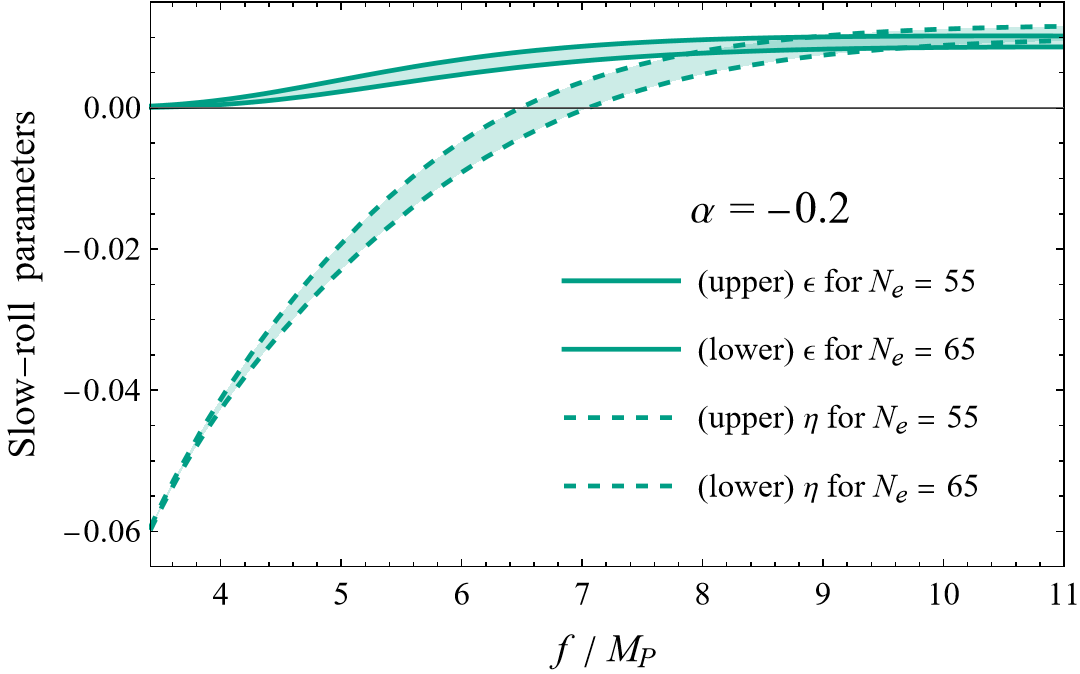}

    \vspace{0.1cm}
    
    \includegraphics[width=0.495\textwidth]{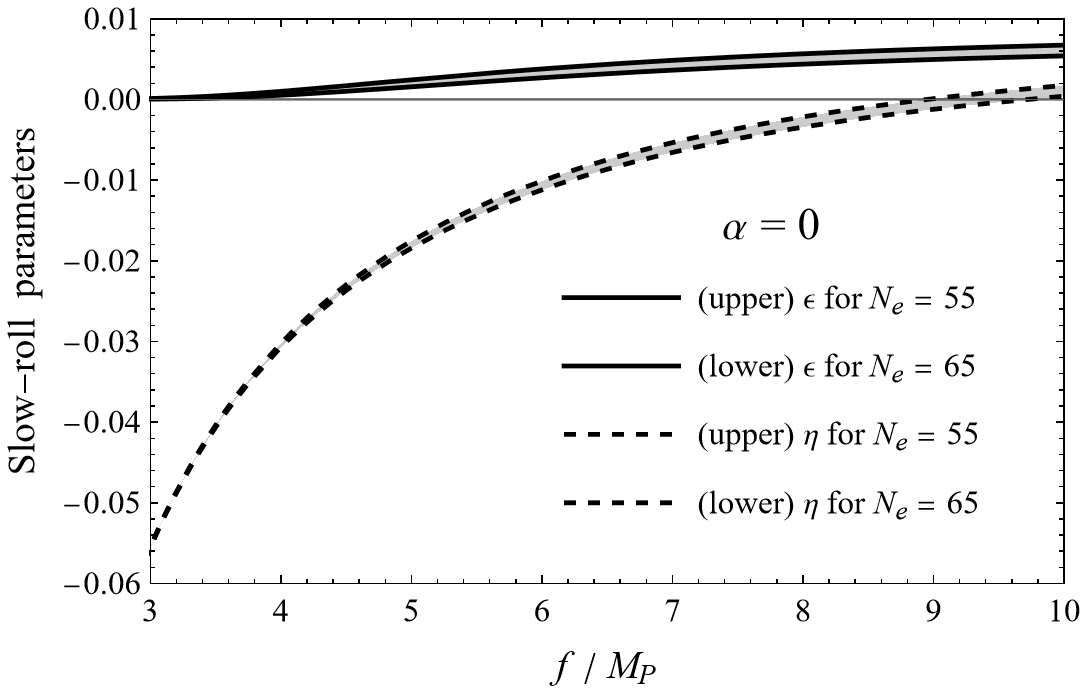}
     \includegraphics[width=0.495\textwidth]{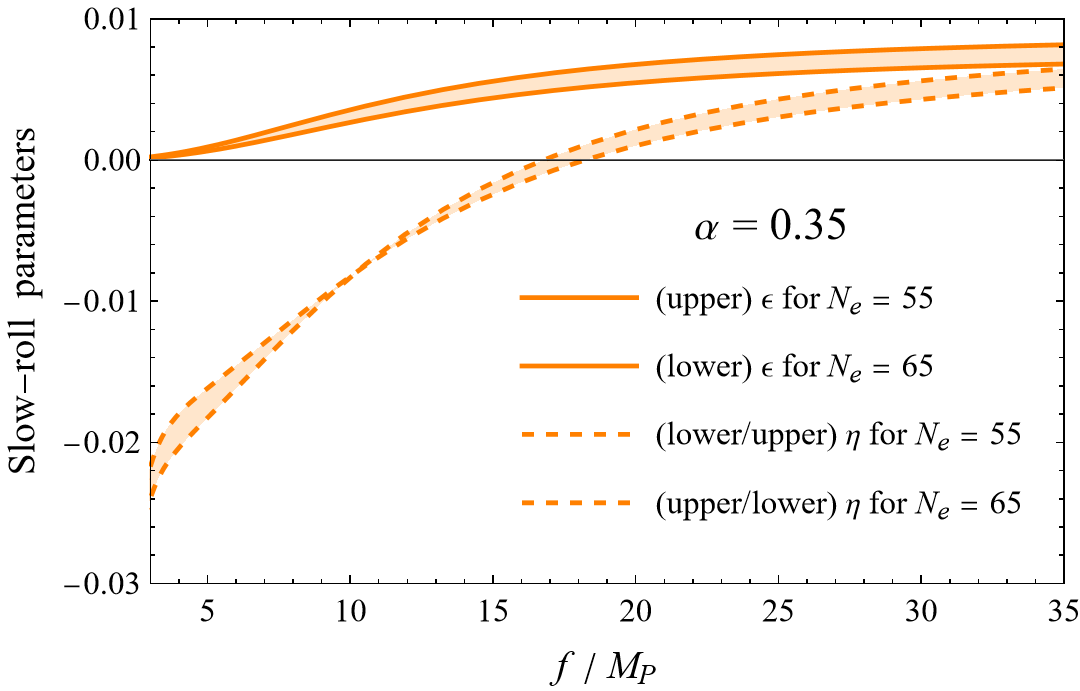}

%
%
    \caption{{\em The slow-roll parameters $\epsilon$ (solid lines) and $\eta$ (dashed lines) as functions of the energy scale $f$ for some values of $\alpha$ between $-1/2$ and $+1/2$. }}
    \label{fig:epsilon(f).eta(f).A}
\end{center}
\end{figure}

\begin{figure}[t!]
\begin{center}
    \includegraphics[width=0.495\textwidth]{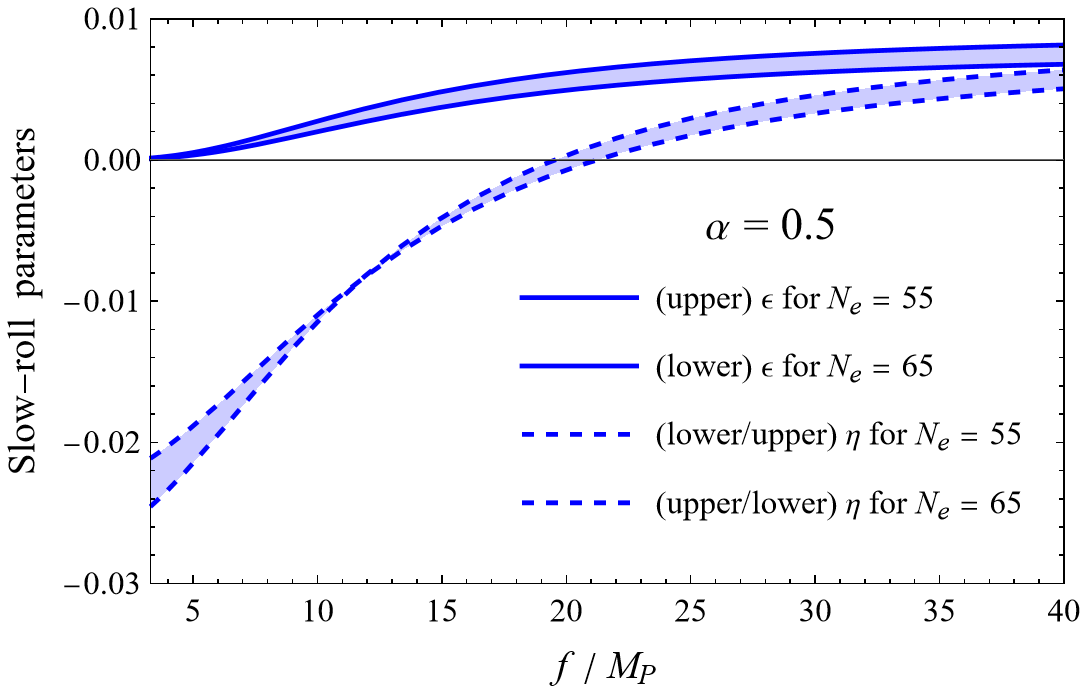}
   \includegraphics[width=0.495\textwidth]{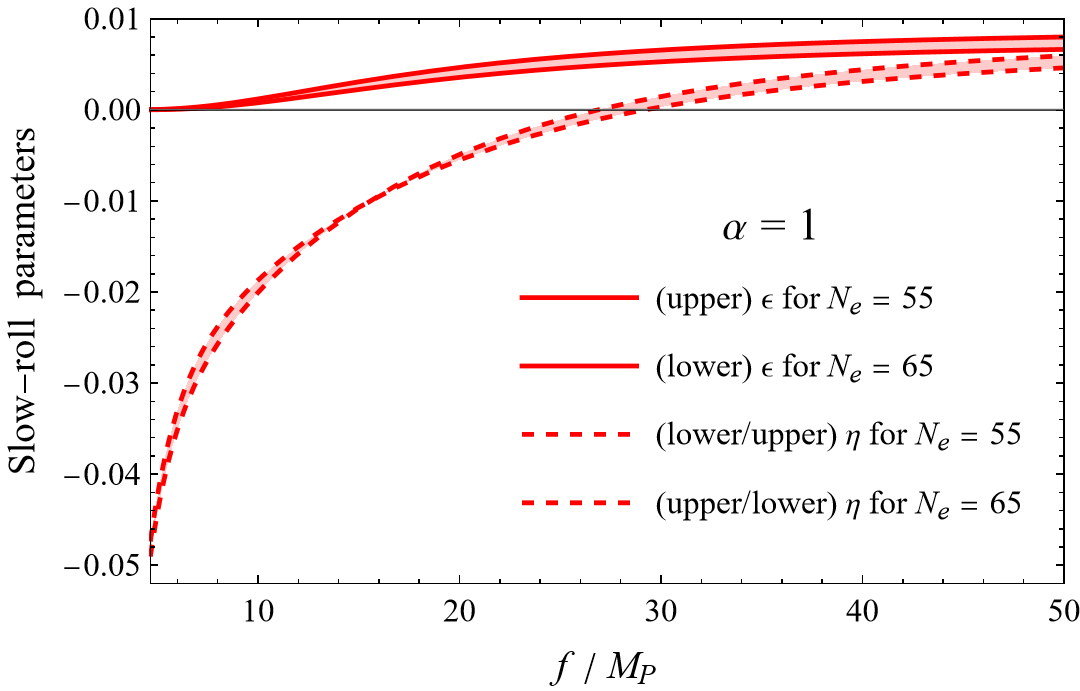}
   
    \vspace{0.1cm}
    


    \includegraphics[width=0.495\textwidth]{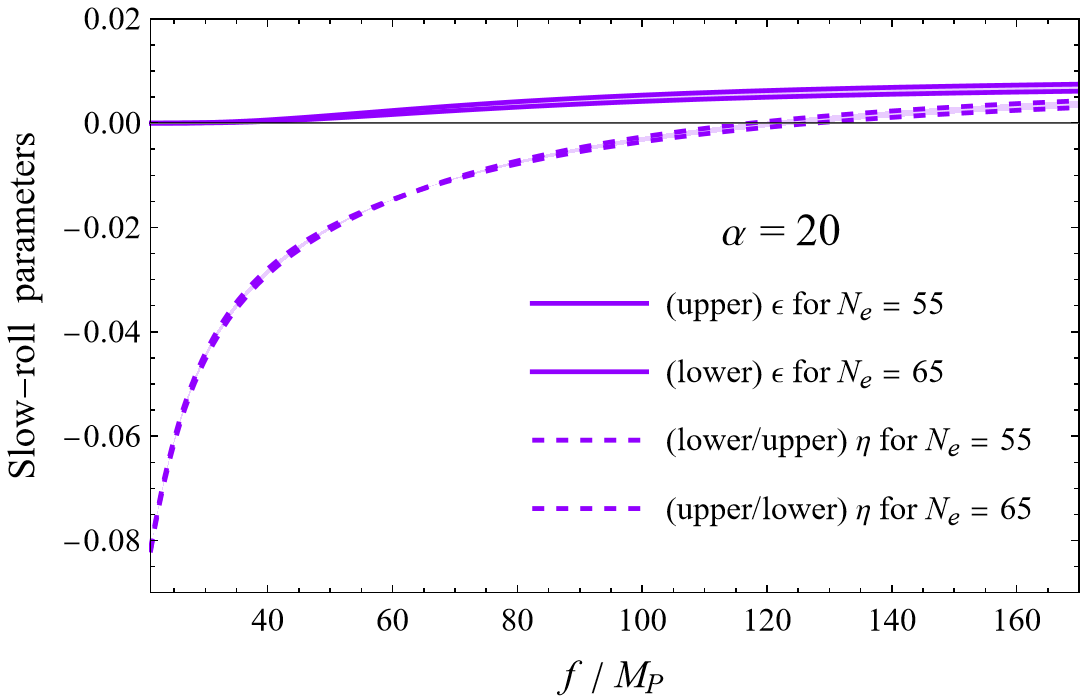}
    \includegraphics[width=0.495\textwidth]{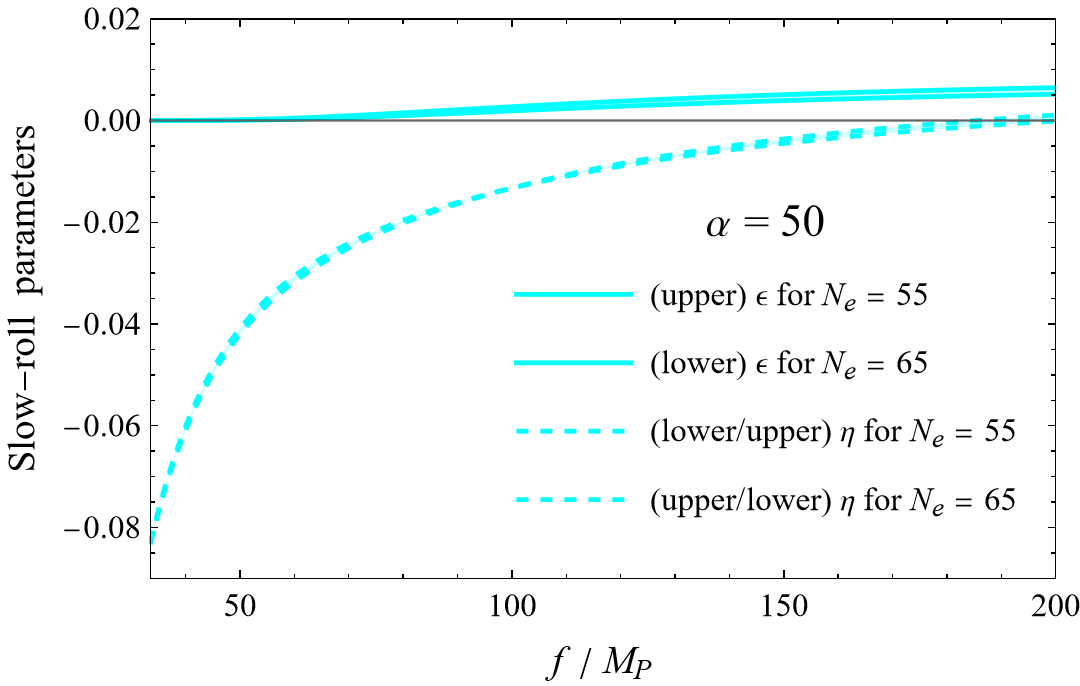}
    \caption{\em The same as in Fig.~\ref{fig:epsilon(f).eta(f).A}, but with $\alpha\geq 1/2$.}
    \label{fig:epsilon(f).eta(f).B}
\end{center}
\end{figure}

\begin{figure}[t!]
\begin{center}
    \includegraphics[width=0.495\textwidth]{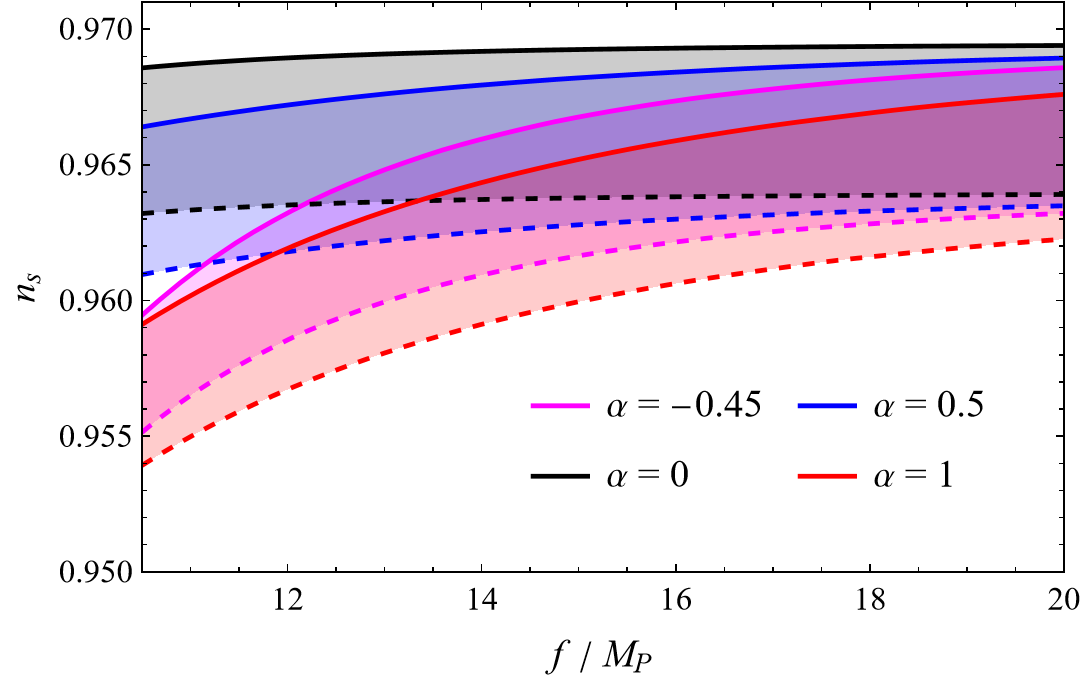}
    \includegraphics[width=0.495\textwidth]{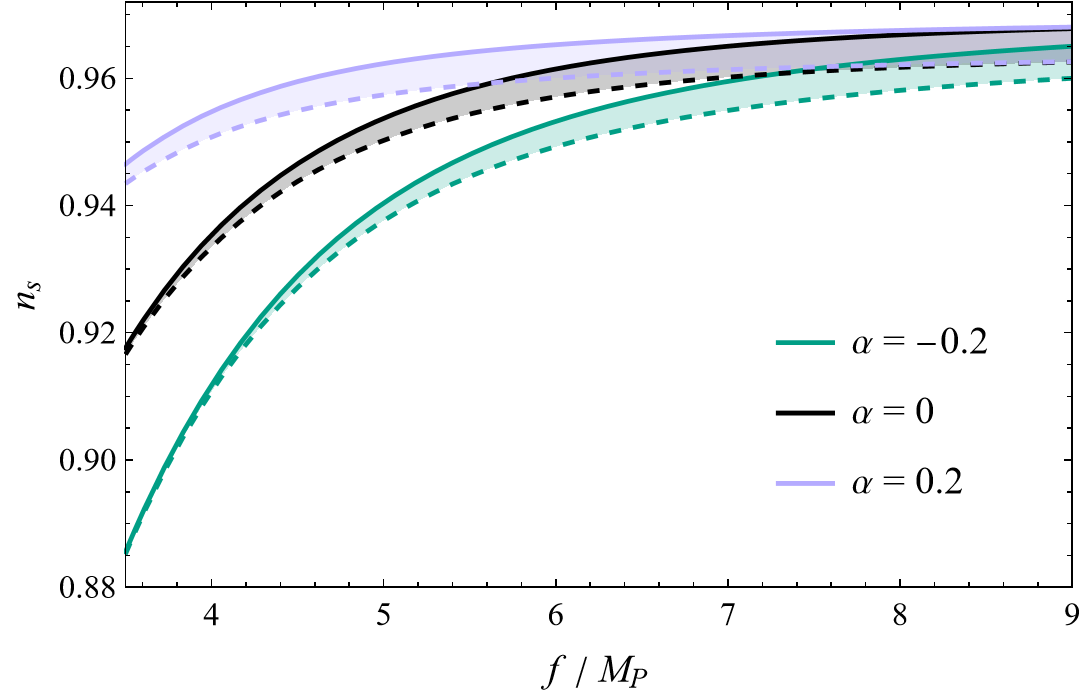}

    \vspace{0.05cm}
    
    \includegraphics[width=0.495\textwidth]{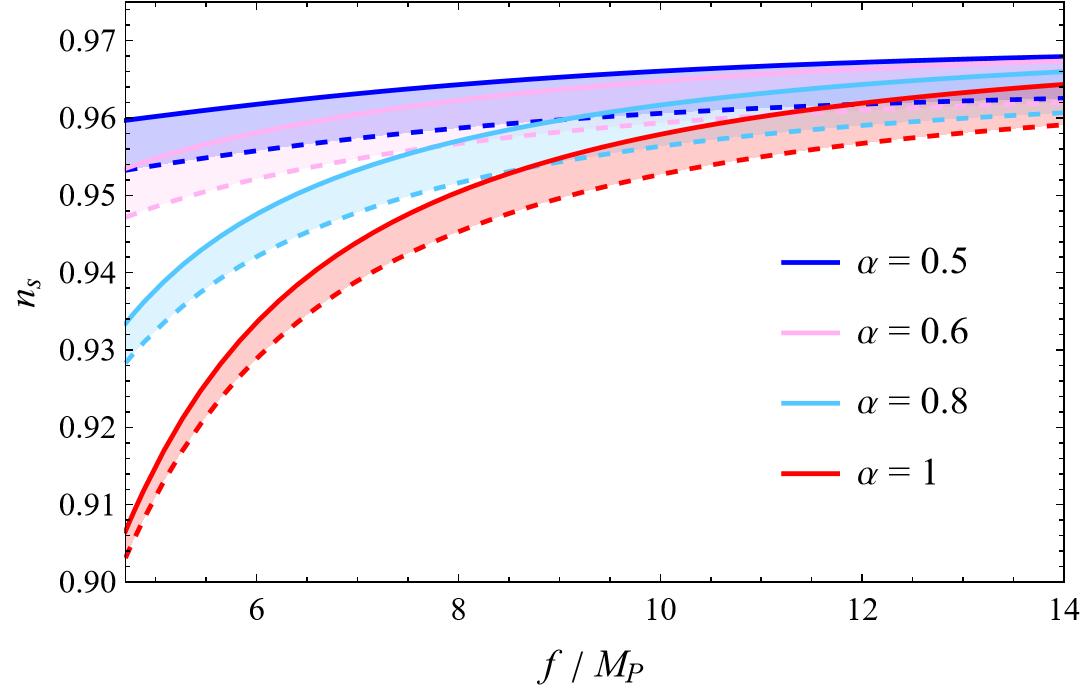}
   \includegraphics[width=0.495\textwidth]{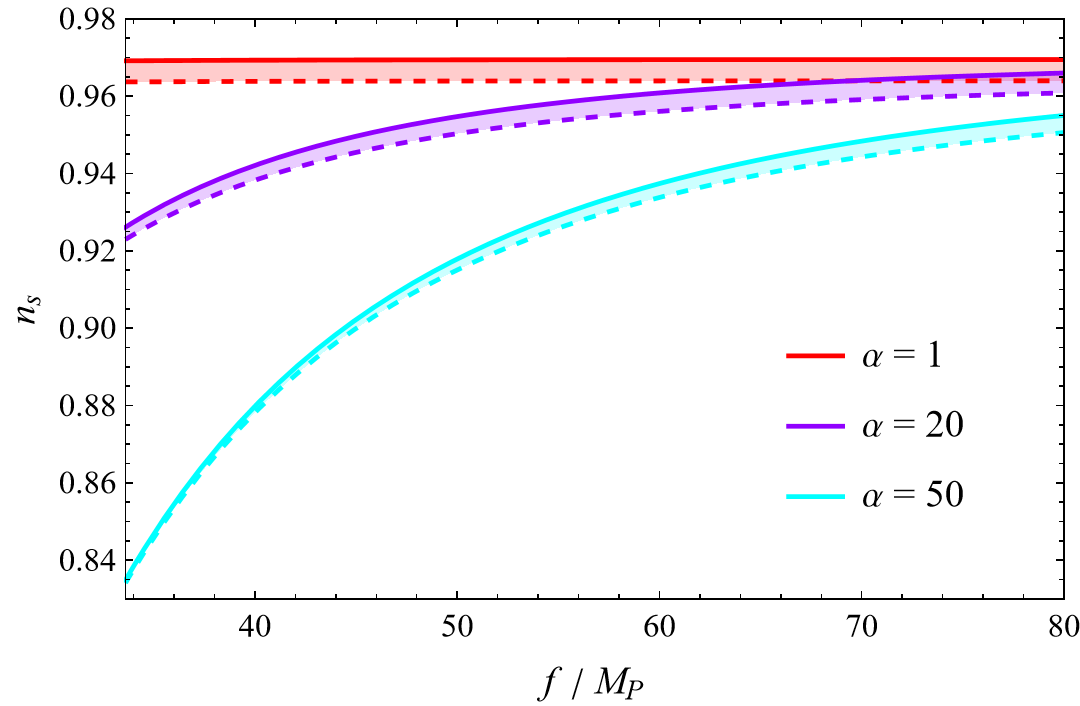}
    \caption{\em The spectral index of the scalar perturbations $n_s$ as a function of $f$ for some values of $\alpha>-1/2$. For each value of $\alpha$ we considered a  number of e-folds between $N_e=55$ (dashed line) and $N_e=65$ (continuous line).}
    \label{fig:ns(f)}
\end{center}
\end{figure}

\begin{figure}[h!]
\begin{center}
    \includegraphics[width=0.495\textwidth]{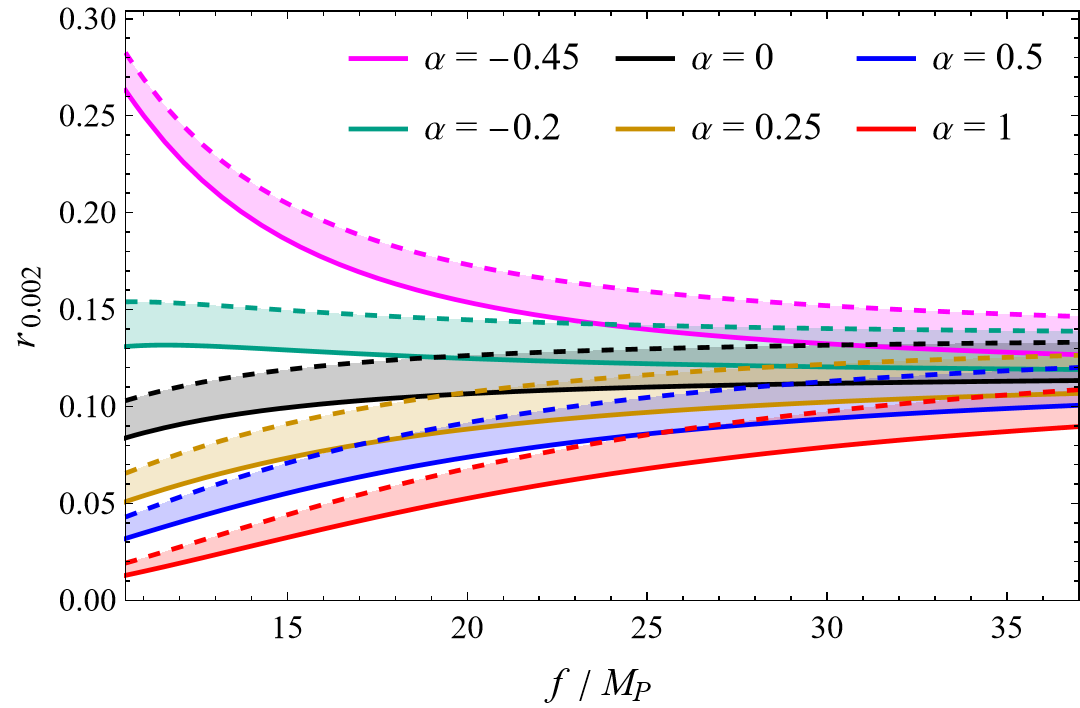}
    \includegraphics[width=0.495\textwidth]{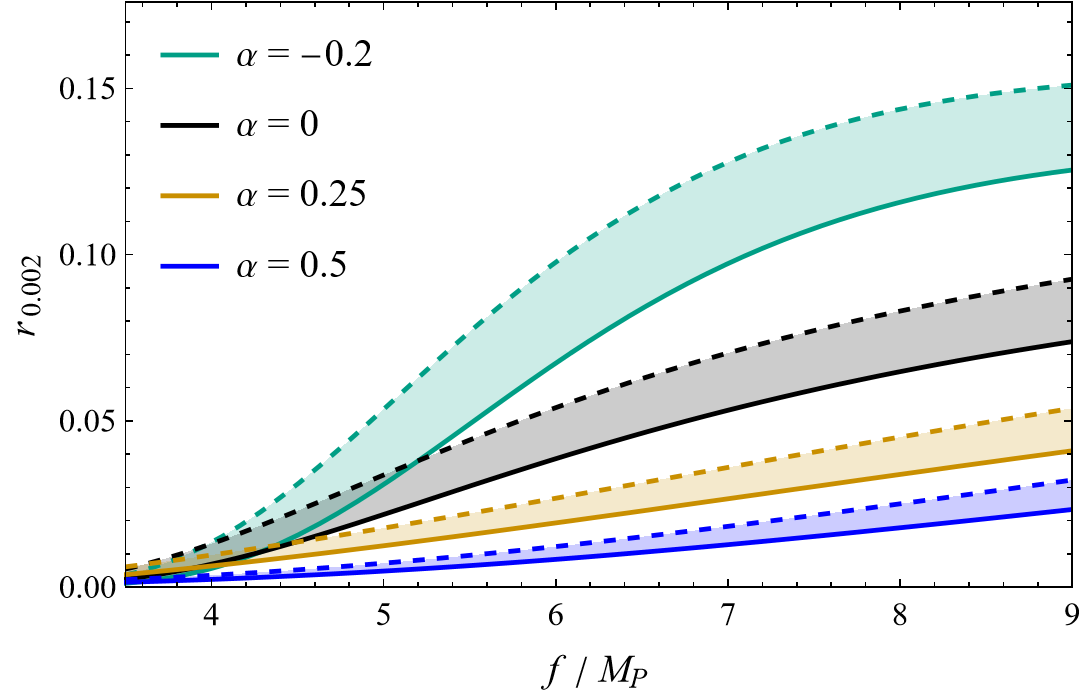}

    \vspace{0.05cm}

    \includegraphics[width=0.495\textwidth]{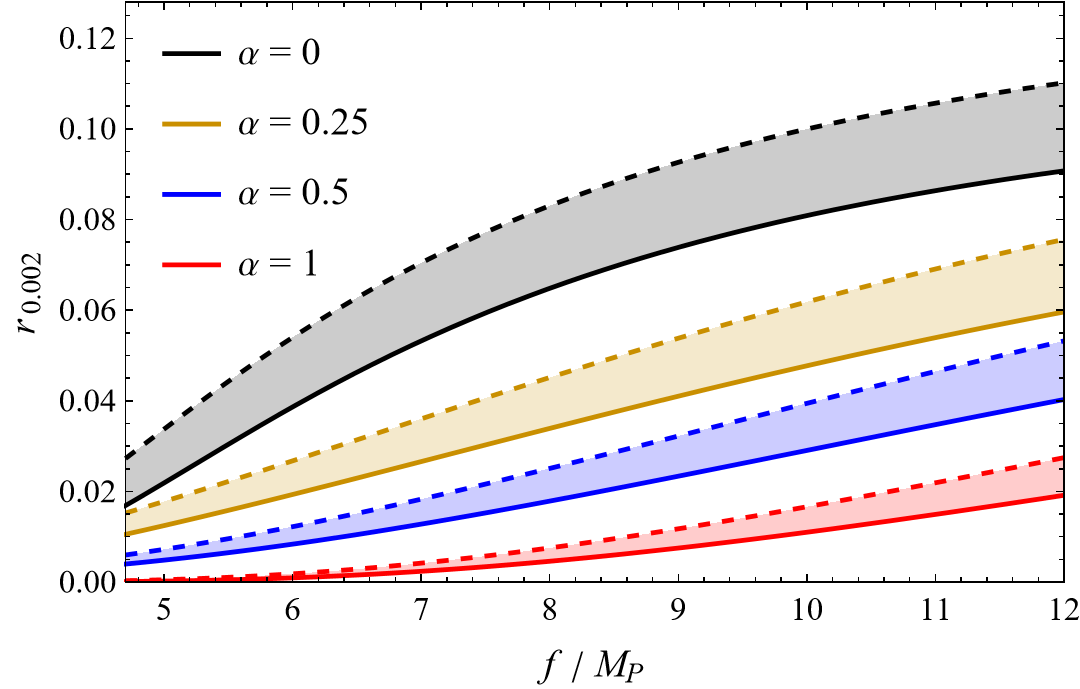}
   \includegraphics[width=0.495\textwidth]{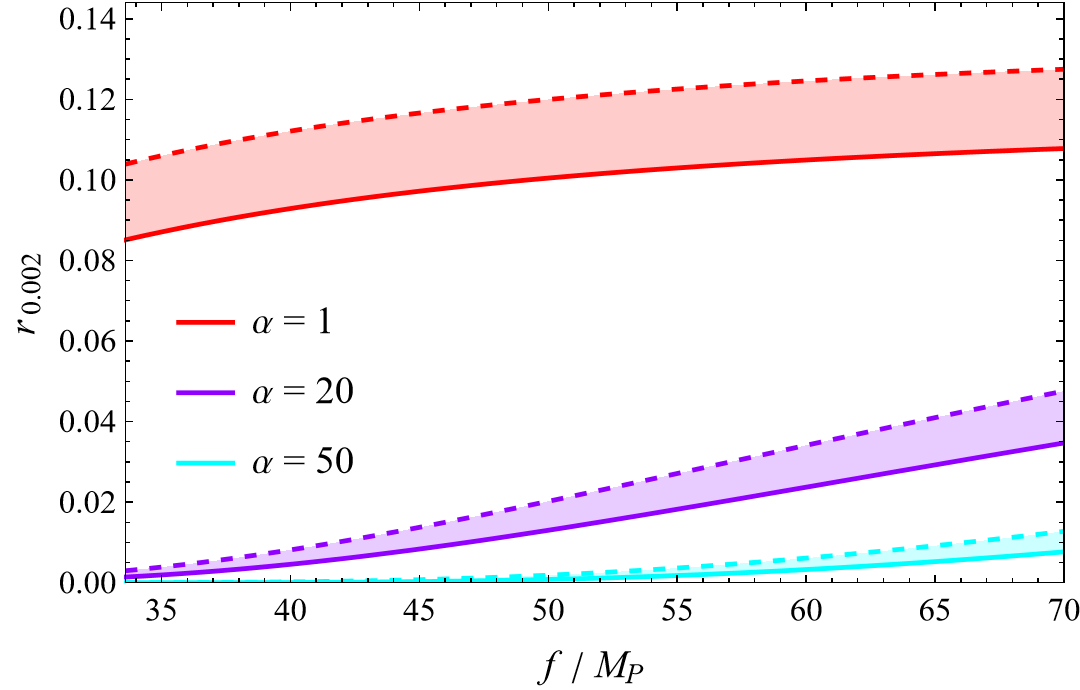}
    \caption{\em The same as in Fig.~\ref{fig:ns(f)}, but with $n_s$ replaced by the tensor-to-scalar ratio $r_{0.002}$.}
    \label{fig:r(f)}
\end{center}
\end{figure}

\begin{figure}[t]
\begin{center}
    \includegraphics[width=0.495\textwidth]{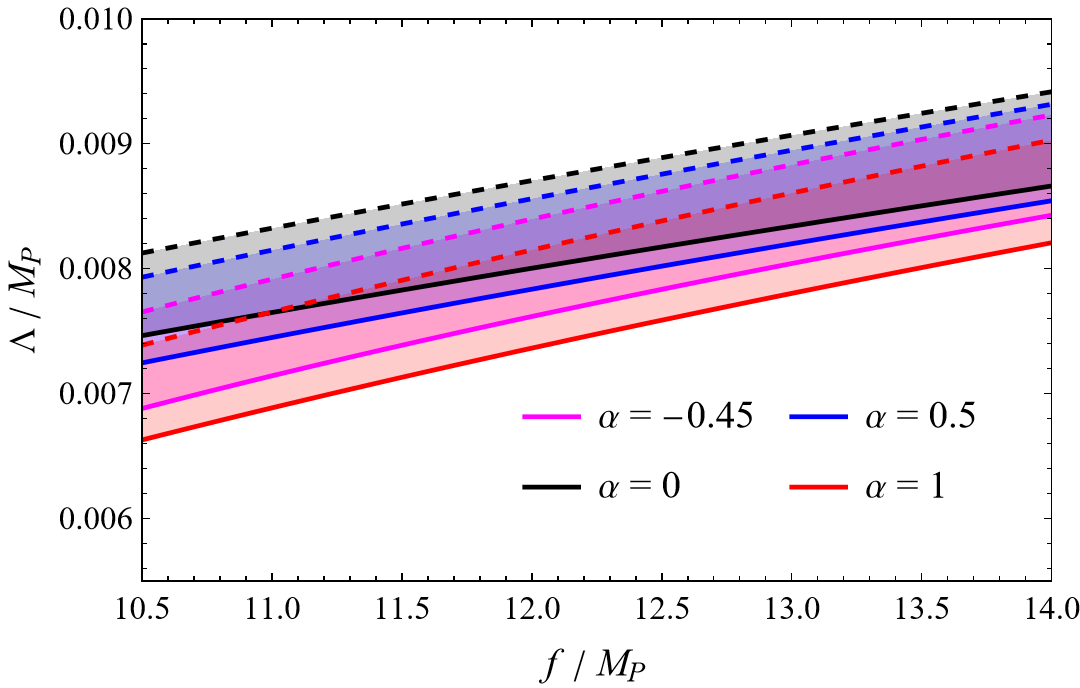}
   \includegraphics[width=0.495\textwidth]{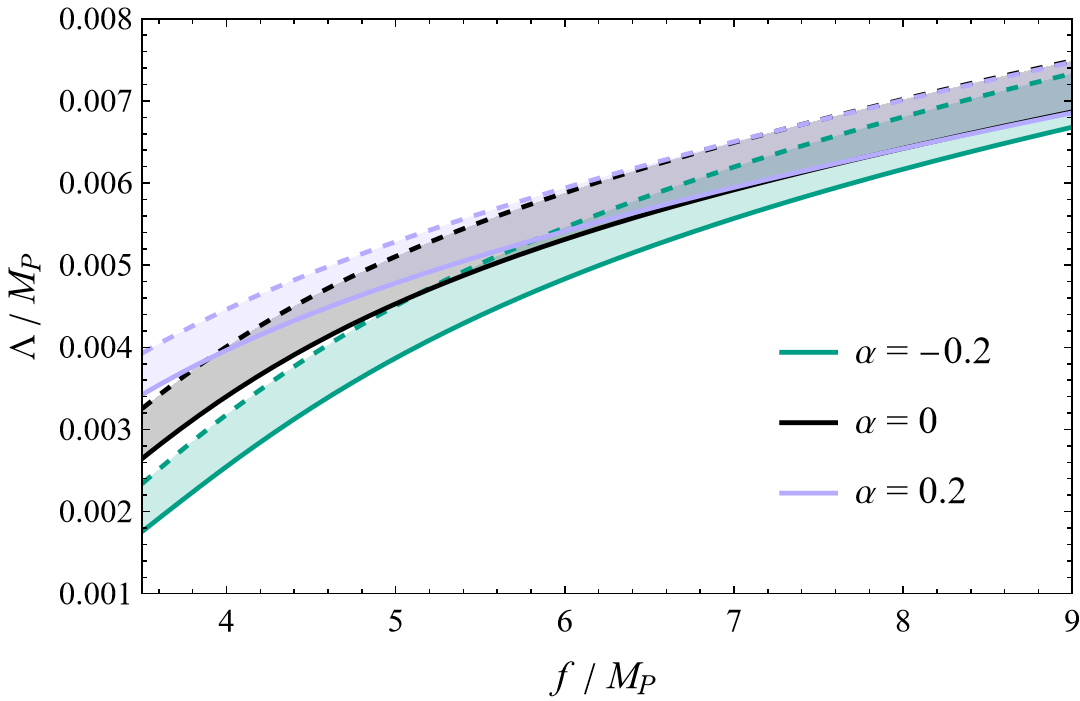}

    \vspace{0.1cm}

    \includegraphics[width=0.495\textwidth]{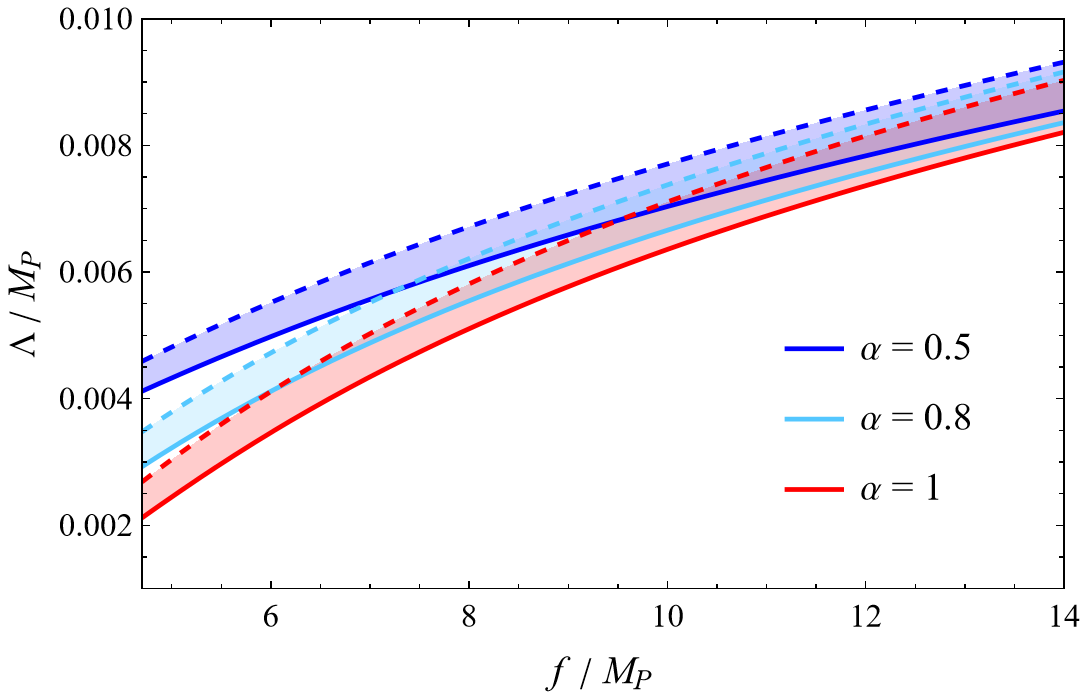}
    \includegraphics[width=0.495\textwidth]{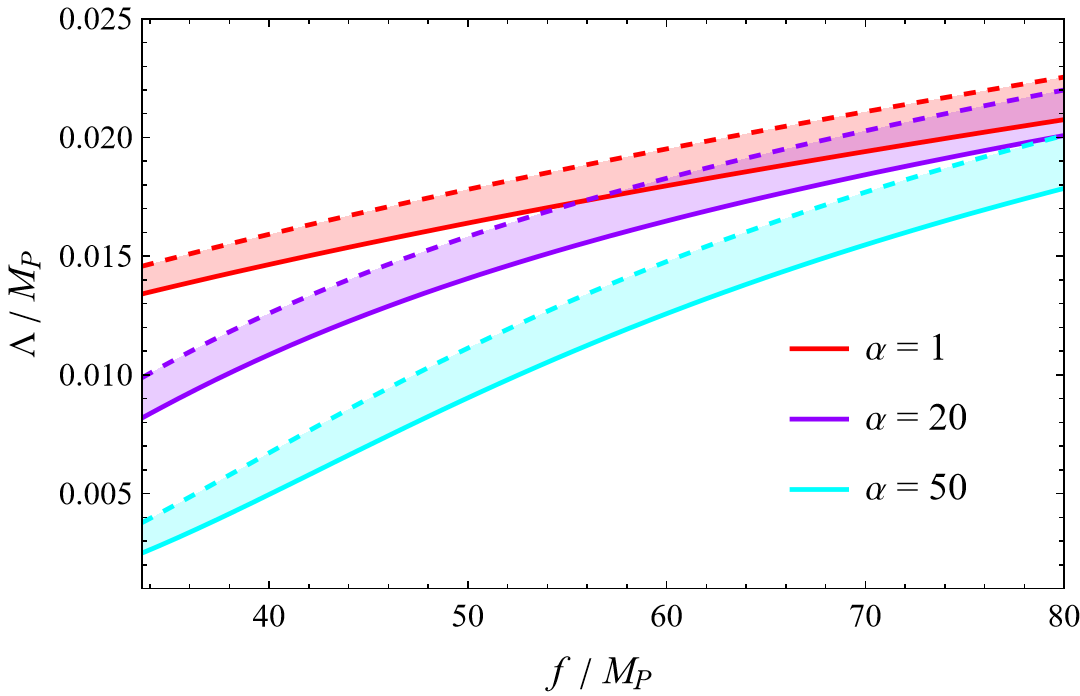}
    \caption{\em The same as in Fig.~\ref{fig:ns(f)}, but with $n_s$ replaced by the energy scale $\Lambda$, which is fixed to reproduce the observed curvature power spectrum (see Eq.~(\ref{PRobserved})).}
    \label{fig:Lambda(f)}
\end{center}
\end{figure}

\begin{figure}[t]
    \centering
  
    \includegraphics[scale=0.45]{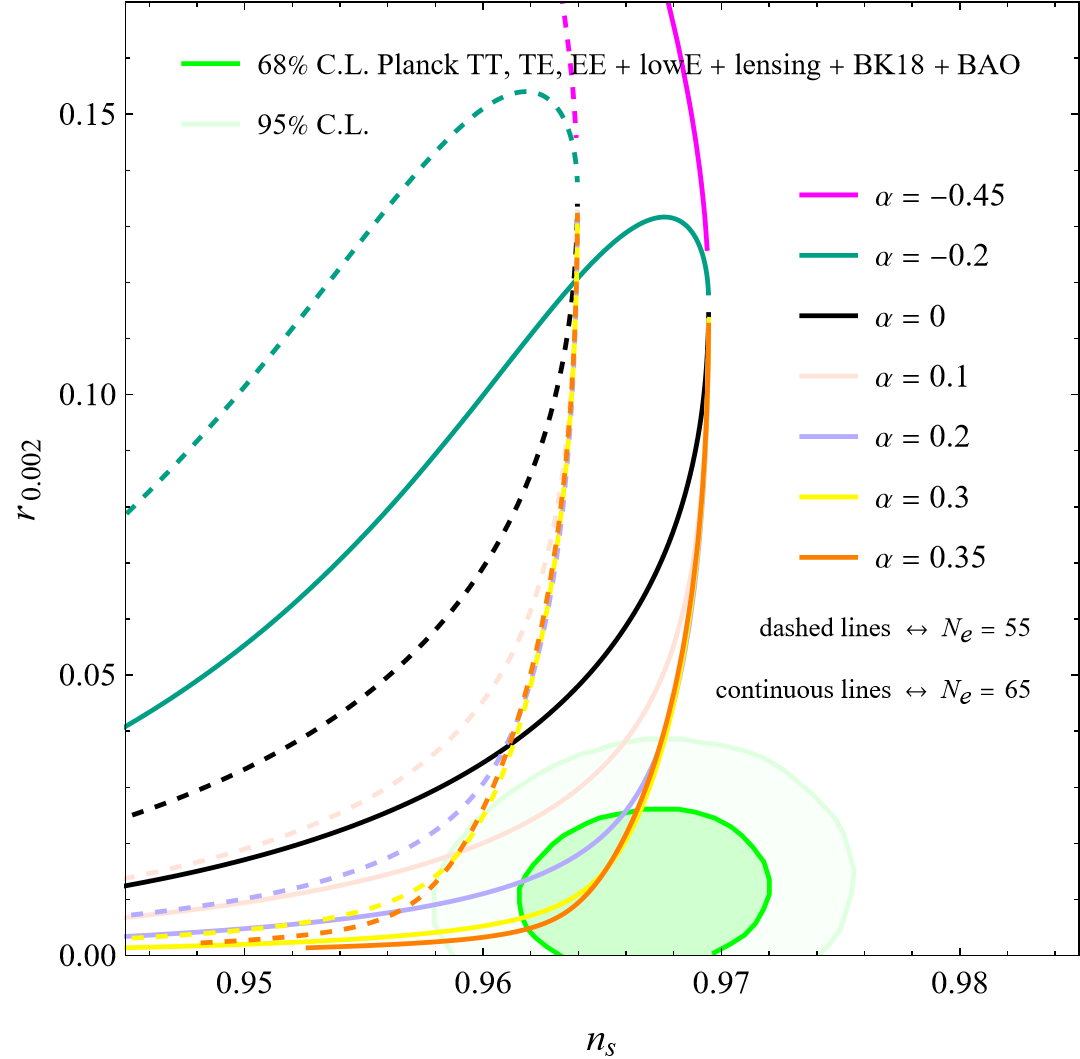}\includegraphics[scale=0.45]{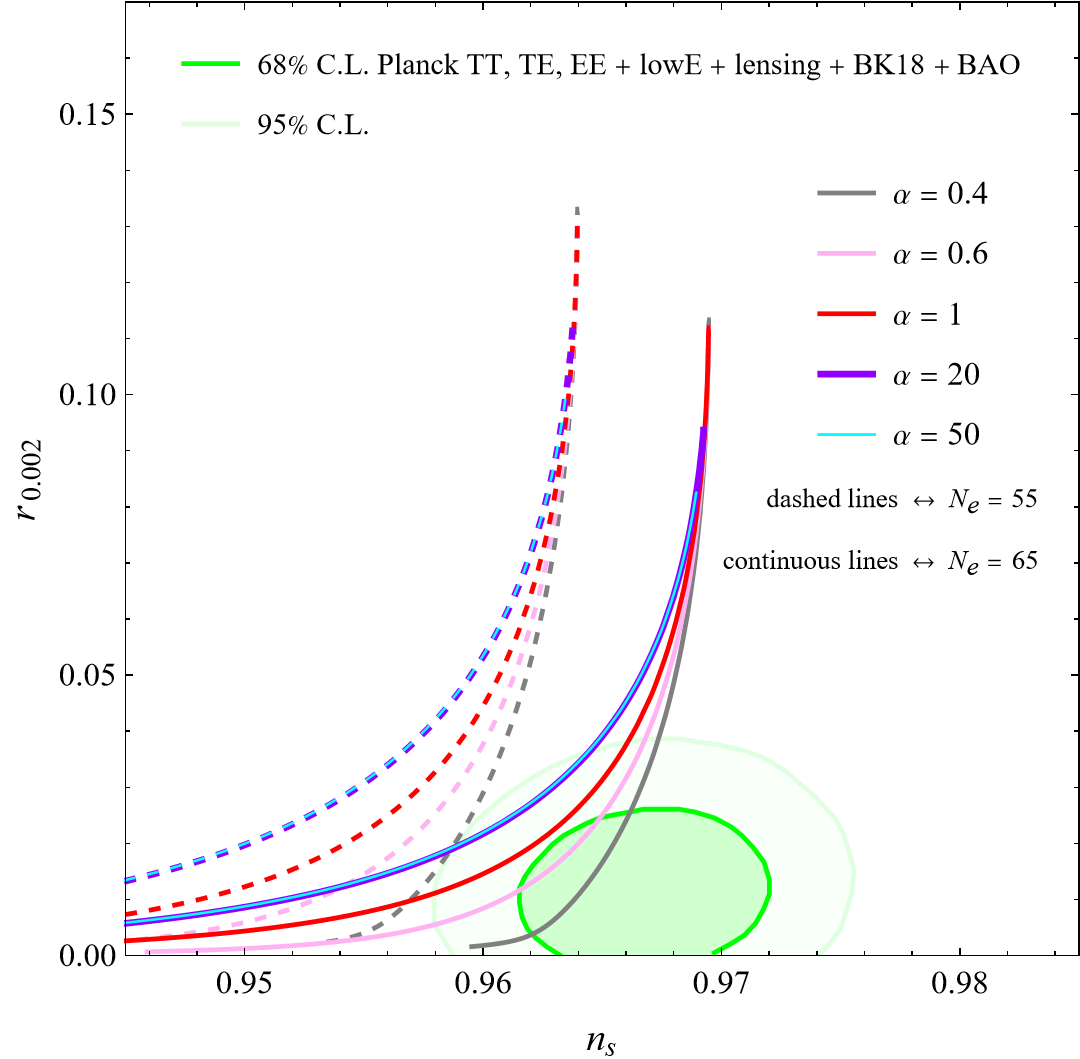}
    
    \caption{\em The theoretical predictions of natural inflation with a periodic non-minimal coupling on the $\{n_s,r_{0.002}\}$ plane compared to the observational constraints from~\cite{Ade:2015lrj,BICEP:2021xfz}. The ranges of $f$ for which the curves are evaluated are: $f/M_P\in[10.21 , 38.21]$ for $\alpha=-0.45$; $f/M_P\in[3.42 , 45.92]$ for $\alpha=-0.2$; $f/M_P\in[2.24 , 44.74]$ for $\alpha=0$; $f/M_P\in[2.02 , 51.02]$ for $\alpha=0.1$; $f/M_P\in[2.24 , 57.24]$ for $\alpha=0.2$; $f/M_P\in[2.61 , 63.61]$ for $\alpha=0.3$; $f/M_P\in[2.79 , 66.79]$ for $\alpha=0.35$; $f/M_P\in[2.96 , 69.96]$ for $\alpha=0.4$; $f/M_P\in[3.6 , 78.6]$ for $\alpha=0.6$; $f/M_P\in[4.66 , 89.66]$ for $\alpha=1$; $f/M_P\in[21.19 , 176.19]$ for $\alpha=20$; $f/M_P\in[33.53 , 218.53]$ for $\alpha=50$.}
    \label{fig:r0.002(ns)}
\end{figure}

In Figs.~\ref{fig:epsilon(f).eta(f).A} 
and~\ref{fig:epsilon(f).eta(f).B} we can see the slow-roll parameters $\epsilon$ and $\eta$ as functions of $f$ for some values of $\alpha > -1/2$. For the ranges of $f$ displayed the slow-roll parameters are indeed small  for each value of $\alpha$.  In Figs.~\ref{fig:ns(f)} and~\ref{fig:r(f)} we can respectively see the scalar spectral index $n_s$ and the tensor-to-scalar ratio $r_{0.002}$ evaluated at the pivot scale $k_*=0.002\,\mathrm{Mpc^{-1}}$ as functions of $f$ for some values of $\alpha> -1/2$. Also, in Fig.~\ref{fig:Lambda(f)} we show the corresponding values of the  energy scale $\Lambda$. Note that $\Lambda^4$ appears as an overall constant in the energy density $U$; one can see that $\Lambda\ll M_P$, so our effective field theory treatment of quantum gravity is reliable although $f$ takes transplankian values. In Fig.~\ref{fig:r0.002(ns)} we provide the predictions of our natural inflation model on the $\{n_s, r_{0.002}\}$ plane together with the observational constraints of Ref.~\cite{Ade:2015lrj,BICEP:2021xfz}.  Finally, in Fig.~\ref{fig:f(alpha)_N=63} we show the corresponding observationally allowed regions on the $\{\alpha,f\}$ plane for $N_e =63$. We have explicitly checked that this number of e-folds is compatible with the bound found in Ref.~\cite{Liddle:2003as} readapted to the natural-inflation case.  One can clearly see that a non-minimal coupling can render natural inflation compatible even with the recent very stringent constraints from Ref.~\cite{BICEP:2021xfz}.

\subsection{Relic natural inflationary background of gravitational waves} \label{ch:the.relic.inflationary.GW.bakground}

As mentioned in the introduction, an inflationary model generates a primordial GW spectrum. These inflationary GWs are produced as quantum tensor fluctuations and form today  a relic stochastic GW background (see Ref.~\cite{Maggiore:2018sht} for a textbook introduction to this topic and Sec.~2 of Ref.~\cite{Salvio:2021kya} for a refined determination of these GW spectrum which is more useful for our purposes).

 In this section we want to investigate the possibility of detecting this background directly with future space-borne interferometers, in the case of natural inflation. 
This will be done by comparing the spectral density of the GWs, defined by 
\be \Omega_{\rm GW}(\nu) \equiv \frac{\nu}{\rho_{\rm cr}}\frac{d\rho_{\rm GW}}{d\nu}, \label{OGWdef}\ee
with the predictions of the sensitivity curves of  DECIGO, BBO and ALIA. In Eq.~(\ref{OGWdef}) $\nu$ is the frequency of the GWs,  $\rho_{\rm cr}\equiv	3H_0^2\bp^2$  is the critical energy density, $H_0$ is the present value of the Hubble rate and $\rho_{\rm GW}$ is the energy density carried by the stochastic background of GWs.

Following Ref.~\cite{Salvio:2021kya} and recalling that GWs that are within the sensitivities of ground-based and space-borne interferometers (such as  DECIGO, BBO and ALIA) as well as pulsar timing arrays all correspond to tensor modes that re-entered the horizon during the radiation dominated era much before the time of radiation-matter equality, we obtain 
\begin{equation}
    h_0^2 \Omega_\mathrm{GW}(\nu) \,\simeq\, \frac{1}{24}\,h_0^2\Omega_R \left( \frac{g_*(T_k)}{\Bar{g}_*} \right) \left( \frac{\Bar{g}_*^S}{g_*^S(T_k)} \right)^{4/3} r_{0.05}\, \mathcal{P}_\mathcal{R}(k_*)\left(\frac{\nu}{\nu_*}\right)^{-\,r/8}. 
    \label{eq:Omega_GW(nu)}
\end{equation}
Here $h_0\equiv H_0/(100~{\rm km\,s^{-1}\,Mpc^{-1}})$ is the dimensionless Hubble constant, $k\equiv(2\pi)\nu$, $k_*=0.05\,\mathrm{Mpc^{-1}}$ is a reference pivot scale, $\nu_*=k_*/(2\pi)$ and $r_{0.05}$ is the tensor-to-scalar ratio at pivot scale $k_*$. In Eq.~(\ref{eq:Omega_GW(nu)}) we used the single-field slow-roll result $n_T=-r/8$. 
Moreover, $T_k$ is the temperature at which the mode with scale $k$ re-enters the horizon, $\Omega_R$ is the cosmological parameter that expresses the fraction between the radiation energy density today and the critical density, and  $g_*^{(S)}(T)$ is the (entropy) effective number of relativistic species at temperature $T$:
\begin{equation}
    g_*(T) \,\equiv\, \sum_{i=\mathrm{bosons}} g_i\left(\frac{T_i}{T}\right)^4 \,+\, \frac{7}{8} \sum_{i=\mathrm{fermions}} g_i\left(\frac{T_i}{T}\right)^4,
    \label{eq:g_*(T).natural.infl}
\end{equation}
\begin{equation}
    g_*^S(T) \,\equiv\, \sum_{i=\mathrm{bosons}} g_i\left(\frac{T_i}{T}\right)^3 \,+\, \frac{7}{8} \sum_{i=\mathrm{fermions}} g_i\left(\frac{T_i}{T}\right)^3.
    \label{eq:g_*^S(T).natural.infl}
\end{equation}
Here $g_i$ is the number of helicity states of the $i$-th bosonic or fermionic species, $T_i$ is its temperature, and we have reserved  $T$ for the photon temperature $T_\gamma$. In Eq.~(\ref{eq:Omega_GW(nu)}) we called $\Bar{g}_*^{(S)}$ the value of $g_*^{(S)}(T_r)$ at a reference temperature $T_r$ below that of $e^{\pm}$ annihilation, but such that the three active neutrinos are still relativistic.

The radiation energy-density ratio $\Omega_R$ can be expressed in terms of the effective number of neutrino species $N_\mathrm{eff}^{(\nu)}$ and the photon energy-density ratio $\Omega_\gamma=\rho_{\gamma,0}/\rho_\mathrm{cr}$, with $\rho_{\gamma,0}=\pi^2 T_0^4/15$ and $T_0$ being today's photon temperature, by means of 
\begin{equation}
    h_0^2 \Omega_R \,=\, \left[\, 1 \,+\,  N_\mathrm{eff}^{(\nu)} \,\frac{7}{8} \left(\frac{4}{11}\right)^{4/3} \,\right] \, h_0^2 \Omega_\gamma,
    \label{eq:Omega_R}
\end{equation}
with
\begin{equation}
    h_0^2 \Omega_\gamma \,=\,  \frac{\pi^2T_0^4}{45M_P^2} \left(\frac{h_0}{H_0}\right)^2.
    \label{eq:Omega_gamma}
\end{equation}
 In order to compute the active neutrino contribution to $N_\mathrm{eff}^{(\nu)}$ and to the parameters $\Bar{g}_*$ and $\Bar{g}_*^S$, one should recall that a relativistic neutrino species after $e^{\pm}$ annihilation features a temperature $T_\nu=(4/11)^{1/3}\,T$.
Note that, by definition, at temperature $T_r$ all active neutrinos are relativistic and would contribute to\footnote{In the SM  $N_\mathrm{eff}^{(\nu)}\simeq3.044$ as recently computed in~\cite{Bennett:2019ewm,Akita:2020szl,Froustey:2020mcq,Bennett:2020zkv}. } $N_\mathrm{eff}^{(\nu)}$. However, recall that near the present epoch at least two active neutrinos are non relativistic (see~\cite{Esteban:2020cvm,deSalas:2020pgw} for recent bounds on their masses). The value of $\Omega_R$, therefore, depends on whether today the lightest neutrino is relativistic or not. Moreover, in general $N_\mathrm{eff}^{(\nu)}$ and   $\Bar{g}_*^{(S)}$ can also receive a contribution from extra species that are relativistic respectively at temperature $T_0$ and $T_r$ (if any).

In this work we consider a standard scenario in which the only light species are those of the SM and the lightest neutrino is non-relativistic today, i.e.~its mass is greater than $T_0$. This means that we consider $N_\mathrm{eff}^{(\nu)}=0$ today. Therefore, we  have
\begin{equation}
h_0^2\Omega_R=h_0^2\Omega_\gamma.
    \label{eq:h_0^2Omega_R=2.473x10^-5}
\end{equation}

In order to use Eq.~(\ref{eq:Omega_GW(nu)}) for the spectral density, we are just left to write down the explicit dependence of the crossing horizon temperature $T_k$ on $k$. Following Ref.~\cite{Salvio:2021kya} we obtain
\begin{equation}
    T_k \,=\, \left( \frac{g_{*0}^S}{\Bar{g}_*^S} \right)^{1/3} \frac{c(T_k)T_0k}{H_0\Omega_R^{1/2}},
    \label{eq:Tk}
\end{equation}
where $g_{*0}^S\equiv g_*^S(T_0)$, and $c(T)$ is a slowly-varying function of $T$ given by 
\begin{equation}
    c(T) \,=\, \left( \frac{g_*(T)}{\Bar{g}_*} \right)^{-1/2} \left( \frac{\Bar{g}_*^S}{g_*^S(T)} \right)^{-1/3}
    \label{eq:c(T).function}
\end{equation}
and we can use Eqs.~(\ref{eq:Tk})-(\ref{eq:c(T).function}) in a iterative way to estimate $T_k$ as a function of $k$
\begin{align}
    T_k^{(0)} \,&\equiv\, T_r \quad\Rightarrow\quad c(\,T_k^{(0)}\,)\,=\,1 \notag\\
    \Rightarrow\quad T_k^{(1)} \,&\equiv\, \left( \frac{g_{*0}^S}{\Bar{g}_*^S} \right)^{1/3} \frac{T_0k}{H_0\Omega_R^{1/2}} \notag\\
    \Rightarrow\quad T_k^{(2)} \,&\equiv\, \left( \frac{g_{*0}^S}{\Bar{g}_*^S} \right)^{1/3} \frac{c(\,T_k^{(1)}\,)T_0k}{H_0\Omega_R^{1/2}}.
    \label{eq:T_k^(2)}
\end{align}
The second order $T_k\simeq T_k^{(2)}$ is already good enough for our purposes. In this case, since $N_\mathrm{eff}^{(\nu)}=0$ at $T=T_0$, from Eq.~(\ref{eq:g_*^S(T).natural.infl}) we get $g_{*0}^S=2$, while at $T=T_r$ all active neutrinos are relativistic and from (\ref{eq:g_*(T).natural.infl})-(\ref{eq:g_*^S(T).natural.infl}) we get
\begin{equation}
    \Bar{g}_* = 2 + \frac{7}{8} \cdot 3 \cdot 2 \cdot \left(\frac{4}{11}\right)^{4/3} \simeq 3.363,
\end{equation}
\begin{equation}
    \Bar{g}_*^S = 2 + \frac{7}{8} \cdot 3 \cdot 2 \cdot \left(\frac{4}{11}\right)^{3/3} \simeq 3.909,
\end{equation}
where we used $g_\nu=2$ and $T_\nu=(4/11)^{1/3}\,T$.  Furthermore, in the SM the functions (\ref{eq:g_*(T).natural.infl})-(\ref{eq:g_*^S(T).natural.infl}) at high temperatures ($T\gtrsim100\,\mathrm{GeV}$) are equal and saturate the constant value $g_*^{(S)} \simeq106.75$. Inserting $g_*(T)=g_*^S(T)=106.75$ in Eq.~(\ref{eq:c(T).function}) we have $c(T)\simeq0.534$, and from Eq.~(\ref{eq:T_k^(2)}) we therefore obtain
\begin{equation}
\begin{aligned}
    T_k\sim T_k^{(2)}&\simeq (9.46\times10^{30}) \cdot k\\
    &\simeq (3.91\times10^7\,\mathrm{GeV})\cdot\frac{\nu}{\mathrm{Hz}}.
\end{aligned}
\end{equation} 
We see that $T_k\gtrsim100\,\mathrm{GeV}$ can indeed be in the frequency range $0.005\,\mathrm{Hz}\lesssim\nu\lesssim1\,\mathrm{Hz}$, which  DECIGO,  BBO and ALIA are most sensitive to. 

We now have all the necessary elements to calculate the GW spectral density: from Eq.~(\ref{eq:Omega_GW(nu)}) we finally obtain
\begin{equation}
    h_0^2\Omega_\mathrm{GW}(\nu)\simeq (8.353\times10^{-16})\cdot r_{0.05}\, \left(2.059\times10^{15}\cdot\frac{\nu}{\mathrm{Hz}}\right)^{-r/8}.
    \label{eq:Omega_GW(nu)_Standard_Model_non-rel.neutrino}
\end{equation}
It is important to underline that Eq.~(\ref{eq:Omega_GW(nu)_Standard_Model_non-rel.neutrino}) is valid under two assumptions: first, that there are no light species other than those of the SM; second, that we can neglect the contribution of reheating as we have done so far. Reheating becomes relevant when the reheating temperature is below $\sim 10^{11}\,\mathrm{GeV}$. It is interesting to note that if you have many weakly coupled scalars coupled to the SM at high enough energies you can have a much higher reheating temperature, even around $10^{15}\,\mathrm{GeV}$~\cite{Salvio:2019wcp}.

\begin{figure}[t]
\begin{center}
    \includegraphics[width=0.75\textwidth]{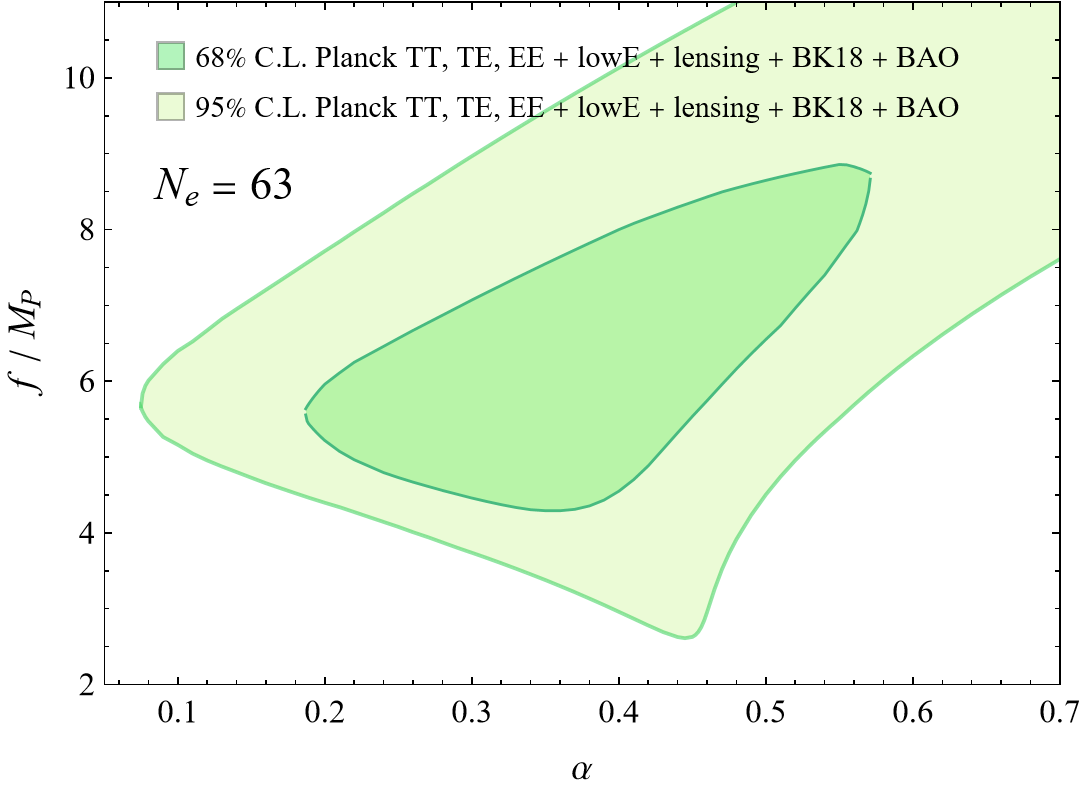}
    \caption{\em Parameter space of slow-roll natural inflation with periodic non-minimal coupling for $N_e=63$ on the $\{\alpha,f\}$  plane that is allowed at $95\%$ $\mathrm{C.L.}$ (light-green region) and at $68\%$ $\mathrm{C.L.}$ (darker green region) by CMB observations. }
    \label{fig:f(alpha)_N=63}
\end{center}
\end{figure}

\begin{figure}[t]
\begin{center}
    \includegraphics[width=0.49\textwidth]{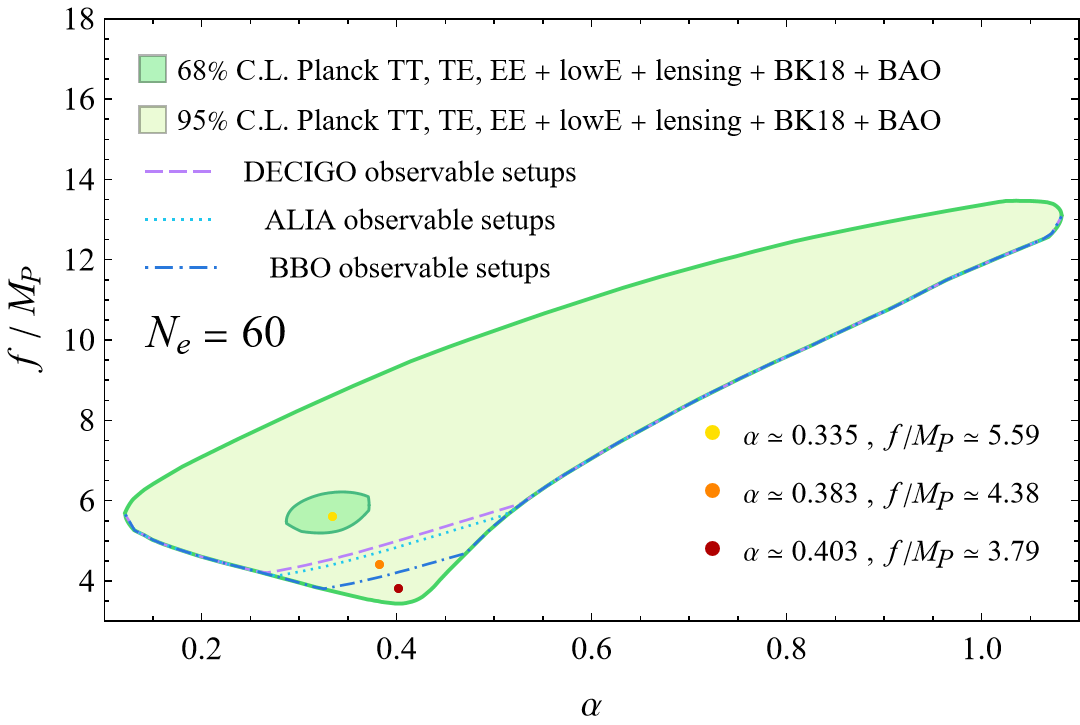} \,\,
\includegraphics[width=0.49\textwidth]{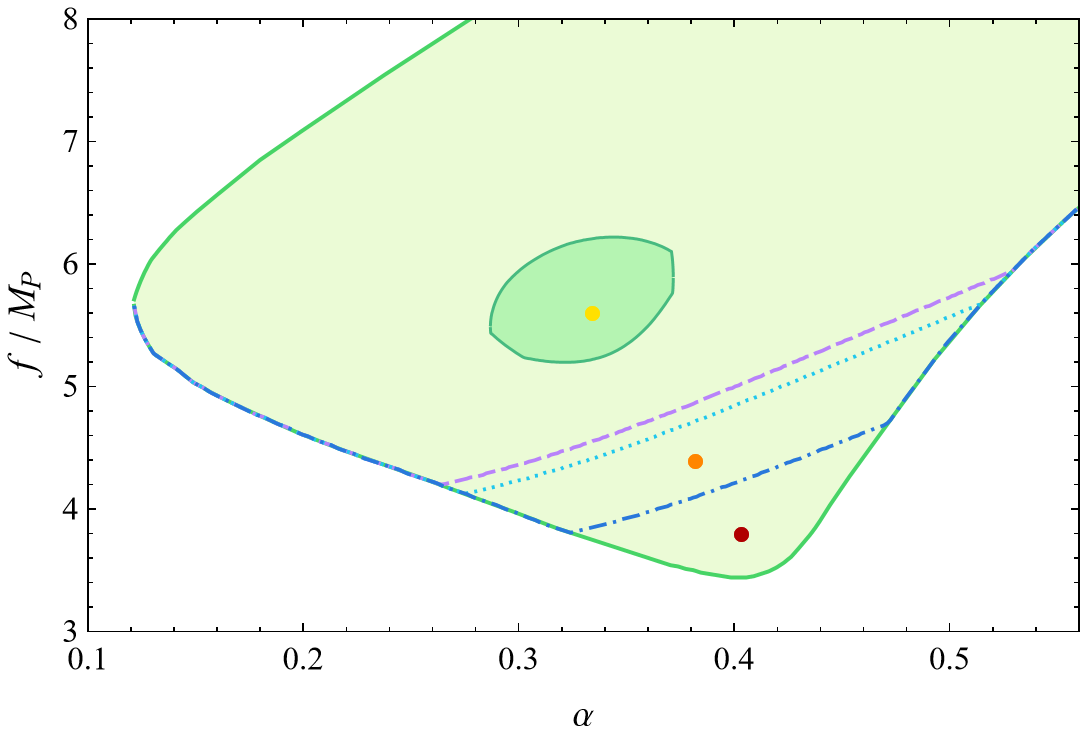}
    \caption{{\em Parameter space of slow-roll natural inflation with periodic non-minimal coupling for $N_e=60$ on the $\{\alpha,f\}$  plane that is allowed at $95\%$ $\mathrm{C.L.}$ (light-green region) and at $68\%$ $\mathrm{C.L.}$ (darker green region) by CMB observations. Moreover, each point on the $\{\alpha,f\}$ plane that lies above the purple dashed line, the light-blue dotted line, or the blue dot-dashed line represents an observationally admitted setup (at least with $95\%$ $\mathrm{C.L.}$) whose produced relic GW background could be observed in the future by DECIGO, ALIA or BBO, respectively. }\textbf{Left}:  {\em full parameter space allowed at $95\%$ $\mathrm{C.L.}$.} \textbf{Right}:  {\em zoom around the  $68\%$ $\mathrm{C.L.}$ region.}   }
   \label{fig:observable.models.N=60}
\end{center}
\end{figure}

\begin{figure}[t]
    \centering
    \includegraphics[scale=0.6]{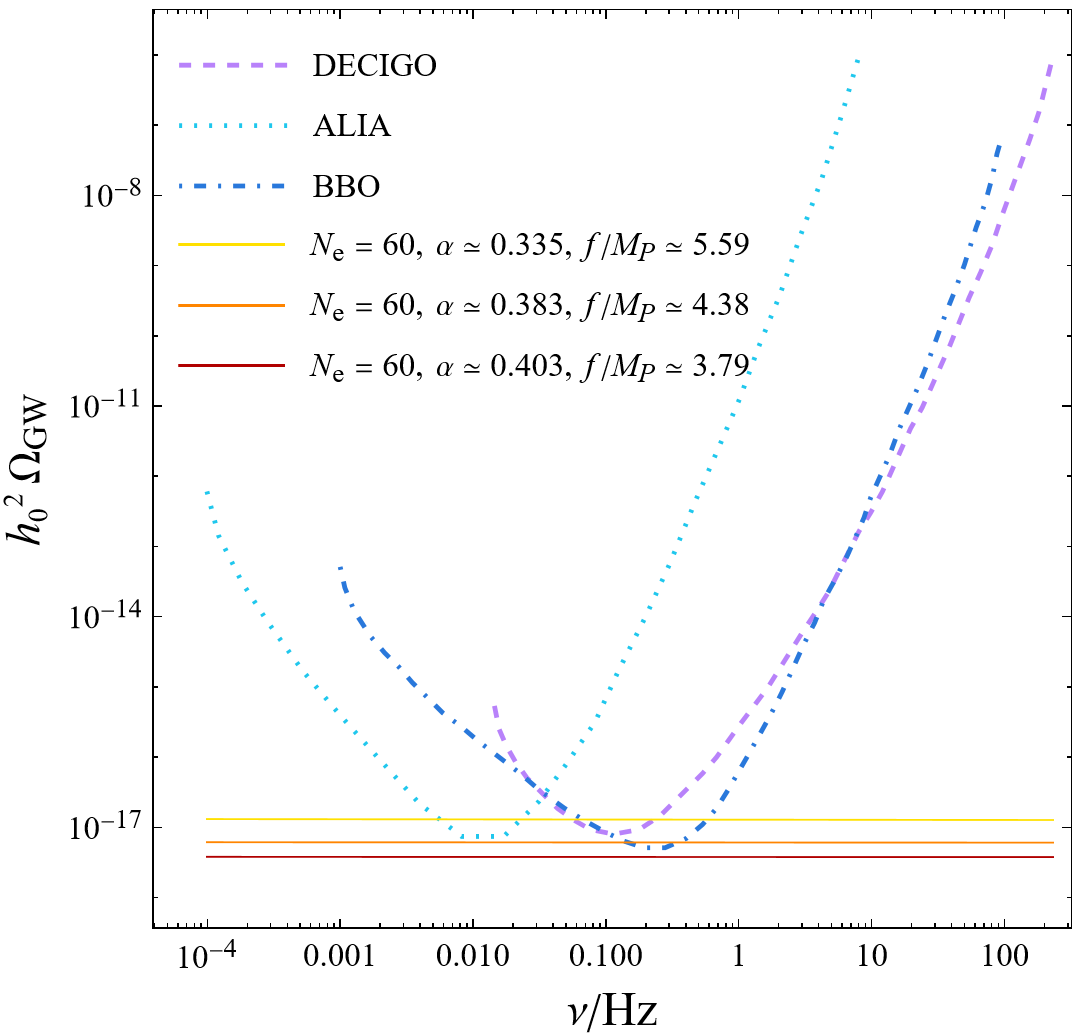}
    \caption{\em The three sensitivity curves of DECIGO, ALIA and BBO  together with the spectral density,  $h_0^2\Omega_\mathrm{GW}(\nu)$, of the relic GW background produced by the three setups depicted by the yellow, orange, and red benchmark points in Fig.~\ref{fig:observable.models.N=60}.  }
    \label{fig:interferometers.sensitivity.curves}
\end{figure}

From Secs.~\ref{The model},~\ref{ch:Slow-roll.inflation} and~\ref{ch:the.natural.inflation.parameter.space} we know that a parameter setup of our natural inflation with non-minimal coupling is uniquely determined by three numbers: the number of e-folds before the end of inflation $N_e$, the non-minimal-coupling parameter $\alpha$ and the energy scale $f$ in the natural-inflation potential. The other scale $\Lambda$ in the natural-inflation potential has been fixed by requiring the curvature power spectrum to reproduce the observed value, Eq.~(\ref{PRobserved}). Then, for each setup $\{N_e,\alpha,f\}$ we can evaluate through Eq.~(\ref{eq:Omega_GW(nu)_Standard_Model_non-rel.neutrino}) the spectral density $\Omega_\mathrm{GW}(\nu)$ associated with the relic GW background. 

Comparing the result with the sensitivity curves of  DECIGO,  BBO and ALIA we can verify whether the considered setup $\{N_e,\alpha,f\}$ produces a directly detectable GW signal. Here those curves are constructed as power-law integrated sensitivity curves and are determined  following the method described in Refs.~\cite{Thrane:2013oya,Dev:2019njv}. For a fixed value of $N_e$, we first found all the observationally admitted setups $\{\alpha,f\}$ (which are compatible with CMB observations) and then, for each setup admitted at least with $95\%$ $\mathrm{C.L.}$, we evaluated $\Omega_\mathrm{GW}$ and observed if this signal falls above the sensitivity curves of the future space-borne interferometers DECIGO, BBO and ALIA for some frequency $\nu$. 

In Fig.~\ref{fig:observable.models.N=60} we can see the result of this procedure for $N_e=60$. The green regions are the observationally admitted setups analogous to those shown in Fig.~\ref{fig:f(alpha)_N=63} for $N_e=63$, while the purple, light blue and blue lines represent the boundaries of GW detectable signals: each point on the $\{\alpha,f\}$ plane which belongs to the green regions and lies above the purple, light blue or blue line represents an observationally admitted setup (at least with $95\%$ $\mathrm{C.L.}$) whose produced relic GW background is potentially detectable by DECIGO, ALIA or BBO, respectively. Fig.~\ref{fig:observable.models.N=60} also shows three benchmark points representing three different  observationally admitted natural-inflation setups:  $\{\alpha,f/M_P\}\simeq\{0.335,5.59\}$ (yellow dot) is admitted within $68\%$ $\mathrm{C.L.}$, while $\{\alpha,f/M_P\}\simeq\{0.383,4.38\}$ (orange dot) and $\{\alpha,f/M_P\}\simeq\{0.403,3.79\}$ (red dot) are both admitted within $95\%$ $\mathrm{C.L.}$. We see that the first setup (yellow) produces a GW background that in principle is observable by all considered interferometers, while the second (orange) can only be  observed by BBO, and the last one (red) produces a GW signal so weak that it is not observable by any interferometer  considered here. In Fig.~\ref{fig:interferometers.sensitivity.curves} we can see the GW spectral density, $h_0^2\Omega_\mathrm{GW}(\nu)$, produced by these three setups together with the three sensitivity curves of the interferometers DECIGO, ALIA and BBO.  We can see that the `yellow' setup is actually above all the minima of the three curves, while the `orange' setup exceeds only the minimum of the  BBO curve, and the `red' one is beneath all the three sensitivity curves. Note that the interferometers are clearly  more sensitive to distinct frequencies: ALIA, DECIGO, and BBO are most sensitive to the frequencies $\nu_\mathrm{ALIA}\simeq0.013\,\mathrm{Hz}$, $\nu_\mathrm{DECIGO}\simeq0.11\,\mathrm{Hz}$, and $\nu_\mathrm{BBO}\simeq0.24\,\mathrm{Hz}$ respectively. 

It is remarkable, as we can see in Fig.~\ref{fig:observable.models.N=60}, that all the $N_e=60$ setups that are observationally admitted within $68\%$ $\mathrm{C.L.}$ (as well as the vast majority of setups admitted within $95\%$ $\mathrm{C.L.}$) generate a GW background that, at least in principle, is well observable by all three space-borne interferometers.

\section{A multi-field natural inflation: the natural-scalaron case}\label{A multi-field natural inflation: the natural-scalaron case}

The purpose of this section is to extend the analysis of Sec.~\ref{Natural inflation with a periodic non-minimal coupling} to a multi-field scenario. Given the motivations of scalaron (a.k.a. Starobinsky) inflation, this will be achieved by adding an $R^2$ term.

\subsection{The model}\label{NSmodel}

The natural-scalaron model has been introduced in Ref.~\cite{Salvio:2021lka}. We refer the reader to this article for the explanation of any non-trivial statement contained in this subsection. In this model the part of the action $S_{\rm infl}$ responsible for inflation is 
\begin{equation}
    S_{\rm infl} = \int d^4x \sqrt{-g} \left[ \frac{F(\phi)}{2}M_P^2R + \beta R^2 - \frac{1}{2}(\partial\phi)^2 - V(\phi) \right]
    \label{eq:natural.inflation.with.R^2}
\end{equation}
where
again $F(\phi)$ and $V(\phi)$ are defined in~(\ref{eq:F(phi)}) and~(\ref{eq:V(phi)}), respectively. The $R^2$ features a real coefficient $\beta$ that must be positive in order for the system to be stable. Since $\Lambda_\mathrm{cc}$ is negligibly small and $M_P$ can be used to set the units of energy, the natural-scalaron model effectively depends on four parameters only: $\beta, \alpha, f, \Lambda$.

After introducing an auxiliary field and performing a field redefinition (a Weyl transformation) one obtains
\begin{equation} S_{\rm infl} = \int d^4x\sqrt{-g}\left[\frac{\bp^2}{2}R-\Lag_{\rm kin}-U \right], \label{Gammast}\end{equation}
where 
\be  \Lag_{\rm kin} \equiv \frac{6\bp^2}{z^2}
 \frac{(\partial \phi)^2 + (\partial z)^2}{2},\nonumber \ee
 \be U(\phi,z)\equiv  \frac{36\bp^4}{z^4}\bigg[{V(\phi)}+   \frac{1}{16\beta}\left(\frac{z^2}{6} -M_P^2F(\phi)\right)^2 \bigg],\nonumber\ee
 and $z$ is an additional scalar, which we will refer to as the scalaron, given by $$z=\sqrt{6M_P^2F(\phi)+24\beta R_J}.$$ 
 Here $R_J$ is the Ricci scalar constructed with the original ``$J$ordan-frame" metric (before performing the Weyl transformation). Thus, the full inflationary system features two scalars, the PNGB $\phi$ and the scalaron $z$; this is the reason why we refer to this model as the natural-scalaron one.
 
 One can go to the pure natural inflation discussed in Sec.~\ref{Natural inflation with a periodic non-minimal coupling} or to the pure scalaron inflation in the $\rho\ll 1$ or $\rho\gg 1$ limits, respectively, where 
\begin{equation}
    \rho \,\equiv\, \frac{\sqrt{\beta}\Lambda^2}{M_P^2}.
    \label{eq:rho.natural-scalaron.parameter}
\end{equation}

Neglecting the tiny $\Lambda_\mathrm{cc}$, the stationary points of the  Einstein-frame potential $U$ are as follows. For $\alpha\leq 1/2$ the only minimum (modulo the $2\pi f$ periodicity) is the absolute minimum, 
\be \phi_1 =\pi f, \qquad z_1 = \sqrt{6}M_P.\ee 
 But for $\alpha>1/2$ there are two non-trivial minima ($\{\phi_1,z_1\}$ and $\{\phi_2,z_2\}$), one of which, $\{\phi_2,z_2\}$, with
 \be \phi_2 = 0, \qquad z_2 = \sqrt{6}M_P \, \sqrt{1+2\alpha + \frac{32\rho^2}{1+2\alpha}}\ee has a value of $U$ (the quantity $U(\phi_2,z_2)$) that is not negligibly small during inflation.  When $\alpha<1/2$ the configuration $\{\phi_2,z_2\}$ is a saddle point and there are no other stationary points apart from $\{\phi_1,z_1\}$ and $\{\phi_2,z_2\}$. When $\alpha>1/2$ there appear  two more stationary points (modulo the $2\pi f$ periodicity) that turn out to be saddle points: 
\begin{align}
    &\phi_3 = f \arccos \left( \frac{1-\alpha}{\alpha} \right), \qquad  z_3 = 2\sqrt{3}M_P\, \sqrt{1+\frac{4\rho^2}{\alpha}} \label{eq:phi_3}\\
    &\phi_4 = - f \arccos \left( \frac{1-\alpha}{\alpha} \right), \qquad z_4 = z_3.
\end{align}
So, like in the pure-natural case, for $\alpha>1/2$ one should be careful not to end the inflationary path in the false minimum $\{\phi_2,z_2\}$.

\subsection{Multi-field slow-roll inflation}

 In order to derive the relevant inflationary formul\ae~it is convenient in this subsection to start with a more general framework (see Refs.~\cite{Salvio:2021lka,Salvio:2022mld} for more details). Notice that the action in~(\ref{Gammast}) belongs to the class 
\be S_I=\int d^4x  \sqrt{-g} \,\bigg[ \frac{\bp^2}{2}R  - 
\frac{K_{ij}(\Phi) }{2}\partial_\mu \phi^i\partial^\mu \phi^j-
U(\Phi)
\bigg], \label{action}
 \ee
 where $\Phi$ is an array of scalar fields with components $\phi^i$ and $K_{ij}$ is a field metric.
  For a generic function $\mathscr{F}$ of $\Phi$, we define $\mathscr{F}_{,i}\equiv \partial \mathscr{F}/\partial \phi^i$
 and $K^{ij}$ represents the inverse of the field metric $K_{ij}$ (which is used to raise and lower the scalar indices $i,j, ...$); for example $\mathscr{F}^{,i}\equiv K^{ij}\mathscr{F}_{,j}$. 
 
  To describe the classical part of inflation we assume a spatially-flat Friedmann-Robertson-Walker metric (during inflation the energy density is dominated by the scalar fields so the curvature contribution can be neglected). 

    In the slow-roll regime  the equations for the scalar and the cosmological scale factor $a$  read 
    \be \dot \phi^i\simeq -\frac{U^{,i}}{3H}, \qquad H^2 \simeq \frac{U}{3  \bp^2} ,\qquad  \implies \dot \phi^i\simeq-\frac{\bp U^{,i}(\Phi)}{\sqrt{3U(\Phi)}}.   \label{slow-roll-eq}\ee
    The number of e-folds $N$ (from a generic time $t$ until the time $t_e$ when inflation ends) is given by
 \be N = \int^t_{t_e} dt' H(t'). \label{Ndef3} \ee 
    When slow roll holds $N$ can be considered as a function of the scalar fields $\Phi$ by requiring $t-t_e$ in Eq.~(\ref{Ndef3}) to be the time it takes for the system to reach the end of inflation starting with initial condition $\Phi$: this is because in the slow-roll approximation the  scalar field equations are of first order (see the last equation in~(\ref{slow-roll-eq})).
 
    Now, it is convenient to introduce the unit vector $\hat\sigma^i$ tangent to the inflationary path $\phi_0^i$,
\be \hat\sigma^i\equiv \frac{\dot\phi_0^i}{\dot\sigma}, \qquad \dot\sigma\equiv \sqrt{K_{ij}(\Phi_0)\dot\phi_0^i\dot\phi_0^j}. \ee 
Besides $\hat\sigma^i$ it is also useful to introduce the set of unit vectors $\hat s^i$ orthogonal to the inflationary path. In the presence of two inflatons
  we have only one of such orthogonal unit vectors   (see e.g.~\cite{Gundhi:2018wyz})
  and, for actions of the form~(\ref{Gammast}), its explicit expression is
 \be  \hat s^1 \equiv  \frac{\dot z_0}{\dot\sigma}, \qquad \hat s^2 \equiv  \frac{-\dot \phi_0}{\dot\sigma}.\ee
  
  When inflation is driven by 
  a generic number of scalar fields
  slow-roll occurs if two conditions are satisfied (see also \cite{Chiba:2008rp} for previous studies):
\be \epsilon \equiv  \frac{  \bp^2 U_{,i}U^{,i}}{2U^2} \ll 1. \label{1st-slow-roll}\ee 
\be |\eta_{\sigma\sigma}|\ll 1, \qquad \mbox{where} \qquad \eta_{\sigma\sigma}\equiv  \hat\sigma^i\hat\sigma^j \eta_{i j}, \quad \mbox{and} \quad\quad \eta^{i}_{\,\,\, j}\equiv \frac{\bp^2 \nabla_j\nabla^iU}
{U}, \label{2nd-slow-roll}\ee
where  $\nabla_i$ is the covariant derivative on the field space computed with the Levi-Civita connection of the field metric $K_{ij}$ in~(\ref{action})


The function of the scalar fields $N$ defined above allows us to compute the curvature power spectrum $\mathcal{P}_\mathcal{R}$, the (curvature)  scalar spectral index  $n_s$ and the tensor-to-scalar ratio $r$. The explicit formul\ae~are~\cite{Sasaki:1995aw,Chiba:2008rp} (here we evaluate the power spectra  at horizon exit, $k=a H$)
\be \mathcal{P}_\mathcal{R}=\left(\frac{H}{2\pi}\right)^2 N_{,i}N^{,i},\label{power-spectrum}\ee
\be  n_s =1-2\epsilon - \frac{ 2 }{ \bp^2 N_{,i}N^{,i}}+\frac{2\eta_{ij}N^{,i}N^{,j}}{N_{,k}N^{,k}},  \label{nsFormula}
\ee
\be r\equiv \frac{\mathcal{P}_T}{\mathcal{P}_\mathcal{R}}=\frac{8}{ \bp^2 N_{,i}N^{,i}}. \label{rW}\ee
Moreover, the tensor power spectrum $\mathcal{P}_T$ and the corresponding spectral index $n_T$ are given by~\cite{Chiba:2008rp}
\be  \mathcal{P}_T =  \frac{8}{M_P^2} \left(\frac{H}{2\pi}\right)^2, \qquad n_T=-2\epsilon. \label{PtspectrumnT} \ee 
independently of the number of inflatons.  This result for $n_T$ reduces to $n_T = -r/8$
in single-field inflation (where the tensor-to-scalar ratio is $16\epsilon$). For an arbitrary number of inflatons we have 
\be n_T\leq -\frac{r}{8}\ee
due to the possible presence of isocurvature modes (those associated with the directions $\hat s^i$). The normalized variation between the tensor spectral index $n_T$ and the corresponding single-field quantity $-r/8$ is, therefore,
    \begin{equation}
        \delta n_T \equiv \frac{(-r/8) - n_T}{|n_T|}\geq0.
        \label{eq:delta.n_T^*}
    \end{equation}
Moreover, the normalized variation between the natural-scalaron tensor-to-scalar ratio $r$ and the corresponding single-field quantity $-8n_T$ is
    \begin{equation}
        \delta r \equiv \frac{(-8n_T) - r}{r}\geq0.
        \label{eq:delta.r_0.05^*}
    \end{equation}
     The quantities $\delta n_T$ and  $\delta r$  give us an estimate of the relevance of isocurvature modes, and, as we will see in Sec.~\ref{ch:the.relic.inflationary.GW.background(natural-scalaron)}, allow us to determine whether it is possible to distinguish between single-field and multi-field inflation  through GW observations performed by future interferometers. Recalling that both in single-field and multi-field slow-roll inflation we have $n_T=-2\epsilon$, we obtain
    \begin{equation*}
        \delta n_T = 1 - \frac{r}{16\epsilon}, \quad\,\,\,\,\quad \delta r = \frac{16\epsilon}{r} - 1
    \end{equation*}
    so that 
    \begin{equation}
        \delta n_T = \frac{\delta r}{1+\delta r} \leq \delta r.
        \label{eq:delta.n_T<delta.r_0.05}
    \end{equation}

In multi-field inflation, just like in single-field inflation, a rescaling of the potential $U\to \lambda U$ affects only the scalar and tensor power spectra 
\be\mathcal{P}_\mathcal{R}\rightarrow\lambda\mathcal{P}_\mathcal{R}, \quad\mathcal{P}_T\rightarrow\lambda\mathcal{P}_T, \quad \epsilon\rightarrow \epsilon, \quad \eta^i_{\,\,j} \rightarrow \eta^i_{\,\,j}, \quad n_s\rightarrow n_s, \quad n_T\rightarrow n_T, \quad r\rightarrow r\label{MultiScale}\ee  (the slow-roll parameters $\epsilon$ and $\eta^i_{\,\,j}$, the scalar and tensor spectral indices $n_s$ and $n_T$, and the tensor-to-scalar ratio $r$ are left invariant). Similarly to what we have done in Sec.~\ref{Natural inflation with a periodic non-minimal coupling}, we will use the scaling property in~(\ref{MultiScale}) to eliminate the dependence of the natural-scalaron model on the energy scale $\Lambda$ by imposing the observational constraint in~(\ref{PRobserved}).  





Following Ref.~\cite{Kaiser:2012ak} (see also~\cite{Gordon:2000hv} for a previous work with a flat field metric), the most important quantities to estimate the size of the isocurvature perturbations in the slow-roll approximation are the elements of the field-dependent covariant squared-mass matrix, $m_{ij}^2 \equiv \nabla_i\nabla_j U$.  The key quantities are in particular the projections of $m_{ij}^2$ on $\hat\sigma^i$ and the $\hat s^i$:
\be \mu^2_\sigma \equiv \hat\sigma^i\hat\sigma^jm_{ij}^2,\qquad \mu^2_s \equiv \hat s^i\hat s^jm_{ij}^2.  \label{muDef}\ee
The effective  mass $\mu_\sigma$ corresponds to the usual curvature perturbations, while
   $ \mu_s$ corresponds to the isocurvature perturbations.

Given an isocurvature mode $I$ with power spectrum $\mathcal{P}_I$, Planck data on isocurvature perturbations~\cite{Ade:2015lrj} constraints the ratio
\be \beta_{\rm iso} \equiv \frac{\mathcal{P}_I}{\mathcal{P}_\mathcal{R}+\mathcal{P}_I}. \ee 
Noting that anyhow $\mathcal{P}_\mathcal{R}$ should be large compared to $\mathcal{P}_I$ to satisfy those observational constraints
\be \beta_{\rm iso}\simeq  \frac{\mathcal{P}_I}{\mathcal{P}_\mathcal{R}}
= \frac{\mathcal{P}_I}{\mathcal{P}_T} \frac{\mathcal{P}_T}{\mathcal{P}_\mathcal{R}}.
\ee
The quantity $\mathcal{P}_T/\mathcal{P}_\mathcal{R}$ gives us the tensor-to-scalar ratio $r$ and, using the same normalization for the tensor and isocurvature perturbations,
 one obtains (at the leading order in the slow-roll approximation, where the universe expands with constant Hubble rate $H$)
\begin{equation}
  \beta_\mathrm{iso}  \,\simeq\, \left|\sqrt{\frac{\pi}{2}}\, (-k\tau)^{3/2}\, H_n^{(1)}(-k\tau) \right|^2r,
    \label{eq:beta_iso.two-field.model}
\end{equation}
 where $\tau$ is the conformal time, $H_n^{(1)}(x)$ is the Hankel function of the first kind and $$n=\sqrt{(9/4)-(\mu_s/H)^2}.$$
In the rest of this paper we will evaluate $\beta_{\rm iso}$ in the superhorizon limit.

 \subsection{Predictions of natural-scalaron inflation and observational constraints}
\label{ch:the.natural-scalaron.parameter.space}

In this section we provide a detailed study of the inflationary predictions of the natural-scalaron model with a periodic non-minimal coupling introduced in Sec.~\ref{NSmodel} and compare those predictions with the most recent observational constraints of Refs.~\cite{Ade:2015lrj} and~\cite{BICEP:2021xfz}.

\begin{figure}[t!]
\begin{center}
    \includegraphics[width=0.50\textwidth]{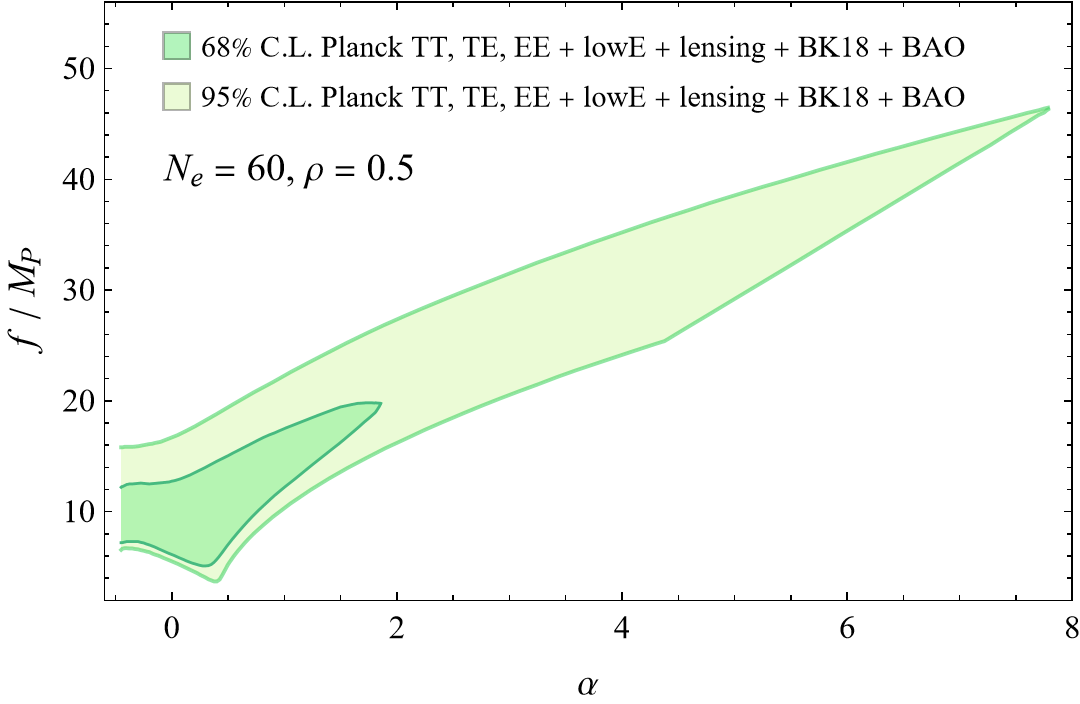} \hspace{0.5cm}
\includegraphics[width=0.45\textwidth]{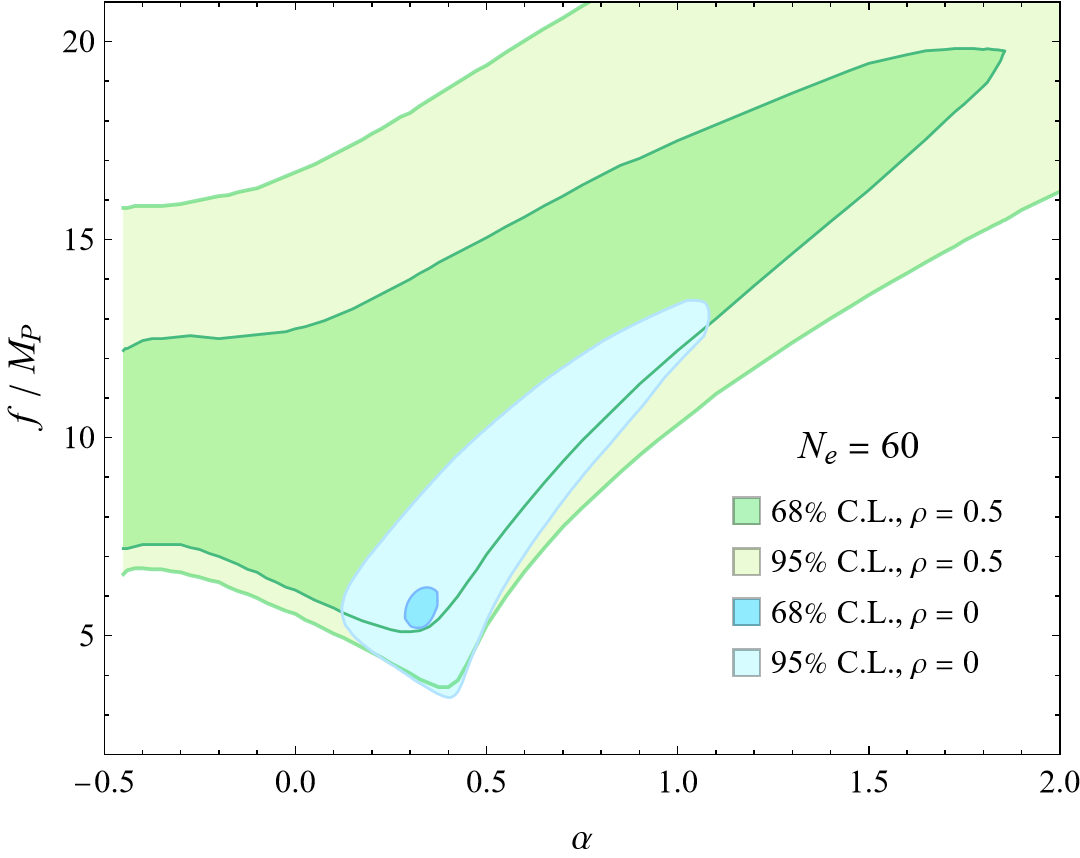}
    \caption{{\em All the observationally admitted setups of natural-scalaron slow-roll inflation for $N_e=60$ and $\rho=0.5$ on the $\{\alpha,f\}$ plane. In this plot we imposed the cutoff $\epsilon<0.05$ and $\eta_{\sigma\sigma}<0.05$ at $N_e$ e-folds before the end of inflation.} \textbf{Left:} {\em the dark and light green regions represent $\{N_e,\rho\}=\{60,0.5\}$ setups allowed by the most recent observational constraints on inflation~\cite{Ade:2015lrj,BICEP:2021xfz} at $68\%$ $\mathrm{C.L.}$ and $95\%$ $\mathrm{C.L.}$, respectively.} \textbf{Right:}  {\em zoom in the $68\%$ $\mathrm{C.L.}$ region where we superimposed in dark and light azure the $\{N_e,\rho\}=\{60,0\}$ setups allowed by the most recent observational constraints at $68\%$ $\mathrm{C.L.}$ and $95\%$ $\mathrm{C.L.}$, respectively (the allowed single-field natural parameter space for $N_e=60$). }}
    \label{fig:f(alpha)_N=60_rho=0.5}
\end{center}
\end{figure}

In Fig.~\ref{fig:f(alpha)_N=60_rho=0.5} we can see all the observationally admitted natural-scalaron setups for $\{N_e,\rho\}=\{60,0.5\}$ on the $\{\alpha,f\}$ plane (the setups corresponding to values of $n_s$ and $r$ in agreement with the current CMB constraints). Moreover, in the right plot we superimposed in azure the observationally admitted $\{N_e,\rho\}=\{60,0\}$ setups, which are just the single-field natural setups for $N_e=60$. This comparison allows us to appreciate how much the presence of the scalaron ($\rho>0$) widens the allowed regions. 
The visible clipping for $\alpha\gtrsim4.4$ on the lower profile of the allowed-setup region (smaller $f$) is the effect of the cutoff $\epsilon<0.05$ and $\eta_{\sigma\sigma}<0.05$ that we imposed at $N_e$ e-folds before the end of inflation. This cutoff has, however, no effect on the most relevant  $68\%$ $\mathrm{C.L.}$ region. 

\begin{figure}[!t]
\begin{center}
   \includegraphics[width=0.43\textwidth]{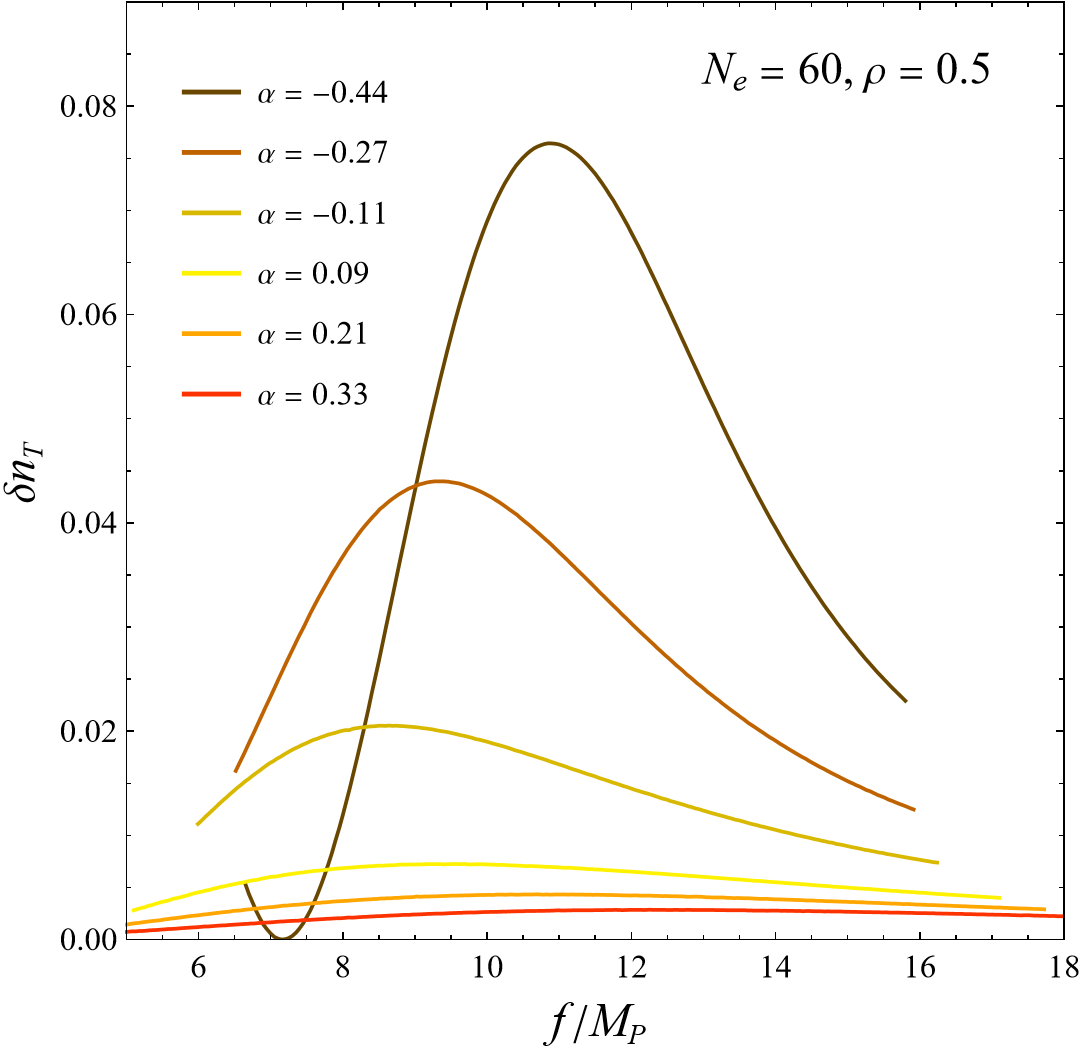} 
    \hspace{0.2cm}
 \includegraphics[width=0.43\textwidth]{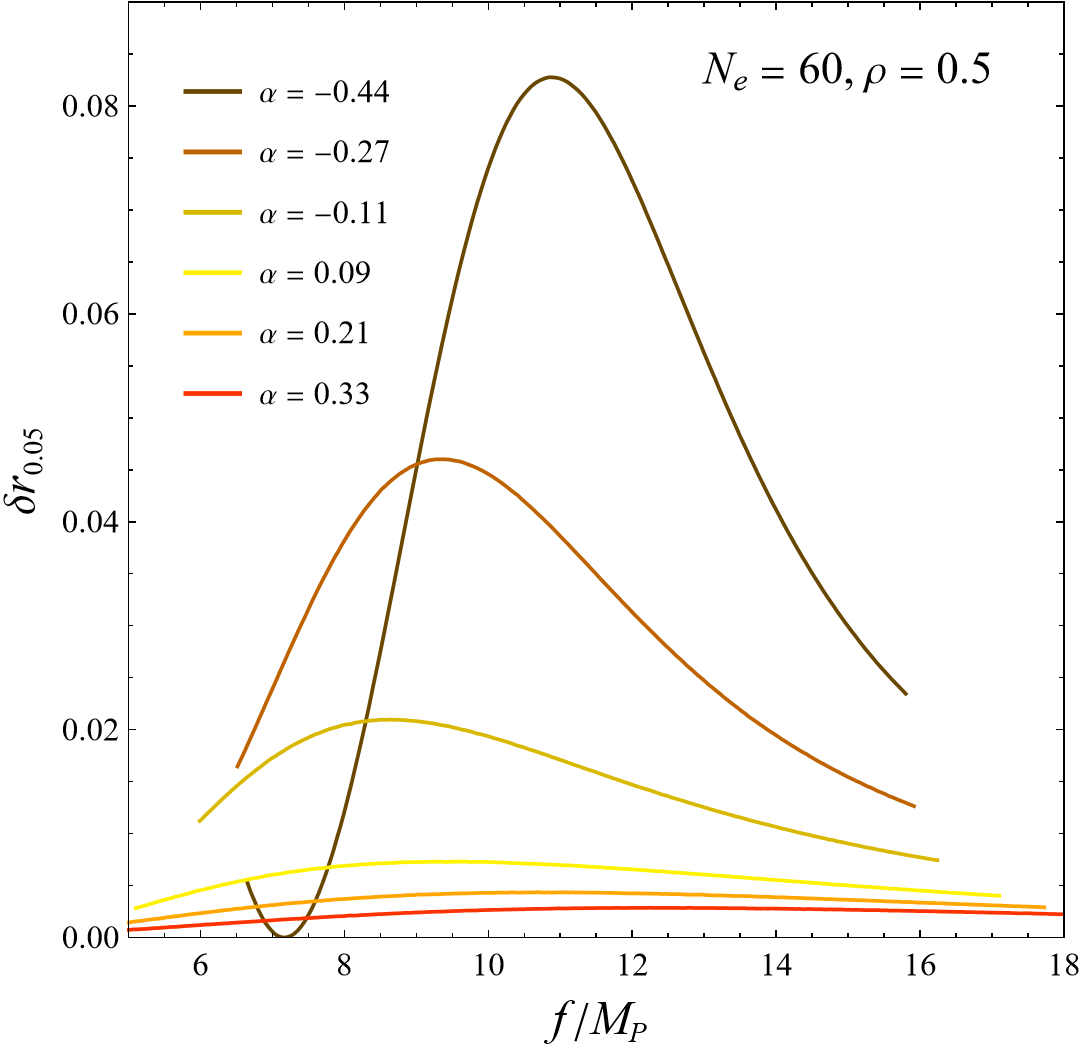}   
    \vspace{0.cm}
    \caption{\em The variation of the tensor spectral index $\delta n_T$ (left plot) and the  variation of the tensor-to-scalar ratio $\delta r_{0.05}$ computed at the pivot scale $k_*=0.05\,\mathrm{Mpc^{-1}}$ (right plot), defined by Eqs.~(\ref{eq:delta.n_T^*})-(\ref{eq:delta.r_0.05^*}), are here shown for some values of $\alpha$ as $f$ varies (both $\alpha$ and $f$ are chosen inside the green regions of Fig.~\ref{fig:f(alpha)_N=60_rho=0.5}). }
    \label{fig:delta.nT&delta.r_N=60_rho=0.5}
\end{center}
\end{figure}

In Fig.~\ref{fig:delta.nT&delta.r_N=60_rho=0.5} we can see the variation of the tensor spectral index $\delta n_T$ (left plot) and the variation of the tensor-to-scalar ratio $\delta r_{0.05}$ (right plot). Both $\delta r_{0.05}$ and $\delta n_T$ express the normalized variation between the natural-scalaron results  and the corresponding pure-natural quantities $n_T=-r/8$. As we predicted in Eq.~(\ref{eq:delta.n_T<delta.r_0.05}), we see that $\delta n_T\leq\delta r_{0.05}$ for each $\alpha$ and $f$. We can see that the maximum difference between single-field pure-natural and multi-field natural-scalaron results occurs at the lowest values of $\alpha$, for which we have a difference of the order of $10\%$. In Sec.~\ref{ch:the.relic.inflationary.GW.background(natural-scalaron)} we will explore the level of sensitivity required for future space-borne interferometers to resolve this difference. As $\alpha$ increases, the curves of Fig.~\ref{fig:delta.nT&delta.r_N=60_rho=0.5} flatten, and the difference between natural-scalaron and pure-natural results becomes increasingly negligible.

\begin{figure}[t!]
\begin{center}
    \includegraphics[width=0.45\textwidth]{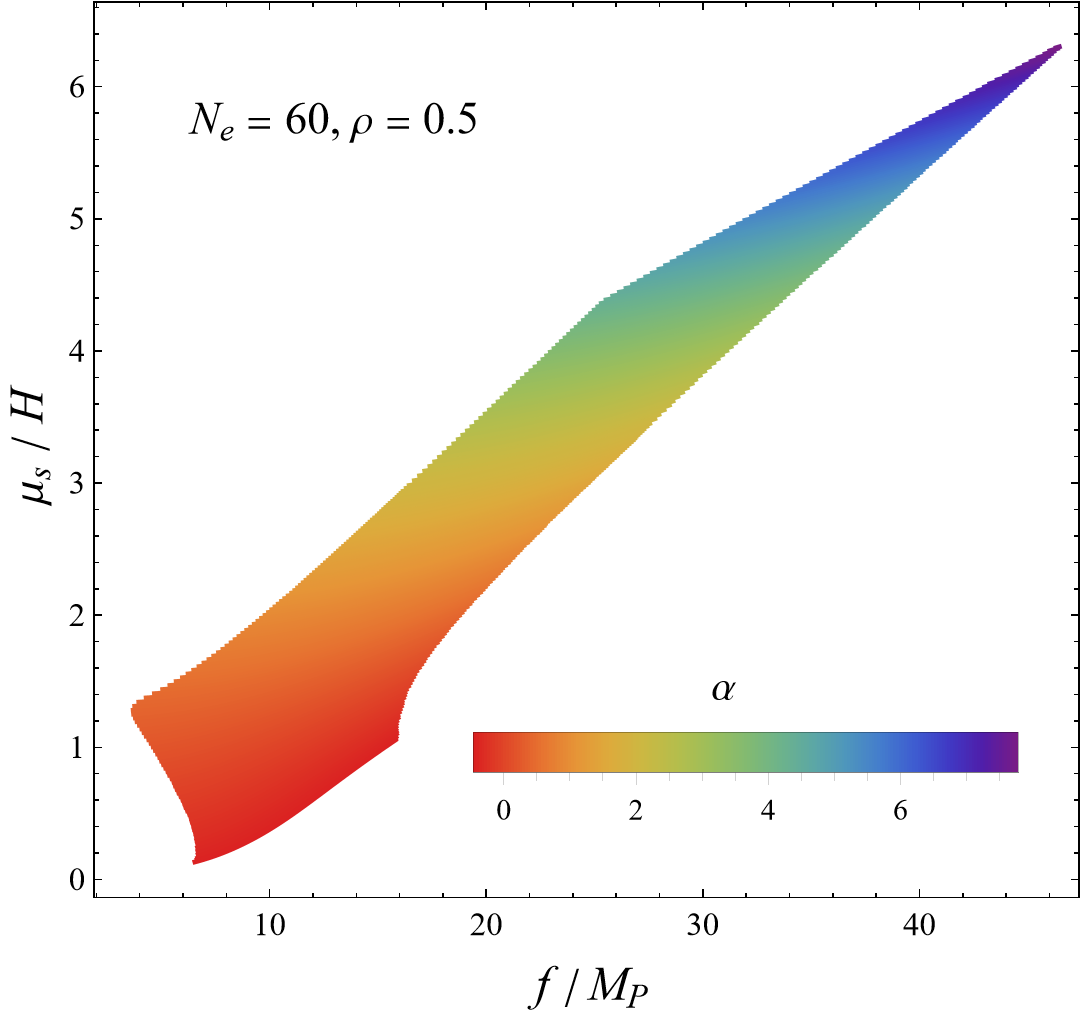}
    \hspace{0.2cm}
    \includegraphics[width=0.52\textwidth]{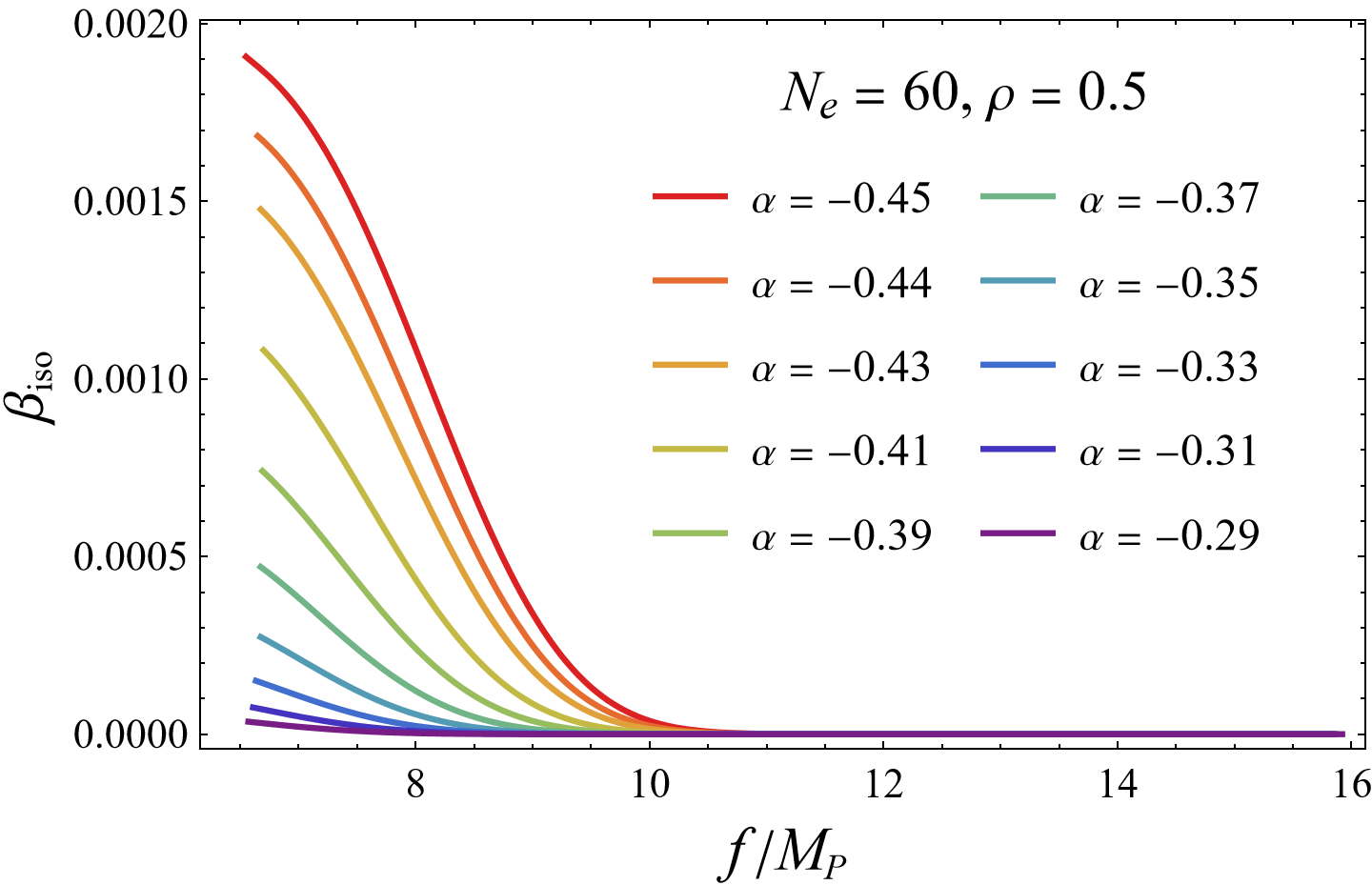}
    \vspace{0.cm}
    \caption{\textbf{Left:} {\em the effective isocurvature mass $\mu_s$  is here shown as $\alpha$ and $f$ range within the entire $95\%$ $\mathrm{C.L.}$ green region of Fig.~\ref{fig:f(alpha)_N=60_rho=0.5}.} \textbf{Right:} {\em the isocurvature ratio $\beta_\mathrm{iso}$ given in Eq.~(\ref{eq:beta_iso.two-field.model}). }}
    \label{fig:mus&biso_N=60_rho=0.5}
\end{center}
\end{figure}

In the left plot of Fig.~\ref{fig:mus&biso_N=60_rho=0.5} we can see the effective isocurvature mass fraction $\mu_s/H$, where $\mu_s$ is defined in Eq.~(\ref{muDef}) and $H$ is the inflationary Hubble rate computed at the leading order in the slow-roll expansion. As we can see, for each value of $\alpha$ there are at least some allowed values of $f$ (if not all of them) such that $\mu_s/H\gtrsim1$. This means that isocurvature perturbations are heavy enough to be safely neglected. In the right plot of Fig.~\ref{fig:mus&biso_N=60_rho=0.5}, instead, we can see the isocurvature ratio $\beta_\mathrm{iso}$ given in Eq.~(\ref{eq:beta_iso.two-field.model}), for some values of $\alpha$ and $f$ inside the green regions of Fig.~\ref{fig:f(alpha)_N=60_rho=0.5}. We see that $\beta_\mathrm{iso}$ is very small in agreement with the Planck constraints~\cite{Ade:2015lrj} for sufficiently large $f$, when the isocurvature power spectrum is below the~$\%$ of the total curvature plus isocurvature power spectrum. The fact that the highest values of $\beta_\mathrm{iso}$ occur for the lowest values of $\alpha$ matches with the isocurvature perturbations being lighter (small $\mu_s$) for low $\alpha$. This trend also explains why in Fig.~\ref{fig:delta.nT&delta.r_N=60_rho=0.5} the maximum of $\delta r_{0.05}$ and $\delta n_T$ increases as $\alpha$ decreases. Indeed, recall  that the single-field relation $n_T=-r/8$ is no longer valid in multi-field inflation precisely because of the  isocurvature modes, which cause $n_T< -r/8$.

\subsection[Relic natural-scalaron background of gravitational waves]{The relic natural-scalaron background of gravitational waves}
\label{ch:the.relic.inflationary.GW.background(natural-scalaron)}

Similarly to Sec.~\ref{ch:the.relic.inflationary.GW.bakground}, here we want  to find out which natural-scalaron setups among those admitted by CMB observations at least at $95\%$ $\mathrm{C.L.}$ (see Fig.~\ref{fig:f(alpha)_N=60_rho=0.5}) produce a primordial GW background that is potentially observable by the future space-borne interferometers, DECIGO, BBO and ALIA. 

This is done again by comparing the inflationary GW spectral density, $h_0^2\Omega_\mathrm{GW}(\nu)$, with the predictions of the sensitivity curves of the future above-mentioned interferometers. In the multi-field case we have
\begin{equation}
    h_0^2 \Omega_\mathrm{GW}(\nu) \,\simeq\, \frac{1}{24}\,h_0^2\Omega_R \left( \frac{g_*(T_k)}{\Bar{g}_*} \right) \left( \frac{\Bar{g}_*^S}{g_*^S(T_k)} \right)^{4/3} r_{0.05}\, \mathcal{P}_\mathcal{R}(k_*)\left(\frac{\nu}{\nu_*}\right)^{n_T}. 
    \label{eq:Omega_GW(nu)MF}
\end{equation}
which differs from Eq.~(\ref{eq:Omega_GW(nu)}) only because the exponent $-r/8$ in Eq.~(\ref{eq:Omega_GW(nu)}) has been replaced by $n_T$, as in the multi-fied case $n_T=-r/8$ generically no longer holds.  We, therefore, have the opportunity to distinguish between multi-field and a corresponding single-field result: we will compare the spectral density in~(\ref{eq:Omega_GW(nu)MF}) with a single-field spectral density defined by
\begin{equation}
    h_0^2\Omega_\mathrm{GW}^{\,\text{s-f}}(\nu) \,\equiv\, \frac{1}{24}h_0^2\Omega_R \left( \frac{g_*(T_k)}{\Bar{g}_*} \right) \left( \frac{\Bar{g}_*^S}{g_*^S(T_k)} \right)^{4/3} r_{0.05}\, \mathcal{P}_\mathcal{R}(k_*)\left(\frac{\nu}{\nu_*}\right)^{-r/8},
    \label{eq:Omega_gw(f)^s-f.natural-scalaron.single-field}
\end{equation}
which is just~(\ref{eq:Omega_GW(nu)MF}) with $n_T$ replaced by $-r/8$, so that one can study
whether the difference between the multi-field GW spectrum, $\Omega_\mathrm{GW}(\nu)$, and its ``single-field equivalent" $\Omega_\mathrm{GW}^{\,\text{s-f}}(\nu)$ is appreciable through observations.

We recall that we consider here a standard scenario in which the only light species are those of the SM and the lightest neutrino is non-relativistic today. So, performing steps similar to those done in Sec.~\ref{ch:the.relic.inflationary.GW.bakground} we obtain
\begin{align}
    h_0^2\Omega_\mathrm{GW}(\nu)&\simeq (8.353\times10^{-16})\cdot r_{0.05}\, \left(2.059\times10^{15}\cdot\frac{\nu}{\mathrm{Hz}}\right)^{n_T},
    \label{eq:Omega_GW(nu)_Standard_Model_natural-scalaron}\\
    h_0^2\Omega_\mathrm{GW}^{\,\text{s-f}}(\nu)&\simeq (8.353\times10^{-16})\cdot r_{0.05}\, \left(2.059\times10^{15}\cdot\frac{\nu}{\mathrm{Hz}}\right)^{-r/8}.
    \label{eq:Omega_GW^s-f(nu)_Standard_Model_natural-scalaron}
\end{align}
Once again, we have neglected  the contribution of reheating, which, as commented in Sec.~\ref{ch:the.relic.inflationary.GW.bakground},  corresponds to assuming a reheating temperature not below the $\sim 10^{11}\,\mathrm{GeV}$ scale like, e.g., in Ref.~\cite{Salvio:2019wcp}.

\begin{figure}[t!]
\begin{center}
   \includegraphics[width=0.52\textwidth]{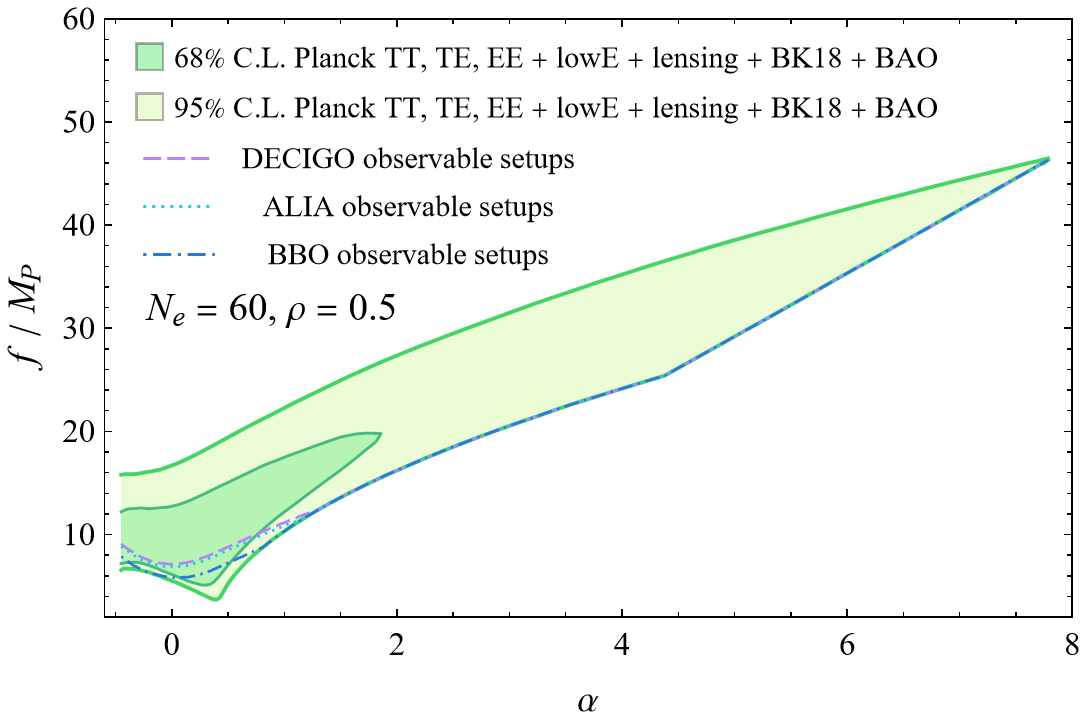}
\hspace{0.5cm}
    \includegraphics[width=0.42\textwidth]{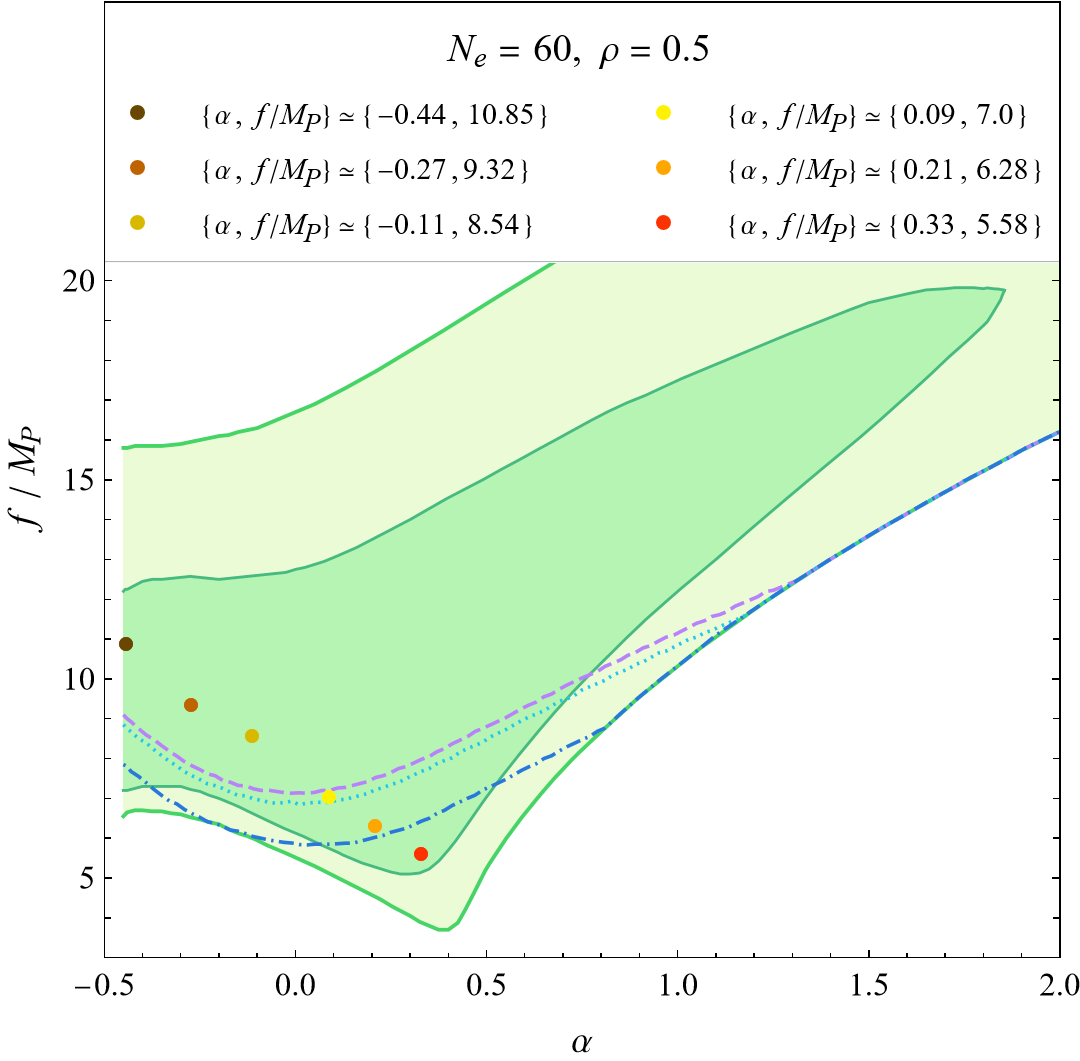}
    \caption{\em For each $95\%$ $\mathrm{C.L.}$ observationally admitted natural-scalaron setup with $\{N_e,\rho\}=\{60,0.5\}$, i.e.~for each setup in the green regions of Fig.~\ref{fig:f(alpha)_N=60_rho=0.5}, we evaluated the GW spectrum and compared it with the sensitivity curves of DECIGO, ALIA and BBO. Each point on this plot which belongs to the green regions and lies above the purple, light blue or blue line represents an observationally admitted natural-scalaron setup (at least at $95\%$ $\mathrm{C.L.}$) whose produced GWs are in principle detectable by DECIGO, ALIA and BBO, respectively. In the right plot we zoom in the $68\%$ $\mathrm{C.L.}$ region and  display  six benchmark points, whose GW spectra are depicted in Fig.~\ref{fig:GW.spectrum.1_N=60_rho=0.5}.}
    \label{fig:f(alpha)_observable.setups_N=60_rho=0.5}
\end{center}
\end{figure}

\begin{figure}[t!]
\begin{center}
    \includegraphics[width=0.45\textwidth]{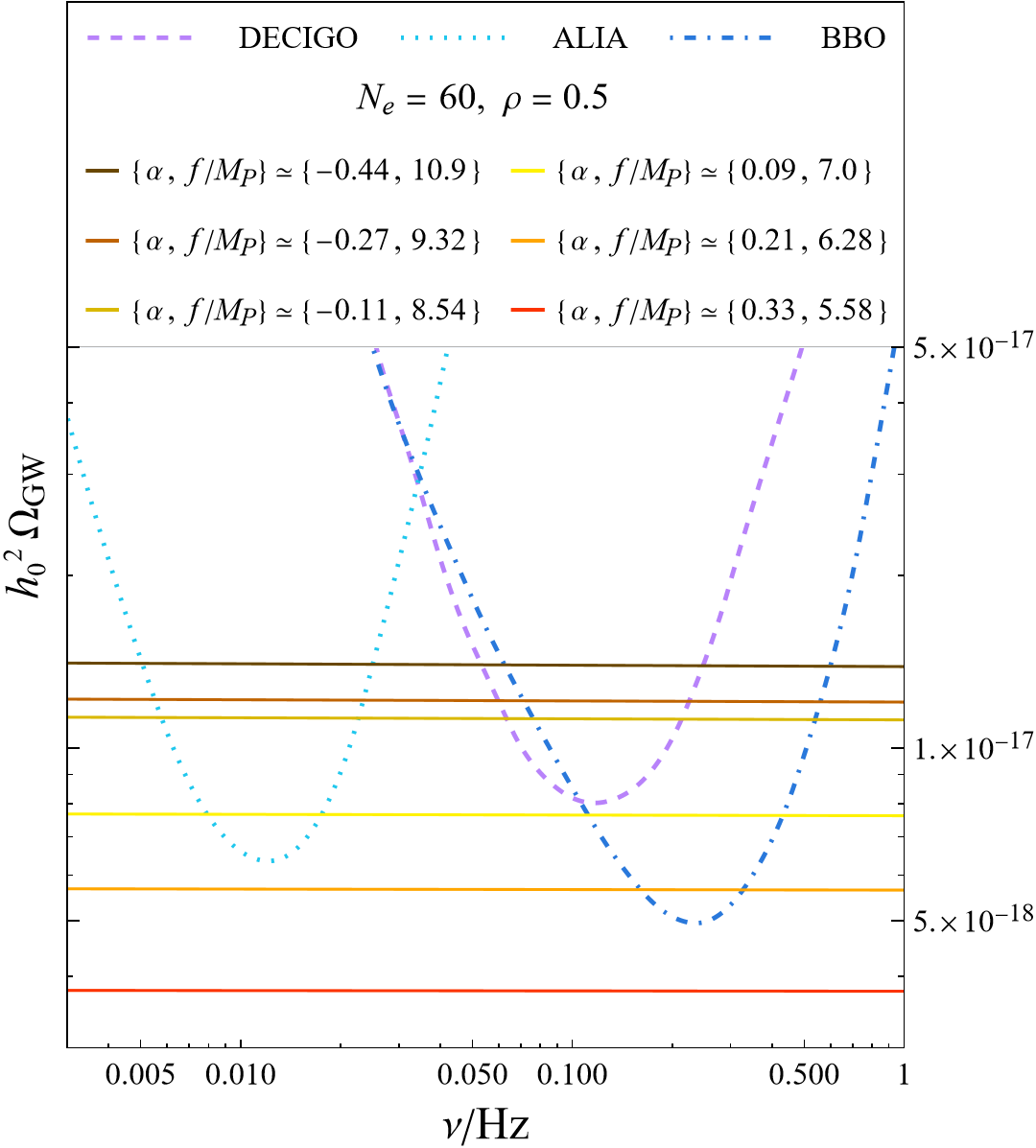}
    \caption{{\em The spectral density $h_0^2\Omega_\mathrm{GW}(\nu)$ of the GW inflationary background produced by the six setups depicted by the six benchmark points in the right plot of  Fig.~\ref{fig:f(alpha)_observable.setups_N=60_rho=0.5}, together with the three sensitivity curves of DECIGO, ALIA and BBO.}}
    \label{fig:GW.spectrum.1_N=60_rho=0.5}
\end{center}
\end{figure}

 For each natural-scalaron setup $\{N_e,\rho,\alpha,f\}$ allowed at least at $95\%$ $\mathrm{C.L.}$ by the most recent CMB observational constraints~\cite{Ade:2015lrj,BICEP:2021xfz} we evaluated $h_0^2\Omega_{\mathrm{GW}}(\nu)$ and checked whether the result falls above the sensitivity curves of DECIGO, BBO and ALIA for some $\nu$ (see Fig.~\ref{fig:GW.spectrum.1_N=60_rho=0.5}). If this is the case, the produced GW background is then potentially observable by DECIGO, BBO and ALIA, respectively. In Fig.~\ref{fig:f(alpha)_observable.setups_N=60_rho=0.5} we can see the result of this procedure for $\{N_e,\rho\}=\{60,0.5\}$. The green regions are just the allowed regions of Fig.~\ref{fig:f(alpha)_N=60_rho=0.5}, while the purple, light blue and blue lines represent the boundaries for GW observable signals: each point on the $\{\alpha,f\}$ plane which belongs to the green regions and lies above the purple, light blue or blue line represents an observationally admitted natural-scalaron setup (at least with $95\%$ $\mathrm{C.L.}$) whose produced relic inflationary GW background is potentially detectable by DECIGO, ALIA and BBO respectively (cf.~Fig.~\ref{fig:observable.models.N=60}).  It is remarkable that almost all the $\{N_e,\rho\}=\{60,0.5\}$ setups that are observationally admitted within $68\%$ $\mathrm{C.L.}$ generate a GW background that, at least in principle, is well observable by all three space-borne interferometers here considered, as well as the vast majority of setups admitted at $95\%$ $\mathrm{C.L.}$.  The right plot of Fig.~\ref{fig:f(alpha)_observable.setups_N=60_rho=0.5} also shows six benchmark points representing six distinct setups, all admitted within $68\%$ $\mathrm{C.L.}$: the first three setups $\{\alpha,f/M_P\}=\{ -0.44, 10.85\}$ (dark brown), $\{\alpha,f/M_P\}=\{-0.27, 9.32\}$ (brown), and $\{\alpha,f/M_P\}=\{ -0.11, 8.54\}$ (light brown) are all observable by all three interferometers; the fourth setup $\{\alpha,f/M_P\}=\{ 0.09, 7\}$ (yellow) is observable by ALIA and BBO; the fifth $\{\alpha,f/M_P\}=\{0.21, 6.28\}$ (orange) could only be observed from BBO; the last $\{\alpha,f/M_P\}=\{0.33, 5.58\}$ (red) cannot be observed by any of the three interferometers. In Fig.~\ref{fig:GW.spectrum.1_N=60_rho=0.5} we can see the GW spectral density $h_0^2\Omega_\mathrm{GW}(\nu)$ produced by these six natural-scalaron setups together with the three sensitivity curves of DECIGO, ALIA and BBO.  
 
 Comparing these findings with those of pure-natural inflation in Secs.~\ref{ch:Slow-roll.inflation}
 and~\ref{ch:the.relic.inflationary.GW.bakground}, we find that increasing $\rho$, on the one hand, improves the agreement with CMB data (i.e.~widens the observationally allowed regions), and, on the other hand, reduces the GW-observable regions of the parameter space (i.e.~the blue, light blue, and purple curves of Fig.~\ref{fig:f(alpha)_observable.setups_N=60_rho=0.5} rise as $\rho$ grows).

\begin{figure}[t!]
\begin{center}
    \includegraphics[width=0.45\textwidth]{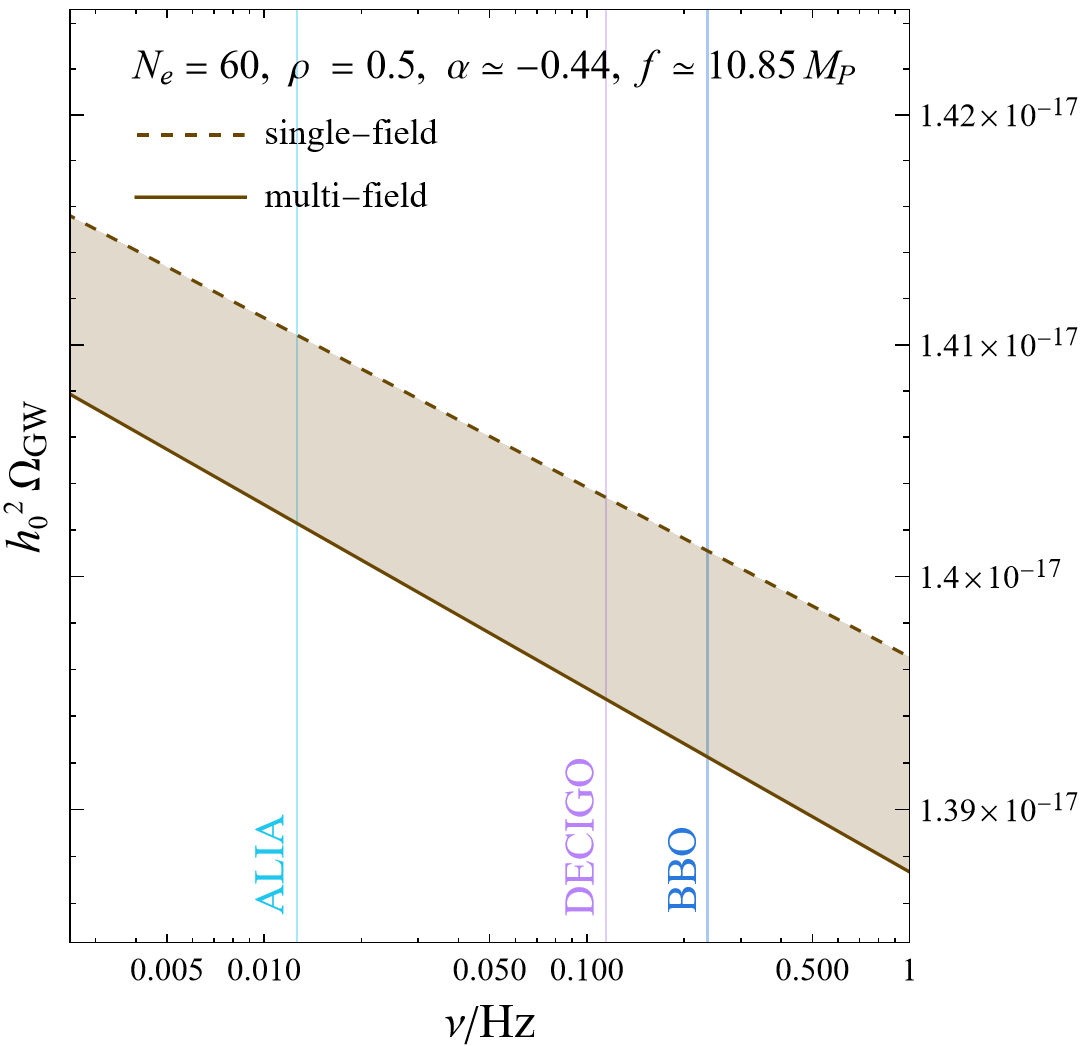} \\\includegraphics[width=0.45\textwidth]{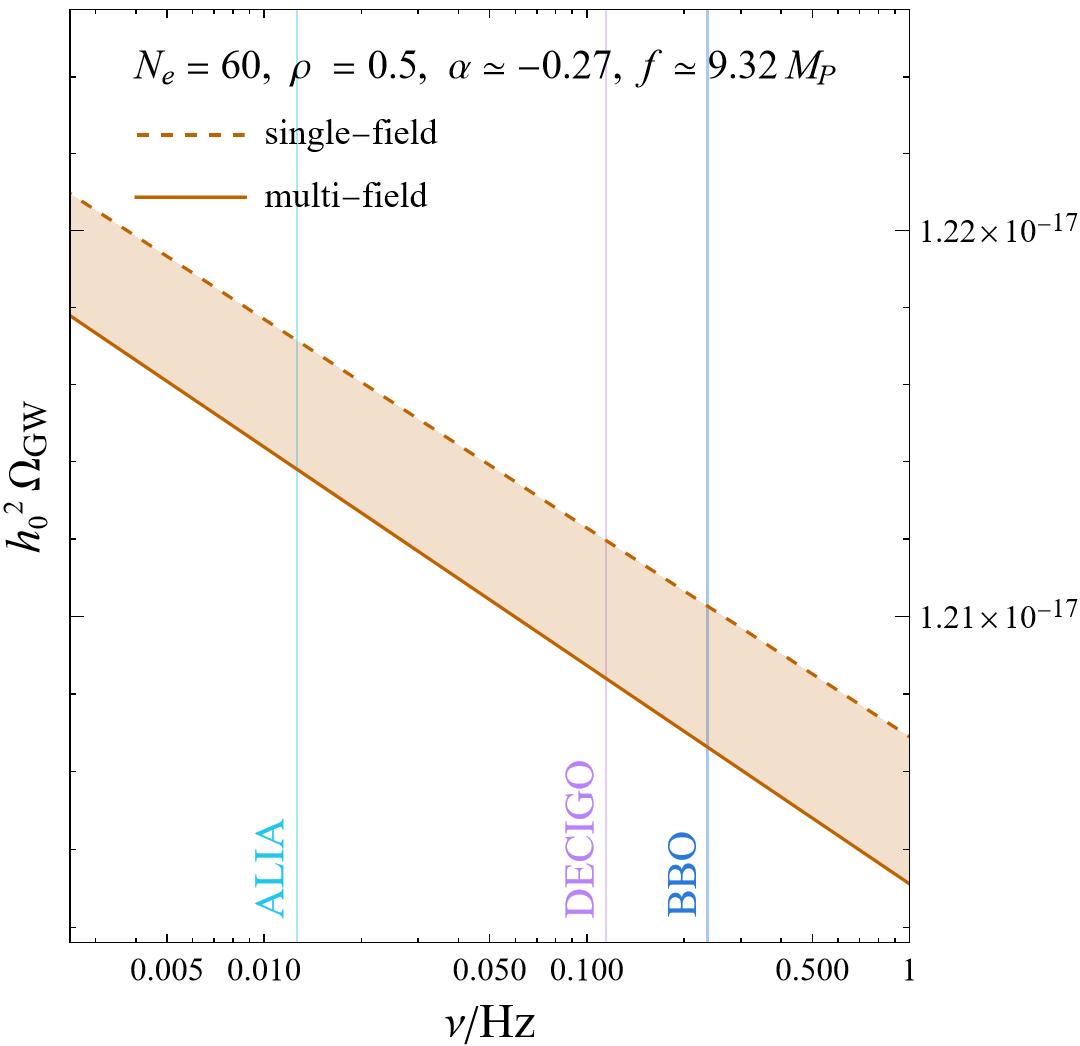}
    \hspace{0.1cm}
    \includegraphics[width=0.45\textwidth]{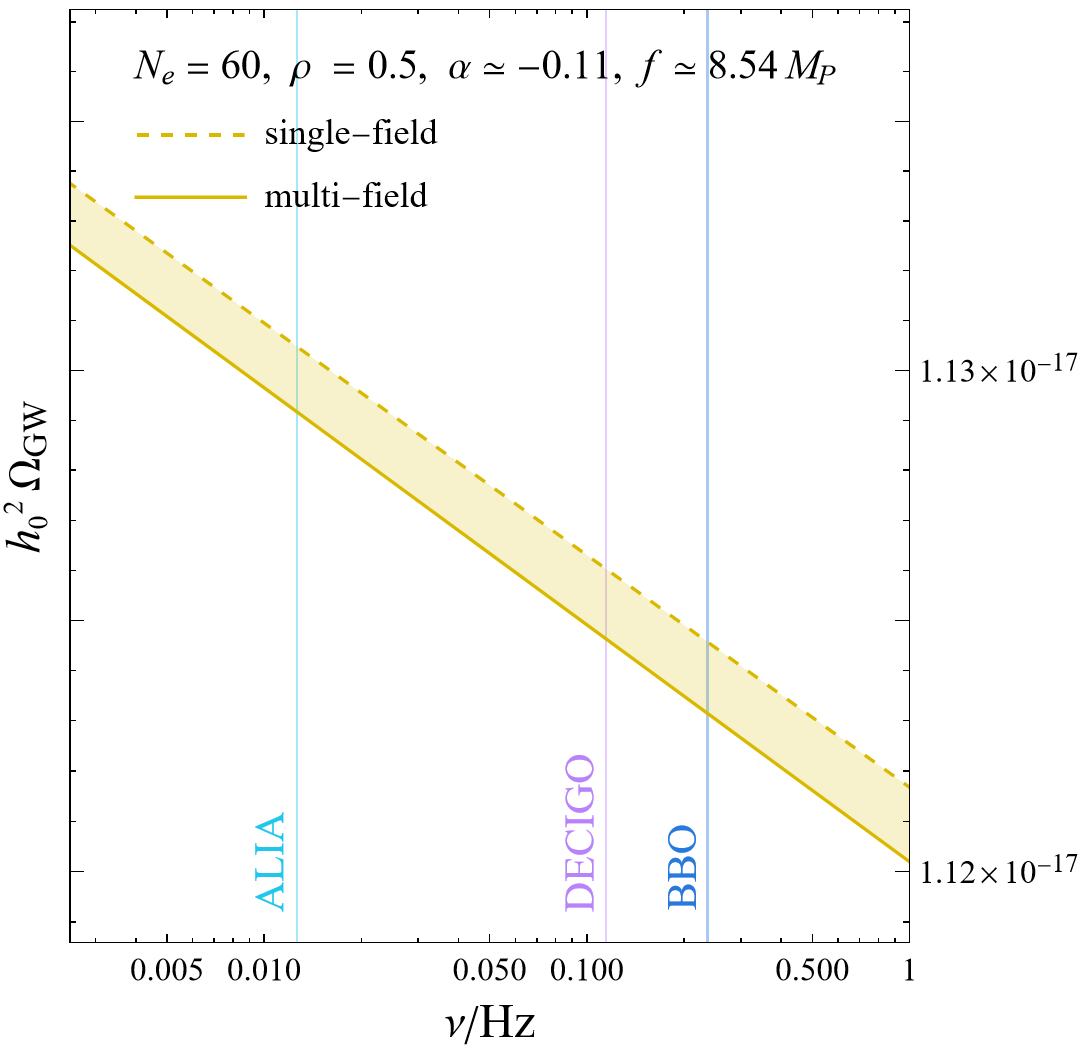}
    \caption{\em Comparison between the natural-scalaron GW spectral density $\Omega_\mathrm{GW}$ (solid line) and its corresponding `single-field' quantity $\Omega_\mathrm{GW}^\mathrm{\text{\,s-f}}$ (dashed line) (i.e.~between (\ref{eq:Omega_GW(nu)_Standard_Model_natural-scalaron}) and (\ref{eq:Omega_GW^s-f(nu)_Standard_Model_natural-scalaron})) for the dark-brown (upper plot) brown (left plot), and light-brown (right plot) setups appearing in Figs.~\ref{fig:f(alpha)_observable.setups_N=60_rho=0.5}-\ref{fig:GW.spectrum.1_N=60_rho=0.5}. The vertical blue, purple and light blue lines respectively mark the frequency at which the sensitivity curves of BBO, DECIGO and ALIA have their minima.}
    \label{fig:GW.spectrum.2_N=60_rho=0.5_1}
\end{center}
\end{figure}

Finally, for the three natural-scalaron setups 
$$\{N_e,\rho,\alpha,f/M_P\}= \quad \{60, 0.5, -0.44, 10.85\}, \quad \{60, 0.5, -0.27, 9.32\}, \quad\{60, 0.5, -0.11, 8.54\}$$ (which are respectively reported in dark brown, brown, and light brown in Figs.~\ref{fig:f(alpha)_observable.setups_N=60_rho=0.5}-\ref{fig:GW.spectrum.1_N=60_rho=0.5}) we also evaluated the corresponding single-field GW spectrum $h_0^2\Omega_\mathrm{GW}^{\,\text{s-f}}(\nu)$. Note that we specifically chose the value of $f$ of these three setups to coincide with the peaks of the corresponding $\delta r_{0.05}$ and $\delta n_T$ curves for each $\alpha$ (see Fig.~\ref{fig:delta.nT&delta.r_N=60_rho=0.5}), in order to maximize the difference between $\Omega_\mathrm{GW}(\nu)$ and $\Omega_\mathrm{GW}^{\,\text{s-f}}(\nu)$. In Fig.~\ref{fig:GW.spectrum.2_N=60_rho=0.5_1} we can see the comparison between $\Omega_\mathrm{GW}(\nu)$ (solid line) and $\Omega_\mathrm{GW}^{\,\text{s-f}}(\nu)$ (dashed line) for each of the three mentioned setups.  The dark brown setup shown in Fig.~\ref{fig:GW.spectrum.2_N=60_rho=0.5_1} is the one that exhibits the largest difference between the solid and dashed line. In fact, the difference between $\Omega_\mathrm{GW}(\nu)$ and $\Omega_\mathrm{GW}^{\,\text{s-f}}(\nu)$ decreases as $\alpha$ increases, as we can understand from Fig.~\ref{fig:delta.nT&delta.r_N=60_rho=0.5}. In the frequency range $0.005\,\mathrm{Hz} \lesssim \nu \lesssim 1\,\mathrm{Hz}$ we are interested in (where BBO, DECIGO and ALIA are expected to have the best sensitivity), the differences between $\Omega_\mathrm{GW}(\nu)$ and $\Omega_\mathrm{GW}^{\,\text{s-f}}(\nu)$ for the considered setups are: $\sim8\times10^{-20}$ (dark brown setup); $\sim3-4\times10^{-20}$ (brown setup); $\sim1-2\times10^{-20}$ (light brown setup). In other words, at most we expect a difference between single-field and multi-field natural inflation of the order of $h_0^2\Omega_\mathrm{GW}^\mathrm{\,diff}\sim10^{-19}$. However, both the solid and dashed lines of Fig.~\ref{fig:GW.spectrum.2_N=60_rho=0.5_1} are all above the sensitivities of all three considered space-borne interferometers (cf.~Fig.~\ref{fig:GW.spectrum.1_N=60_rho=0.5}) and 
$\Omega_\mathrm{GW}^\mathrm{\,diff}/\Omega_\mathrm{GW}\sim10^{-2}$ is the relative resolution 
the interferometers should have to distinguish {\it different} spectra which both fall within their sensitivity. We can, therefore, hope that future technological advancements will allow us to observationally distinguish between the single-field and multi-field cases. Now, another remark is in order: Fig.~\ref{fig:GW.spectrum.2_N=60_rho=0.5_1} shows that $\Omega_\mathrm{GW}(\nu)-\Omega_\mathrm{GW}^{\,\text{s-f}}(\nu)$ depends on the parameters $\rho$, $\alpha$, $f$ and $N_e$; therefore, in order to hope to distinguish between single- and multi-field inflation it is important that the  observations of BBO, DECIGO and ALIA will be accompanied by refined CMB observations that will be available (e.g.~through LiteBIRD~\cite{LiteBIRD:2022cnt}) and will be able to further constrain the parameter space.

 \section{Conclusions}\label{Conclusions}
 
 Let us now provide  a detailed summary of the novel results  obtained in this work.
 \begin{itemize}
 \item First, in Secs.~\ref{The model},~\ref{ch:Slow-roll.inflation} and~\ref{ch:the.natural.inflation.parameter.space} we performed a complete analysis of slow-roll inflation in the natural single-field model with a periodic potential and non-minimal coupling that admit a UV completion. This has allowed us to identify the  parameter space regions that are in agreement with the latest CMB observational constraints provided by the Planck, BICEP and Keck collaborations. While natural inflation without non-minimal coupling ($\alpha=0$) appears to be excluded, a non-zero value of $\alpha$ can lead to the agreement with the latest observational constraints.
 \item Second, in Sec.~\ref{ch:the.relic.inflationary.GW.bakground} we studied the relic natural inflationary background of GWs. There, we identified the parameter space regions that are allowed by the current CMB constraints and, remarkably at the same time,  correspond to a GW background accessible to future space-borne interferometers (DECIGO, BBO and ALIA) and those regions that cannot be probed by these future experiments.
 \item Moreover, in Sec.~\ref{A multi-field natural inflation: the natural-scalaron case} 
 we extended the analysis performed in the previous sections to a two-field version of the natural inflation model (the natural-scalaron one) obtained by adding an $R^2$ term. 
 \item Finally, in Sec.~\ref{ch:the.relic.inflationary.GW.background(natural-scalaron)} we examined the level of resolution future interferometers should possess to differentiate between GW signals, providing insights into the number of scalar fields contributing to natural inflation.  We found that the relative difference between the single-field  and two-field predictions for the GW spectral density can be $\Omega_\mathrm{GW}^\mathrm{\,diff}/\Omega_\mathrm{GW}\sim10^{-2}$ compatibly with Planck constraints on isocurvature modes.
  \end{itemize}

   We hope that the findings discussed in the last point will pave the way for distinguishing between single-field and multi-field inflation through GW observations in more general contexts, extending beyond natural inflation and its scalaron extension. In order for this hope to become reality, it is important that the considered models will still be allowed by future CMB constraints, such as those that will be performed by LiteBIRD, which will tell us more precisely the parameter values.

 \subsection*{Acknowledgments}
 This work has been partially supported by the grant DyConn from the University of Rome Tor Vergata. We thank Gianfranco Pradisi for useful discussions.
  
  \vspace{1cm}
 
\footnotesize
\begin{multicols}{2}

\end{multicols}


\begin{thebibliography}{}
\small{


  \bibitem{Abbott:2016blz} 
  B.~P.~Abbott {\it \textit{et al.}} [LIGO Scientific and Virgo Collaborations],
  ``Observation of Gravitational Waves from a Binary Black Hole Merger,''
  Phys.\ Rev.\ Lett.\  {\bf 116}, 061102 (2016)
  [\hhref{1602.03837}].

  
  
  
\bibitem{TheLIGOScientific:2016wyq1}
B.~Abbott \textit{\textit{et al.}} [LIGO Scientific and Virgo],
``GW150914: Implications for the stochastic gravitational wave background from binary black holes,''
Phys. Rev. Lett. \textbf{116}, no.13, 131102 (2016)
[\hhref{1602.03847}]. 

\bibitem{LIGOScientific:2017ync}
B.~P.~Abbott \textit{et al.},
``Multi-messenger Observations of a Binary Neutron Star Merger,''
Astrophys. J. Lett. \textbf{848} (2017) no.2, L12
[\hhref{1710.05833}].

\bibitem{NANOGrav:2023gor}
G.~Agazie \textit{et al.} [NANOGrav],
``The NANOGrav 15 yr Data Set: Evidence for a Gravitational-wave Background,''
Astrophys. J. Lett. \textbf{951} (2023) no.1, L8
[\hhref{2306.16213}].

\bibitem{Antoniadis:2023ott}
J.~Antoniadis, P.~Arumugam, S.~Arumugam, S.~Babak, M.~Bagchi, A.~S.~B.~Nielsen, C.~G.~Bassa, A.~Bathula, A.~Berthereau and M.~Bonetti, \textit{et al.}
``The second data release from the European Pulsar Timing Array III. Search for gravitational wave signals,''
[\hhref{2306.16214}].

\bibitem{Reardon:2023gzh}
D.~J.~Reardon, A.~Zic, R.~M.~Shannon, G.~B.~Hobbs, M.~Bailes, V.~Di Marco, A.~Kapur, A.~F.~Rogers, E.~Thrane and J.~Askew, \textit{et al.}
``Search for an Isotropic Gravitational-wave Background with the Parkes Pulsar Timing Array,''
Astrophys. J. Lett. \textbf{951} (2023) no.1, L6
[\hhref{2306.16215}].

\bibitem{Xu:2023wog}
H.~Xu, S.~Chen, Y.~Guo, J.~Jiang, B.~Wang, J.~Xu, Z.~Xue, R.~N.~Caballero, J.~Yuan and Y.~Xu, \textit{et al.}
``Searching for the Nano-Hertz Stochastic Gravitational Wave Background with the Chinese Pulsar Timing Array Data Release I,''
Res. Astron. Astrophys. \textbf{23} (2023) no.7, 075024
[\hhref{2306.16216}].

\bibitem{Salvio:2023qgb}
A.~Salvio,
``Model-independent radiative symmetry breaking and gravitational waves,"
JCAP \textbf{04} (2023), 051
[\hhref{2302.10212}].

\bibitem{Salvio:2023ynn}
A.~Salvio,
``Supercooling in Radiative Symmetry Breaking: Theory Extensions, Gravitational Wave Detection and Primordial Black Holes,''
[\hhref{2307.04694}].

\bibitem{Kajantie:1993ag}
K.~Kajantie, K.~Rummukainen and M.~E.~Shaposhnikov,
``A Lattice Monte Carlo study of the hot electroweak phase transition,''
Nucl. Phys. B \textbf{407} (1993), 356-372
[\hhref{hep-ph/9305345}].

\bibitem{Kajantie:1995kf}
K.~Kajantie, M.~Laine, K.~Rummukainen and M.~E.~Shaposhnikov,
``The Electroweak phase transition: A Nonperturbative analysis,''
Nucl. Phys. B \textbf{466} (1996), 189-258
[\hhref{hep-lat/9510020}].
%

\bibitem{Kajantie:1996mn}
K.~Kajantie, M.~Laine, K.~Rummukainen and M.~E.~Shaposhnikov,
``Is there a~ hot electroweak phase transition at $m_H \gtrsim m_W$?,''
Phys. Rev. Lett. \textbf{77} (1996), 2887-2890
[\hhref{hep-ph/9605288}].

\bibitem{Gould:2019qek}
O.~Gould, J.~Kozaczuk, L.~Niemi, M.~J.~Ramsey-Musolf, T.~V.~I.~Tenkanen and D.~J.~Weir,
``Nonperturbative analysis of the gravitational waves from a first-order electroweak phase transition,''
Phys. Rev. D \textbf{100} (2019) no.11, 115024
[\hhref{1903.11604}].

\bibitem{Gould:2022ran}
O.~Gould, S.~G\"uyer and K.~Rummukainen,
``First-order electroweak phase transitions: A nonperturbative update,''
Phys. Rev. D \textbf{106} (2022) no.11, 114507
[\hhref{2205.07238}].

\bibitem{Vilenkin:2000jqa}
A.~Vilenkin and E.~P.~S.~Shellard,
``Cosmic Strings and Other Topological Defects,''
Cambridge University Press, 2000,
ISBN 978-0-521-65476-0.


\bibitem{DECIGO}
        S.~Kawamura \textit{\textit{et al.}},
Class. Quant. Grav. \textbf{23} (2006), S125-S132.
M. Musha [DECIGO Working group], ``Space gravitational wave detector DECIGO/pre-DECIGO," Proc. SPIE Int. Soc. Opt. Eng. 10562, 105623T (2017).
S.~Kawamura  \textit{\textit{et al.}},
``Current status of space gravitational wave antenna DECIGO and B-DECIGO,''
[\hhref{2006.13545}].
See also the \href{https://decigo.jp/index_E.html}{DECIGO webpage}.



\bibitem{Crowder:2005nr}
J.~Crowder and N.~J.~Cornish,
``Beyond LISA: Exploring future GW missions,''
Phys. Rev. D \textbf{72}, 083005 (2005)
[\hhref{gr-qc/0506015}].

\bibitem{BBO2}
G.~Harry, P.~Fritschel, D.~Shaddock, W.~Folkner and E.~Phinney,
``Laser interferometry for the big bang observer,''
Class. Quant. Grav. \textbf{23}, 4887-4894 (2006).
V.~Corbin and N.~J.~Cornish,
``Detecting the cosmic GW background with the big bang observer,''
Class. Quant. Grav. \textbf{23}, 2435-2446 (2006)
[\hhref{gr-qc/0512039}].

\bibitem{Gong:2014mca}
X.~Gong, Y.~K.~Lau, S.~Xu, P.~Amaro-Seoane, S.~Bai, X.~Bian, Z.~Cao, G.~Chen, X.~Chen and Y.~Ding, \textit{et al.}
``Descope of the ALIA mission,''
J. Phys. Conf. Ser. \textbf{610} (2015) no.1, 012011
[\hhref{1410.7296}].

\bibitem{Bezrukov:2007ep}
  F.~L.~Bezrukov and M.~Shaposhnikov,
  ``The Standard Model Higgs boson as the inflaton,''
  Phys.\ Lett.\ B {\bf 659} (2008) 703
  [\hhref{0710.3755}].
  
  
\bibitem{Hamada:2014iga}
  Y.~Hamada, H.~Kawai, K.~y.~Oda and S.~C.~Park,
  ``Higgs Inflation is Still Alive after the Results from BICEP2,''
  Phys.\ Rev.\ Lett.\  {\bf 112} (2014) no.24,  241301
  [\hhref{1403.5043}].
  
\bibitem{Bezrukov:2014bra}
  F.~Bezrukov and M.~Shaposhnikov,
  ``Higgs inflation at the critical point,''
  Phys.\ Lett.\ B {\bf 734} (2014) 249
  [\hhref{1403.6078}].

\bibitem{Hamada:2014wna}
  Y.~Hamada, H.~Kawai, K.~y.~Oda and S.~C.~Park,
  ``Higgs inflation from Standard Model criticality,
  Phys.\ Rev.\ D {\bf 91} (2015) 053008
  [\hhref{1408.4864}].

\bibitem{Salvio:2018rv}
A.~Salvio,
``Critical Higgs inflation in a Viable Motivated Model,''
Phys. Rev. D \textbf{99} (2019) no.1, 015037
[\hhref{1810.00792}].


\bibitem{Salvio:2021kya}
A.~Salvio,
``Hearing Higgs with gravitational wave detectors,''
JCAP \textbf{06} (2021), 040
[\hhref{2104.12783}].

\bibitem{Freese:1990rb}
  K.~Freese, J.~A.~Frieman and A.~V.~Olinto,
  ``Natural inflation with pseudo - Nambu-Goldstone bosons,''
  Phys.\ Rev.\ Lett.\  {\bf 65} (1990) 3233.
F.~C.~Adams, J.~R.~Bond, K.~Freese, J.~A.~Frieman and A.~V.~Olinto,
  ``Natural inflation: Particle physics models, power law spectra for large scale structure, and constraints from COBE,''
  Phys.\ Rev.\ D {\bf 47} (1993) 426
  [\hhref{hep-ph/9207245}].
  
\bibitem{Pajer:2013fsa}
E.~Pajer and M.~Peloso,
``A review of Axion Inflation in the era of Planck,''
Class. Quant. Grav. \textbf{30}, 214002 (2013)
[\hhref{1305.3557}]

\bibitem{Salvio:2021lka}
A.~Salvio,
``Natural-Scalaron Inflation,''
JCAP \textbf{10} (2021), 011
[\hhref{2107.03389}].

\bibitem{AlHallak:2022gbv}
M.~AlHallak, N.~Chamoun and M.~S.~Eldaher,
``Natural Inflation with non minimal coupling to gravity in R $^{2}$ gravity under the Palatini formalism,''
JCAP \textbf{10} (2022), 001
[\hhref{2202.01002}].

\bibitem{Bostan:2022swq}
N.~Bostan,
``Non-minimally coupled Natural Inflation: Palatini and Metric formalism with the recent BICEP/Keck,''
JCAP \textbf{02} (2023), 063
[\hhref{2209.02434}].



\bibitem{AlHallak:2022haa}
M.~AlHallak, K.~K.~A.~Said, N.~Chamoun and M.~S.~El-Daher,
``On Warm Natural Inflation and Planck 2018 Constraints,''
Universe \textbf{9} (2023) no.2, 80
[\hhref{2211.07775}].

\bibitem{Bostan:2023ped}
N.~Bostan and S.~Roy Choudhury,
``First constraints on Non-minimally coupled Natural and Coleman-Weinberg inflation in the light of massive neutrino self-interactions and Planck+BICEP/Keck,''
[\hhref{2310.01491}].


\bibitem{Ade:2015lrj}
  P.~A.~R.~Ade {\it et al.} [Planck Collaboration],
  ``Planck 2015 results. XX. Constraints on inflation,''
  Astron.\ Astrophys.\  {\bf 594} (2016) A20 
  [\hhref{1502.02114}].
  Y.~Akrami {\it et al.} [Planck Collaboration],
  ``Planck 2018 results. X. Constraints on inflation,''
  Astron. Astrophys. \textbf{641} (2020), A10
  [\hhref{1807.06211}].

\bibitem{BICEP:2021xfz}
P.~A.~R.~Ade \textit{et al.} [BICEP and Keck],
``Improved Constraints on Primordial Gravitational Waves using Planck, WMAP, and BICEP/Keck Observations through the 2018 Observing Season,''
Phys. Rev. Lett. \textbf{127} (2021) no.15, 151301
[\hhref{2110.00483}].

\bibitem{Salvio:2019wcp}
A.~Salvio,
``Quasi-Conformal Models and the Early Universe,''
Eur. Phys. J. C \textbf{79} (2019) no.9, 750
[\hhref{1907.00983}].

\bibitem{Salvio:2020axm}
A.~Salvio,
``Dimensional Transmutation in Gravity and Cosmology,''
Int. J. Mod. Phys. A \textbf{36} (2021) no.08n09, 2130006
[\hhref{2012.11608}].

\bibitem{Ferreira:2018nav}
  R.~Z.~Ferreira, A.~Notari and G.~Simeon,
  ``Natural Inflation with a periodic non-minimal coupling,''
  JCAP {\bf 1811} (2018) no.11,  021
  [\hhref{1806.05511}].
  G.~Simeon,
``Scalar-tensor extension of Natural Inflation,''
JCAP \textbf{07}, 028 (2020)
[\hhref{2002.07625}].



\bibitem{Utiyama:1962sn}
  R.~Utiyama and B.~S.~DeWitt,
  ``Renormalization of a classical gravitational field interacting with quantized matter fields,''
  J.\ Math.\ Phys.\  {\bf 3} (1962) 608.
  %
  
\bibitem{Salvio:2014soa}
A.~Salvio and A.~Strumia,
``Agravity,''
JHEP \textbf{06} (2014), 080
[\hhref{1403.4226}].

\bibitem{Salvio:2015kka}
A.~Salvio and A.~Mazumdar,
``Classical and Quantum Initial Conditions for Higgs Inflation,''
Phys. Lett. B \textbf{750} (2015), 194-200
[\hhref{1506.07520}].

\bibitem{Salvio:2017qkx}
A.~Salvio and A.~Strumia,
``Agravity up to infinite energy,''
Eur. Phys. J. C \textbf{78} (2018) no.2, 124
[\hhref{1705.03896}].

    \bibitem{Starobinsky:1980te} 
  A.~A.~Starobinsky,
  ``A New Type of Isotropic Cosmological Models Without Singularity,''
  Phys.\ Lett.\ B {\bf 91}, 99 (1980).
  
\bibitem{Liddle:2003as}
A.~R.~Liddle and S.~M.~Leach,
``How long before the end of inflation were observable perturbations produced?,''
Phys. Rev. D \textbf{68} (2003), 103503
[\hhref{astro-ph/0305263}].

\bibitem{Maggiore:2018sht}
M.~Maggiore,
``Gravitational Waves. Vol. 2: Astrophysics and Cosmology,'' Oxford University Press, 3, 2018.
  





\bibitem{Bennett:2019ewm}
J.~J.~Bennett, G.~Buldgen, M.~Drewes and Y.~Y.~Y.~Wong,
``Towards a precision calculation of the effective number of neutrinos $N_{\rm eff}$ in the Standard Model I: the QED equation of state,''
JCAP \textbf{03} (2020), 003
[\hhref{1911.04504}].

\bibitem{Akita:2020szl}
K.~Akita and M.~Yamaguchi,
``A precision calculation of relic neutrino decoupling,''
JCAP \textbf{08} (2020), 012
[\hhref{2005.07047}].

\bibitem{Froustey:2020mcq}
J.~Froustey, C.~Pitrou and M.~C.~Volpe,
``Neutrino decoupling including flavour oscillations and primordial nucleosynthesis,''
JCAP \textbf{12} (2020), 015
[\hhref{2008.01074}].


\bibitem{Bennett:2020zkv}
J.~J.~Bennett, G.~Buldgen, P.~F.~De Salas, M.~Drewes, S.~Gariazzo, S.~Pastor and Y.~Y.~Y.~Wong,
``Towards a precision calculation of $N_{\rm eff}$ in the Standard Model II: Neutrino decoupling in the presence of flavour oscillations and finite-temperature QED,''
JCAP \textbf{04} (2021), 073
[\hhref{2012.02726}].

\bibitem{Esteban:2020cvm}
I.~Esteban, M.~C.~Gonzalez-Garcia, M.~Maltoni, T.~Schwetz and A.~Zhou,
``The fate of hints: updated global analysis of three-flavor neutrino oscillations,''
JHEP \textbf{09}, 178 (2020)
[\hhref{2007.14792}].

\bibitem{deSalas:2020pgw}
P.~F.~de Salas, D.~V.~Forero, S.~Gariazzo, P.~Mart\'\i{}nez-Mirav\'e, O.~Mena, C.~A.~Ternes, M.~T\'ortola and J.~W.~F.~Valle,
``2020 Global reassessment of the neutrino oscillation picture,''
[\hhref{2006.11237}].



\bibitem{Thrane:2013oya}
E.~Thrane and J.~D.~Romano,
``Sensitivity curves for searches for gravitational-wave backgrounds,''
Phys. Rev. D \textbf{88} (2013) no.12, 124032
[\hhref{1310.5300}].

\bibitem{Dev:2019njv}
P.~S.~B.~Dev, F.~Ferrer, Y.~Zhang and Y.~Zhang,
``Gravitational Waves from First-Order Phase Transition in a Simple Axion-Like Particle Model,''
JCAP \textbf{11} (2019), 006
[\hhref{1905.00891}].

\bibitem{Salvio:2022mld}
A.~Salvio,
``BICEP/Keck data and quadratic gravity,''
JCAP \textbf{09} (2022), 027
[\hhref{2202.00684}].

\bibitem{Gundhi:2018wyz}
A.~Gundhi and C.~F.~Steinwachs,
``Scalaron-Higgs inflation,''
Nucl. Phys. B \textbf{954}, 114989 (2020)
[\hhref{1810.10546}].

\bibitem{Chiba:2008rp}
  T.~Chiba and M.~Yamaguchi,
  ``Extended Slow-Roll Conditions and Primordial Fluctuations: Multiple Scalar Fields and Generalized Gravity,''
  JCAP {\bf 0901} (2009) 019
  [\hhref{0810.5387}].

\bibitem{Sasaki:1995aw}
  M.~Sasaki and E.~D.~Stewart,
  ``A General analytic formula for the spectral index of the density perturbations produced during inflation,''
  Prog.\ Theor.\ Phys.\  {\bf 95} (1996) 71
  [\hhref{astro-ph/9507001}].
    
\bibitem{Kaiser:2012ak}
D.~I.~Kaiser, E.~A.~Mazenc and E.~I.~Sfakianakis,
``Primordial Bispectrum from Multifield Inflation with Nonminimal Couplings,''
Phys. Rev. D \textbf{87}, 064004 (2013)
[\hhref{1210.7487}].


\bibitem{Gordon:2000hv}
C.~Gordon, D.~Wands, B.~A.~Bassett and R.~Maartens,
``Adiabatic and entropy perturbations from inflation,''
Phys. Rev. D \textbf{63}, 023506 (2000)
[\hhref{astro-ph/0009131}].

\bibitem{LiteBIRD:2022cnt}
E.~Allys \textit{et al.} [LiteBIRD],
``Probing Cosmic Inflation with the LiteBIRD Cosmic Microwave Background Polarization Survey,''
[\hhref{2202.02773}].

   }
  

\end{thebibliography}
\end{document}